\documentclass[11pt]{article}
\pdfoutput=1

\usepackage{jheppub}
\usepackage[T1]{fontenc} 

\usepackage{longtable}
\usepackage[vcentermath]{youngtab}
\usepackage{array,multirow}
 \usepackage{multirow}
\usepackage{amssymb}
\usepackage{amsfonts}
\usepackage{amsbsy}
\usepackage{amsmath}
\usepackage{amsthm}
\usepackage{graphicx}
\usepackage{epstopdf}
\usepackage{color}
\usepackage{hyperref}
\usepackage{wasysym}

\makeatletter
\g@addto@macro\bfseries{\boldmath}
\def\NAT@sort{\z@}
\makeatother

\newtheorem{conjecture}{Conjecture}

\newcommand{\set}[1]{\{#1\}}

\def \be {\begin{equation}}
\def \ee {\end{equation}}
\def \bsp {\begin{split}}
\def \esp {\end{split}}
\def \bea {\begin{eqnarray}}
\def \eea {\end{eqnarray}}

\def\Z{\mathbb{Z}}
\def\F{\mathbb{F}}
\def\Q{\mathbb{Q}}

\def\R{\mathbb{R}}

\def\P{\mathbb{P}}

\def\n3a{t}

\DeclareMathOperator{\Hom}{Hom}

\newcommand{\ds}{\ensuremath{\Delta}}
\newcommand{\dd}{\ensuremath{\nabla}}

\newcommand{\vb}{\ensuremath{v^{(B)}} }
\newcommand{\m}{\ensuremath{m^{(2)}} }
\newcommand{\vF}{\ensuremath{v^{(F)}} }
\newcommand{\mF}{\ensuremath{m^{(F)}} }

\definecolor{Red}{rgb}{1.00, 0.00, 0.00}

\newcommand{\yw}[1]{{\color{white}#1}}

\title{
On the prevalence of elliptic and genus one fibrations among toric
hypersurface Calabi-Yau threefolds
}
\author{Yu-Chien Huang,}
\author{Washington Taylor}

\affiliation{Center for Theoretical Physics\\
Department of Physics\\
Massachusetts Institute of Technology\\
77 Massachusetts Avenue\\
Cambridge, MA 02139, USA}

\emailAdd{{\tt yc\_huang} {\rm at} {\tt mit.edu}}
\emailAdd{{\tt wati} {\rm at} {\tt mit.edu}}

\preprint{MIT-CTP-4993}

\abstract{
We systematically analyze
the  fibration structure of toric hypersurface Calabi-Yau threefolds
with large and small Hodge numbers.
We show that there are only four such Calabi-Yau threefolds with
$h^{1, 1} \geq 140$ or $h^{2, 1} \geq 140$ that do not have manifest
elliptic or genus one fibers  arising from a fibration  of the
associated 4D polytope.  There is a
genus one fibration whenever
either Hodge number is 150 or greater, and an elliptic fibration when
either Hodge number is   228 or greater. We  find that for small
$h^{1, 1}$ the fraction of polytopes in the KS database that do not
have a genus one or elliptic fibration drops exponentially
as $h^{1,1}$  increases.
We also consider the different toric fiber types that arise in the
polytopes of elliptic Calabi-Yau threefolds.}

\begin{document}
\maketitle

\flushbottom


\section{Introduction}

Calabi-Yau manifolds play a central role in string theory; these
geometric spaces can describe extra dimensions of space-time in
supersymmetric ``compactifications'' of the theory.
The analysis of Calabi-Yau manifolds has been a major focus of the work of
mathematicians and physicists since this connection was first
understood
\cite{chsw}.
Nonetheless, it is still not known whether the number of distinct
topological types of Calabi-Yau threefolds is finite or infinite.
A large class of Calabi-Yau threefolds can be described as
hypersurfaces in toric varieties; these were systematically classified
by Kreuzer and Skarke \cite{Kreuzer:2000xy, database} and represent most of the
explicitly known Calabi-Yau threefolds at large Hodge numbers.

A specific class of Calabi-Yau manifolds that are of particular
mathematical and physical interest are those that admit
a genus one or elliptic
fibration (an elliptic fibration is a genus one fibration
with a global section).  
Elliptically fibered Calabi-Yau manifolds have additional structure
that makes them easier to understand mathematically, and they play a
central role in the approach to string theory known as ``F-theory''
\cite{Vafa-F-theory, Morrison-Vafa}.
Genus one fibrations are also relevant in  F-theory in the context of
discrete gauge groups, as described in
e.g. \cite{Braun:2014oya, Morrison-WT-sections, aggk,
  Mayrhofer:2014opa, Cvetic:2015moa}; see
\cite{Timo-TASI, Cvetic:2018bni} for further background and references on
this and other F-theory-related issues.
Unlike  the general class of Calabi-Yau threefolds, it is known that
the number of distinct topological types of elliptic  and genus one
Calabi-Yau
threefolds is finite \cite{Gross} (See also \cite{Grassi} for earlier
work that laid the foundation for this proof, and \cite{KMT-2} for a
more constructive and explicit argument for finiteness).
In recent years, an
increasing body of circumstantial evidence has suggested that in fact
a large fraction of the known Calabi-Yau manifolds admit an elliptic
or genus one fibration.  A direct analysis of the related structure of
K3 fibrations for many of the toric hypersurface constructions in the
Kreuzer-Skarke database was carried out in \cite{Candelas-cs},
demonstrating directly the prevalence of fibrations by
smaller-dimensional Calabi-Yau fibers among known Calabi-Yau
threefolds.  The study of F-theory has led to a systematic methodology
for constructing and classifying elliptic Calabi-Yau threefolds
\cite{clusters, toric, Hodge, Wang-WT, Johnson-WT,
  Johnson:2016qar}. Comparing the structure of geometries constructed
in this way to the Kreuzer-Skarke database shows that at large Hodge
numbers, virtually all Calabi-Yau threefolds that are known are in
fact elliptic.  In a companion paper to this one
\cite{Huang-Taylor-long}, we use this approach to show that all Hodge
numbers with $h^{1, 1}$ or $h^{2, 1}$ greater or equal to 240 that
arise in the Kreuzer-Skarke database are realized explicitly by
elliptic fibration constructions over toric or related base surfaces.
Finally, from a somewhat different point of view the analysis of
complete intersection Calabi-Yau manifolds and generalizations thereof
has shown that these classes of Calabi-Yau threefolds and fourfolds
are also overwhelmingly dominated by elliptic and genus one fibrations
\cite{Gray-hl1, Gray-hl, Anderson-aggl, Anderson-ggl, aggl-2, aggl-3}.

In this paper we carry out a direct analysis of the toric hypersurface
Calabi-Yau manifolds in the Kreuzer-Skarke database.  There are 16
reflexive 2D polytopes that can act as fibers of a 4D polytope
describing a Calabi-Yau threefold; the presence of any of these fibers
in the 4D polytope indicates that the corresponding Calabi-Yau
threefold hypersurface is genus one or elliptically fibered.  We systematically consider all
polytopes in the Kreuzer-Skarke database that are associated with
Calabi-Yau threefolds with one or both Hodge numbers at least 140.  We
show that with only four exceptions these Calabi-Yau threefolds all
admit an explicit elliptic or more general genus one fibration that can be seen
from the toric structure of the polytope.  We furthermore find that
for toric hypersurface Calabi-Yau threefolds with small $h^{1,1}$, the fraction that lack a
genus one or elliptic fibration decreases roughly exponentially with
$h^{1, 1}$.  Together these
results strongly support the notion that genus one and elliptic
fibrations are quite generic among Calabi-Yau threefolds.

The outline of this paper is as follows: In Section \ref{sec:fibers}
we describe the 16 types of toric fibers of the polytope that can lead
to a
genus one or elliptic fibration of the hypersurface Calabi-Yau and our
methodology for analyzing the fibration structure of the polytopes.
In Section \ref{sec:results}, we give our results on those Calabi-Yau
threefolds with the largest Hodge numbers that do not admit an
explicit elliptic or genus one fibration in the polytope description, as well as
some results on the
 distribution of fiber types  and multiple
 fibrations.
In Section  \ref{sec:prevalence} we discuss some simple aspects of the
likelihood of the existence of fibrations and compare to the observed
frequency of fibrations in the KS database at small $h^{1,1}$.
 Section \ref{sec:conclusions} contains some concluding
remarks.

Along with this paper, we are making
the results of the fiber analysis of polytopes in the Kreuzer-Skarke
database associated with Calabi-Yau threefolds having Hodge numbers
$h^{1, 1}\geq 140$ or $h^{2, 1}\geq 140$  available in Mathematica
form
\cite{data}.

\section{Identifying toric fibers} 
\label{sec:fibers}

A  fairly comprehensive introductory review of the toric hypersurface
construction and how elliptic fibrations are described in this context
is given in the companion paper \cite{Huang-Taylor-long}, in which we
describe in much more detail the structure of the elliptic fibrations
for Calabi-Yau threefolds $X$ with very large Hodge numbers ($h^{1, 1}
(X)\geq 240$ or $h^{2, 1} (X)\geq 240$).  Here we give only a very brief summary of the
essential points.

\subsection{Toric hypersurfaces and the 16 reflexive 2D fibers}

The basic framework for understanding Calabi-Yau manifolds through
hypersurfaces in toric varieties was developed by Batyrev
\cite{Batyrev}.
A {\it lattice polytope} 
$\dd$
is defined to be the set of lattice points in
$N =\Z^n$ that are contained within the convex hull of a finite set of
vertices $v_i \in N$.
 The dual of a polytope $\nabla$ is defined to be 
\begin{equation}
\nabla^*=\{u\in M_\R=  M\otimes \R: \langle u,v\rangle\geq-1, \forall v\in \nabla\},
\label{dual}
\end{equation}
where $M = N^*=\Hom(N,\Z)$ is the dual lattice.  A lattice polytope
$\nabla\subset N$ containing the origin is {\it reflexive} if its dual
polytope is also a lattice polytope.
When $\dd$ is reflexive, we denote the dual polytope by $\ds =\dd^*$.
The elements of the dual polytope $\ds$ can be associated with
monomials in a section of the anti-canonical bundle of a toric variety
associated to $\dd$.   A section of this bundle defines a hypersurface
in $\dd$ that is a Calabi-Yau manifold of dimension $n-1$.

When the polytope $\dd$ has a 2D subpolytope $\dd_2$ that is also
reflexive, the associated Calabi-Yau manifold has a genus one
fibration.  There are 16 distinct reflexive 2D polytopes, listed in
Appendix~\ref{sec:appendix-fibers}. 
These fibers are analyzed in the language of polytope ``tops''
\cite{tops} in \cite{Bouchard-Skarke}.
The structure of the genus one and elliptic fibrations
associated with each of these 16 fibers is studied in some detail in
the F-theory context in \cite{Braun:2011ux, BGK-geometric, Klevers-16}.

Of the 16 reflexive 2D polytopes listed in Appendix \ref{sec:appendix-fibers}, 13 are
always associated with elliptic fibrations.  This can be seen, following
\cite{BGK-geometric},  by observing that the anticanonical class $- K_2$
of the toric 2D variety associated with a given $\dd_2$ is $\sum C_i$
where $C_i$ are the toric curves associated with
rays in a toric fan for $\dd_2$.  The intersection of the curve $C_i$ with
the genus one fiber associated with the vanishing locus of a section
of $- K_2$ is thus $C_i \cdot (- K_2) = 2+ C_i \cdot C_i$, so $C_i$
defines a section associated with a single point on a generic fiber
only for a curve of self-intersection $C_i \cdot C_i = -1$.  The three
fibers $F_1, F_2, F_4$ are associated with the
weak Fano surfaces $\P^2,\F_0 =\P^1 \times\P^1$, and $\F_2 =\P^2 [1, 1, 2]$,
which have no $- 1$ curves, while the other 13 fibers $F_i$ all have $- 1$ curves.  Thus, polytopes $\dd$ with any fiber
$\dd_2$ that is $F_n, n \notin\{1, 2, 4\}$ give CY3s with elliptic
fibrations, while those $\dd$ with only fibers of types $F_1, F_2,
F_4$ are genus one fibered but may not be elliptically fibered.

The basic goal of this paper is a systematic scan through the
Kreuzer-Skarke database to determine which reflexive polytopes
associated with Calabi-Yau threefolds that have large  Hodge numbers
or small $h^{1,1}$
have toric reflexive 2D fibers that indicate the existence of an
elliptic or genus one fibration for the associated Calabi-Yau
threefold.
Note that this analysis only identifies elliptic and genus one
fibrations that are manifest in the polytope structure.  As discussed
further in \S\ref{sec:prevalence}, a more comprehensive analysis of
the fibration structure of a given Calabi-Yau threefold can be carried
out using methods analogous to those used in \cite{aggl-3}.

\subsection{Algorithm for checking a polytope for fibrations}
\label{sec:algorithm}

We use a similar algorithm to that we used in
\cite{Huang-Taylor-long} to check for reflexive 2D fibers of a 4D reflexive polytope.  Except for a small tweak to optimize
efficiency, this is essentially the approach outlined in
\cite{BGK-geometric}.
The basic idea is to check a given polytope
for each of the possible 16 reflexive subpolytopes.  For a given
polytope $\dd$ and potential fiber polytope $\dd_2$, we
proceed in the
following two steps:

\begin{enumerate}
\item  To increase the efficiency of the analysis we start by determining
the subset $S$
of the lattice points in $\dd$ that could possibly be contained in a
fiber of the form $\dd_2$, using a simple criterion.  For each fixed
fiber type $\dd_2$, there is a maximum possible value $I_{\rm max}$
of the inner
product $\vF \cdot m$ for any $\vF \in\dd_2, m
\in\ds_2$.  
For example, for the 2D $\P^{2,3,1}$ polytope $(F_{10})$,
$I_\text{max} = 5$.
The values of $I_\text{max}$ for each of the reflexive 2D polytopes
$\nabla_2$ are listed in Appendix~\ref{sec:appendix-fibers}.  When
$\dd_2$ is a fiber of $\dd$, which implies that there is a
projection from $\ds$ to $\ds_2$, $I_{\rm max}$ is also the maximum
possible value of the inner product $\vF \cdot m$ for any $m \in\ds$.
Thus,
we define the set $S$ to be the set of lattice points $v \in\dd$ such
that $v \cdot m \leq I_\text{max}$ for all vertices $m$ of $\ds$.  Particularly
for polytopes $\dd$ that contain many lattice points, generally
associated with Calabi-Yau threefolds with large $h^{1, 1}$, this step
significantly decreases the time needed for the algorithm.

\item  We then consider each pair of vectors $v, w$ in $S$ and check if the
intersection of $\dd$ with the plane spanned by $v, w$ consists of
precisely a set of lattice points that define the 2D polytope
$\dd_2$.  If such a pair of vectors exists then $\dd$ has a fiber
$\dd_2$ and the associated Calabi-Yau threefold has an  elliptic fibration
structure defined by this fiber type. 
\end{enumerate}

In practice,
we implement
these steps directly only to check for the presence of the minimal
2D subpolytopes $F_1, F_2, F_4$ within a 2D plane; all the other 2D
reflexive polytopes contain the points of $F_1$ as a subset (in some
basis).  
These three cases use the values $I_{\rm max} = 2, 1, 3$ respectively
as shown in
Appendix~\ref{sec:appendix-fibers}.  
The three minimal 2D polytopes do not contain any other 2D reflexive
polytopes, and it requires a minimal number of linear equivalence
relations among the toric divisors to check if these minimal polytopes
are present as a subset of the points in $\dd$
that are in a plane defined by
a non-colinear pair $v, w \in S$:
 \begin{itemize}
\item $F_1$: $-(v+w)\in S$
\item $F_2$: $-v, -w\in S$
\item $F_4$: $-(v+w)/2\in S$
\end{itemize}
We could in principle use
this kind of direct check to determine the presence of the larger
subpolytopes as well, though this becomes more complicated for the
other fibers and we proceed slightly more indirectly. 
After identifying all the 2D
planes that are spanned by non-colinear
pairs $v, w$ and contain one of the three
minimal 2D subpolytopes, we calculate the intersection of the 4D
polytope with the 2D plane to obtain the full subpolytope that
contains the minimal 2D subpolytope.  This intersection can be
determined by identifying all lattice points $x \in \dd$ that
give rise to a $4\times 4$ matrix of rank two with another three
non-colinear vectors in the 2D plane. 
Note that this
intersection must give a 2D reflexive polytope, since there can only
be one interior point in the 2D fiber polytope as any other interior
point besides the origin would also be an interior point of the full
4D polytope, which is not possible if the 4D polytope is reflexive.

Let
us call the sets of subpolytopes containing $F_1, F_2$, and $F_4$
respectively ${\cal S}_1, {\cal S}_2$, and ${\cal S}_4$.
We can then efficiently determine which fiber type arises in each case
by some simple checks. 
 Observing that all the 2D polytopes other than the three minimal ones contain the $F_1$ polytope, we  immediately have
 \begin{itemize}
\item   $\set{\dd_2^{F_2}}={\cal S}_2 \setminus {\cal S}_1$,
\item   $\set{\dd_2^{F_4}}={\cal S}_4 \setminus {\cal S}_1$.
\end{itemize}
Then we group the fibers associated with the
rest of the 2D polytopes,
which are all in ${\cal S}_1$, by the number of lattice points:
\begin{itemize}
\item 5 points: $F_3$
\item 6 points: $F_5, F_6$
\item 7 points: $F_7, F_8, F_9, F_{10}$
\item 8 points: $F_{11}, F_{12}$
\item 9 points: $F_{13}, F_{14}, F_{15}$
\item 10 points: $F_{16}$
\end{itemize}
This immediately fixes the fibers $F_3$ and $F_{16}$.  To distinguish
the specific fiber types for the remaining four groups a number of approaches could be used.  We
have simply used a projection to compute the self-intersections of
each curve in a given fiber and the sequence of these
self-intersections.  (Note that in a toric surface, the self
intersection of the curve associated with the vector $v_i$ is $m$,
where $v_{i-1} + v_{i +1} = -mv_i$.)
By simply counting the numbers
of $-2$ curves we can identify $F_{5-13}$.
Finally, $F_{14}, F_{15}$ have the same
numbers of curves of each self-intersection, so we use the order of
the self-intersections of the curves in the projection to distinguish
these two subpolytopes.

\subsection{Stacked fibrations and negative self-intersection curves
  in the base}

\label{ncb}

\begin{table}[]
\centering
\begin{tabular}{|c|c|cccccccccccccccc|}
\hline
$C^2$ & ord $\sigma_{n=1,2,3,4,(5),6}$                     & 1 & 2 & 3 & 4 & 5 & 6 & 7 & 8 & 9 & 10 & 11 & 12 & 13 & 14 & 15 & 16 \\ \hline
$-3$        & $\set{1, 1, 1, 2, (2), 2}$ &3& 4&  4& 3& 5& 4& 6& 4& 5& 3 & 4 & 5  & 3  & 4  & 4  & 3  \\ \hline
$-4$        & $\set{1, 1, 2, 2, (3), 3}$ &3&4 & 4& 3& 5& 4& 6& 4& 5& 3& 4  & 5  & 3  & 4  & 4  & 3  \\ \hline
$-5$        & $\set{1, 2, 2, 3, (3), 4}$ &3&   &  2& 2& 1& 3&   & 1& 2 & 2 & 2 & 1   & 2  & 2  &    &  3 \\ \hline
$-6$        & $\set{1, 2, 2, 3, (4), 4}$ &3&   &  2& 2& 1& 3&   & 1 &2 & 2 & 2 & 1   & 2  & 2  &    &   3  \\ \hline
$-7$        & $\set{1, 2, 3, 3, (4), 5}$ &   &   &   & 2&   & 1 &   & 1 &   & 1 & 1  &    &   2  &    &    &    \\ \hline
$-8$        & $\set{1, 2, 3, 3,( 4), 5}$ &   &   &   & 2&   & 1 &   & 1 &   & 1 & 1  &    & 2    &    &    &    \\ \hline
$-9$        & $\set{1, 2, 3, 4,( 4), 5}$ &   &   &   &   &   &   &   &   &   & 1   &    &    &    &    &    &    \\ \hline
$-10$       & $\set{1, 2, 3, 4, (4), 5}$ &   &   &   &   &   &   &   &   &   & 1  &    &    &    &    &    &    \\ \hline
$-11$       & $\set{1, 2, 3, 4, (5), 5}$ &   &   &   &   &   &   &   &   &   & 1 &    &    &    &    &    &    \\ \hline
$-12$       & $\set{1, 2, 3, 4,( 5), 5}$ &   &   &   &   &   &   &   &   &   & 1  &    &    &    &    &    &    \\ \hline
$-13$       & $\set{1, 2, 3, 4, (5), 6}$ &   &   &   &   &   &   &   &   &   &   &    &    &    &    &    &    \\ \hline
\end{tabular}
\caption{\footnotesize  Curves $C$ with self-intersection
  $C\cdot C$ that are allowed in the base of a stacked $F$-fibered
  polytope for the 16 fiber
  types $F$. The numbers below the labels of the 16 fiber types count the
  numbers of  the vertices of $F$
  that give vertex stacked-form fibrations where the corresponding curve
  can appear in the base. (Note that $-3$ and $-4$ curves are allowed
  in all cases, so the first and second rows give the total number
  of the vertices of a given fiber, and the most negative curve that
  can occur for a given fiber corresponds to the position of the last non-empty entry in
  the column.)  The second column gives the orders of vanishing of
  $\sigma_n\in\mathcal{O}(-n K)$ along $C$,  $n=1,2,3,4,(5),6$  (none of
  the fibered polytopes has $\mathcal{O}(-n K)$ for either $n\geq 7$
  or $ n=5$). A $(4,6)$ singularity arises along the whole curve
  unless there exists a section  $\sigma_n\in\mathcal{O}(-n K)$ such that
  ord$_C(\sigma_n)< n$. The existence of such a section depends on the
    fiber type and the specified vertex of the base used for the stacking.
    Curves with $-13\leq C\cdot C\leq -3$ are considered (while curves
    $C^2\geq-2$ are always allowed since $\set{\text{ord}_C(\sigma_n)\rvert
      n=1,2,3,4,(5),6}=\set{0,0,0,0,(0),0}$,  there is always a
    (4,6)  singularity along
    the whole curve when $C^2\leq-13$ since ord$_C(\sigma_n)=n$ for all
    $n=1,2,3,4,5,6$).
  } 
\label{allowed}
\end{table}

\begin{table}[]
\centering
\begin{tabular}{|c|c|c|c|c|c|c|c|c|c|c|c|}
\hline
$m$ & $-3$ & $-4$ & $-5$ & $-6$ & $-7$ & $-8$ & $-9$ & $-10$ & $-11$ & $-12$ & $-13$       \\ \hline
min($n$)  & 2 & 2 & 3 & 3 & 4 & 4 & 6 & 6  & 6  & 6  & - \\ \hline
\end{tabular}
\caption{\footnotesize For each $m$, 
  the minimal value of $n$ such that a section
  $\sigma_n\in\mathcal{O}(-nK_B)$ exists preventing $(4,6)$ points over a
  curve of self-intersection $m$. Note that since there are no $\sigma_5$s
  in any cases (see the third column in Table \ref{models}), min($n$)
  jumps from $4$ to $6$ between $m=-8$ and $m=-9$.}
\label{translatentom}
\end{table}

In the companion paper \cite{Huang-Taylor-long}, we have found that at
large Hodge numbers many of the polytopes in the KS database  belong to
a particular ``standard stacking'' class of $\P^{2,3,1}$ fiber type
($F_{10}$) fibrations over toric base surfaces, which are 
$F_{10}$ fibrations where all rays in the base are stacked over  a
specific vertex $v_s$ of $F_{10}$.  This simple class of  fibrations corresponds naturally to Tate-form Weierstrass
models over the given base, which take the form $y^2 + a_1 yx + a_3y =
x^3 + a_2x^2 + a_4x + a_6$.  In this paper we systematically consider
the distribution of different fiber types, and also analyze which of
the $\P^{2,3,1}$ fibrations are of the ``standard stacking''
type.  As background for these analyses, we describe in this section
the more general ``stacked'' form of polytope fibrations and perform some further
analysis of which stacked fibration types can occur over bases with
curves of given-intersection; since certain fibers cannot arise in
fibrations over bases with extremely negative self-intersection curves
(at least in simple stacking fibrations),
 this helps to explain the dominance
of $\P^{2,3,1}$ fibers at large $h^{1, 1}$.

\subsubsection{Stacked fibrations}
\label{sec:stacked}

As discussed in more detail in \cite{Kreuzer:1997zg, Huang-Taylor-long},
the presence of a reflexive fiber $F =\nabla_2 \subset\nabla$ gives
rise to a projection map $\pi: \nabla \rightarrow\Z^2$, where $\pi (F)
= 0$, associated with a genus one or elliptic fibration of the
Calabi-Yau hypersurface $X$ over a
toric complex surface base $B$.
The ``stacked'' form of a fibration refers to a polytope in which the
rays of the base all have pre-images under $\pi$ that
lie in a plane in $\nabla$ passing through one of the
points in the fiber polytope $\dd_2$.  Specifically, a polytope
$\dd$ that is in the stacked form can always be put into coordinates
so that the lattice points in $\dd$ contain a subset
\begin{equation}
\{(\vb_i)_{1,2};(\vF_s)_{1,2})\rvert v_{i}^{(B)}
\in \{\text{vertex rays in } \Sigma_B\} \}\cup
\{(0,0,(\vF_i)_{1,2})\rvert \vF_i 
\in \{\text{vertices of $\dd_2$\}}\},
\label{pile}
\end{equation}
where $\Sigma_B$ is the toric fan of the base $B$ and $\vF_s$ is a specified
point of the fiber subpolytope $\dd_2$.
We refer to such polytopes as $\vF_s$ stacked $F$-fibered polytopes.

In some contexts it may be useful to focus attention on
the stacked fibrations where the point $\vF_s$ is a vertex of $\dd_2$,
 as these represent the extreme cases of
stacked fibrations, and have some particularly simple
properties\footnote{In particular, the analysis of \S 6.2 of
  \cite{Huang-Taylor-long} can be easily generalized to show that a
  fibration has a vertex stacking on $\vF_s \in\dd_2$ iff there is a single
  monomial over every point in the dual face of $\ds_2$ and these
  monomials all lie in a linear subspace of $\ds$.}.  
We
can refer to these as ``vertex stacked'' fibrations.
The {\it standard}  $\P^{2,3,1}$ fibrations discussed in
\cite{Huang-Taylor-long}
(sometimes there  called ``standard stacking'' fibrations)
refer to the cases of stacked fibrations where the fiber is $F_{10}$ and the
specified stacking point is the vertex $v_s^{(F)} = (-3, -2)$.\footnote{Note
  that in \cite{Huang-Taylor-long}, we have a different convention for
$\P^{2, 3, 1}$ which uses slightly different coordinates
 from those one we use here, so that the
vertex in the notation of that paper is $v_s^{(F)} = (2,
3)$.}
These are based on a standard type of construction
in the toric hypersurface literature (see e.g. \cite{Skarke:1998yk}).
As described in
detail in  \cite{Huang-Taylor-long}, in the case of a standard  stacking, the monomials in $\ds$ match
naturally
 with the set of monomials in the Tate-form Weierstrass
model. Generalizing this analysis  gives bounds on
what kinds of curves can be present in the base supporting a stacked
fibration with different fiber types.

\subsubsection{Negative curve bounds}

For any stacked fibration with a given fiber type $F$ and specified point
$v_s^{(F)}$ for the stacking, the monomials in the dual polytope
$\ds$ are sections of various line bundles ${\cal O} (-nK_B)$.
By systematically analyzing the possibilities we see that many fibers
cannot be realized
in stacked fibrations over bases with curves of very negative
self-intersection without giving rise to singularities in the
fibration  over these curves that go outside the Kodaira classification
and have no Calabi-Yau resolution.

We analyze this explicitly as follows.  To begin with, the lattice
points of the dual polytope $\ds$ of an $F$-fibered polytope $\dd$
are
of the form
\begin{equation}
\set{((\m)_{1,2};(\mF_j)_{1,2})\rvert \mF_j\in \ds^{(F)}_2; (\m)_1,
  (\m)_2\in\Z} \,,
\label{dsform}
\end{equation}
where
$\ds_2^{(F)}$  is one of
the 16 dual subpolytopes that are given in detail in 
Appendix~\ref{sec:appendix-fibers-dual}.
For a given base $B$, we have the condition
\begin{equation}
\m\cdot \vb_i\geq -n, \forall i \Leftrightarrow \m\text{ gives a
  section in } \mathcal{O}(-nK_B) \,.
\end{equation}
Given that $((\vb_i)_{1,2},(\vF_s)_{1,2})\in\dd$ for all $i$ in a
fibration that has the ``stacked'' form (\ref{pile}), the reflexive condition $m \cdot v\geq -1, m
\in\ds, v\in\dd$ implies that a lattice point
$m=((\m)_{1,2},(\mF_j)_{1,2})\in \ds$ gives a section in
$\mathcal{O}(-(\vF_s\cdot\mF_j+1)K_B)$. (See Figure~\ref{sections} for 
examples with the $F_{10}$ fiber type, using the three different
vertices $\vF_s$ of $\dd_2$
as the specified points for  three different stackings, including
the ``standard  stacking'' in which the monomials over the different
lattice points in $\ds_2$ correspond to sections $a_n$ of different line bundles in the
Tate-form Weierstrass
model.) 
Note that the lattice points in $\ds$ that
project to the same lattice point in $\ds_2$ always
give sections that belong to
the same line bundle, since the line bundle depends only on
$m^{(F)}_j$.

This shows that the allowed monomials in any polytope dual to a
stacked
fibration construction over a base $B$ take values as sections of
various line bundles ${\cal O} (-nK_B)$.
For each vertex $\vF_s$ of the 2D polytope $\dd_2$, and for each fiber type $F$,
the number of lattice points in $\ds_2$ corresponding to the resulting line
bundle $\mathcal{O}(-nK)$ is listed in the third column in Table
\ref{models}.
For points $\vF_s$ in $\dd_2$ that are not vertices, the numbers of such
points will  interpolate between the vertex values; the largest values
of $n$ are found from vertex stackings.

The line bundles in which the monomials take sections place
constraints on the structure of the base.
The
order of vanishing of a section $\sigma_n\in \mathcal{O}(-nK_B)$ over a
generic point in a rational curve $C$ with self-intersection $m=C\cdot C \leq - 3$
is\footnote{This calculation can be simply done by using the Zariski
  decomposition, along the lines of  \cite{clusters}.}
\begin{equation}
\text{ord}_C(\sigma_n)=\left\lceil \frac{n (m+2)}{m} \right\rceil.
\end{equation}
The orders of vanishing  $\set{\text{ord}_C(\sigma_n)\rvert
  n=1,2,3,4,(5),6}$ for each $m$, $-3\geq m\geq -13$, are listed in
the second column in Table \ref{allowed}. 
Note that none of
  the 16 fiber types gives a section of $\mathcal{O}(-5K_B)$ (see the
  third column in Table \ref{models}).

For a Weierstrass model, where the coefficients $f, g$ are sections of
the line bundles ${\cal O} (- 4 K_B)$ and ${\cal O} (- 6 K_B)$, the
Kodaira condition that a singularity have a Calabi-Yau resolution is
that $f, g$ cannot vanish to orders 4 and 6.  For the more general
class of fibrations we are considering here, the necessary condition
is that at least one section $\sigma_{n=1,2,3,4, (5), 6}$ exists with
$\text{ord}_C(\sigma_n)<n$.  This condition is necessary so
that when the sections are combined to make a Weierstrass form, the
resulting $f, g$ give either a section in $\mathcal{O}(-4K_B)$ or a
section in $\mathcal{O}(-6K_B)$, respectively, whose order of vanishing does not
exceed $4$ or $6$.  Note that as the absolute value
$|m|$ of the self-intersection of the curve $C$ increases, the minimal
$n$ that satisfies $\text{ord}_C(\sigma_n)<n$ is non-decreasing.  The
minimum value min($n$) so that this condition is satisfied is listed
for each $m$ in Table \ref{translatentom}. Therefore, given a fiber
type $F$ with a specified point $\vF_s$, the allowed negative curves
in the base that are allowed for a  stacking construction using
the point $\vF_s$ that gives a resolvable Calabi-Yau construction are
such that the following two conditions are satisfied: the existence of
a section $\sigma_{n=1,2,3,4,\text{ or } 6}$ such that (1)
$\sigma_n\in \mathcal{O}(-(\vF_s\cdot\mF_j+1)K_B)$ and (2)
ord$_C(\sigma_n)<n$.  For each fiber type $\dd_2$, we have considered
the stacking constructions over each vertex. The most negative
self-intersection curve that is allowed in the base for each fiber
type is tabulated in the last non-empty entry in the corresponding column in
Table \ref{allowed}, and a $\vF_s$ that gives rise to stacked
fibrations in which the most
negative curve is allowed, and the corresponding line bundles associated with
lattice points in $\ds_2$ are given in Appendix
\ref{sec:appendix-fibers-dual}.
Note that since for any lattice point in $\ds_2$, the largest value of
$n$ such that for any choice of stacking point $\vF_s$ the corresponding points in $\ds$ are sections of ${\cal
  O}(-nK_B)$ 
arises from a vertex, it is sufficient to consider the maximum $n$
across the possible choices of vertices $\vF_s$.

This analysis shows that any polytope that has the stacked form with a
given fiber type $F$ gives a genus one fibration over a base $B$ in
which the self-intersection of the curves has a lower bound given by
the last nonempty entry in the corresponding column of
Table~\ref{allowed}.  For the fiber $F_{10}$, this bound is more
general.  It is not possible to find any elliptic fibration with a
smooth Calabi-Yau resolution over a base that contains curves of
self-intersection $C \cdot C <-12$.  While we have not proven it for
  polytopes that do not have the stacking form described here, it
  seems plausible to conjecture that the bounds on curves in the base
  for each fiber type given in Table~\ref{allowed} will also hold for
  arbitrary fibrations (i.e. for general ``twists'' of the fibration
  that do not have the stacking type).  We have not encountered any
  cases in our analysis that would violate this conjecture.  And it is
  straightforward to see using the analysis done here already that
  these curve bounds will still hold when there is a coordinate system
  where each ray of the base has a pre-image living over some ray $v_F
  \in \nabla_2$, even when these rays are not all the same $\vF_s$ as
  in the stacking case, since the bound applying for each curve will
  match that of some choice of $\vF_s$.  
If the more general conjecture is correct, then, for example, 
it would follow in general that any
  reflexive polytope with a fiber $F_4$ can only have curves in the
  base of self-intersection $\geq - 8$, those with a fiber $F_1$ can
  only have curves in the base of self-intersection $\geq - 6$,
  etc. We leave, however, a general proof of this assertion to further
  work.

\subsection{Explicit construction of reflexive polytopes from stackings}

In \cite{Huang-Taylor-long}, we showed that the standard
stacking construction with the fiber $\P^{2,3,1}$, combined with a
large class of Tate-form Weierstrass tunings, can be used to
explicitly construct a large fraction of the reflexive polytopes in
the Kreuzer-Skarke database at large Hodge numbers.  The stacking construction
with other fibers can be used similarly to construct other reflexive
polytopes in the KS database.

Explicitly, given the negative curve bounds on the base determined
above, we can construct a stacked $F$-fibered polytope
over $B$ as follows, following a parallel procedure to that
described in \cite{Huang-Taylor-long} for the $\P^{2,3,1}$-fibered
standard stackings: Given a fiber  $F$ with a specified  ray
$\vF_s$, and a smooth 2D toric base $B$ in which the self-intersections
of all curves are  not lower than the negative curve bound
associated with $\vF_s$, we start with the minimal fibered polytope
$\tilde{\dd}\subset N$ (which may not be reflexive) that is the convex hull
of the set in equation (\ref{pile}).  
If $\tilde{\dd}$ is reflexive, then we are done; otherwise we adopt
the ``dual of the dual'' procedure used in \cite{Huang-Taylor-long} to
resolve $\tilde{\dd}$: define $\ds^\circ=\text{ convex
  hull}((\tilde{\dd})^*\cap M)$. As long as the negative curve bound
is satisfied (no $(4,6)$ curves), $\ds^\circ$ is a reflexive polytope,
and the resolved polytope in $N$ is $\dd\equiv
(\ds^\circ)^*$.

\begin{figure}[]
\centering
\begin{tabular}{ccc}
$\vF_s=(1,0)$                                         & $\vF_s=(0,1)$                                           & $\vF_s=(-3,-2)$                                           \\
\includegraphics[width=4.8cm]{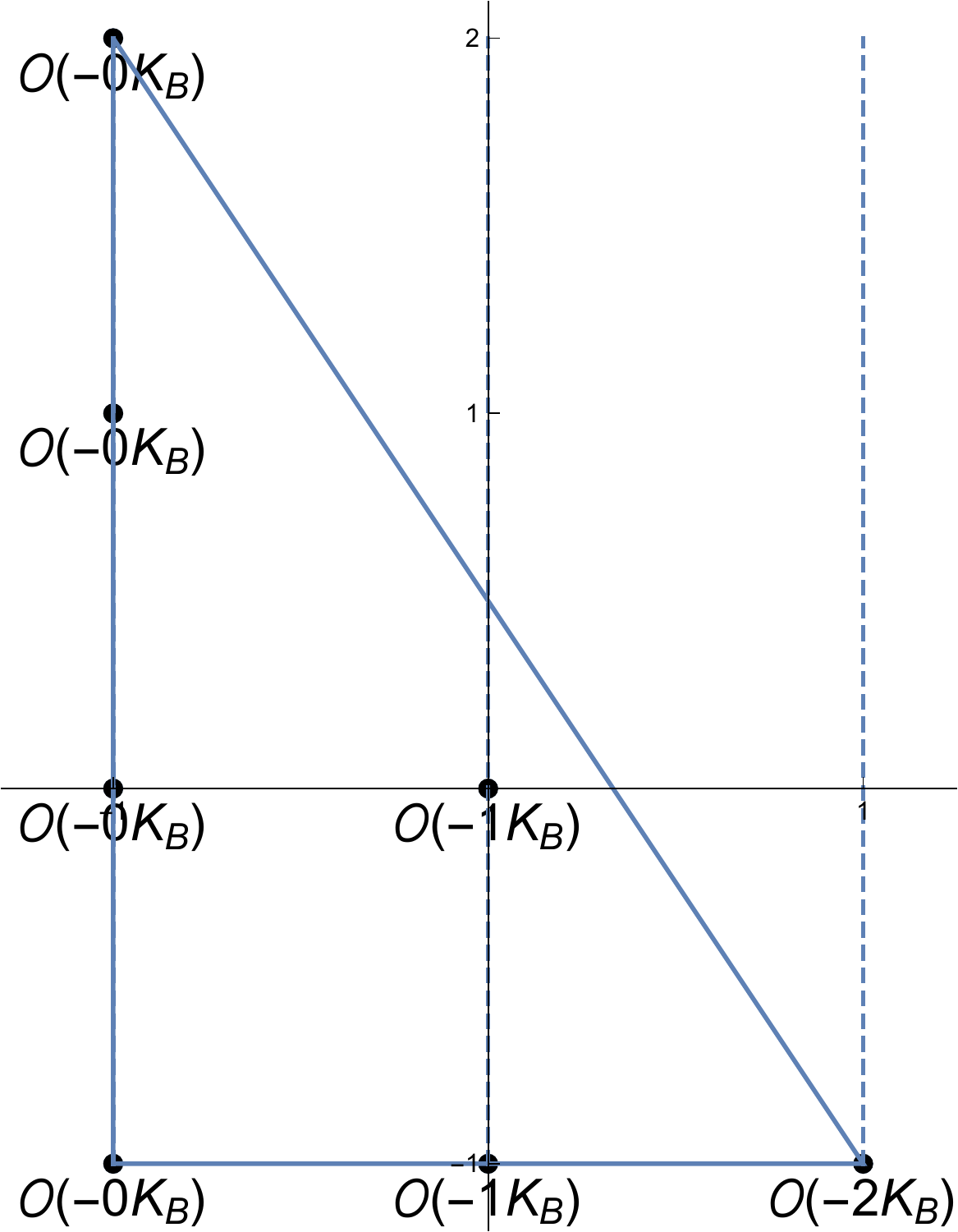} &\includegraphics[width=4.8cm]{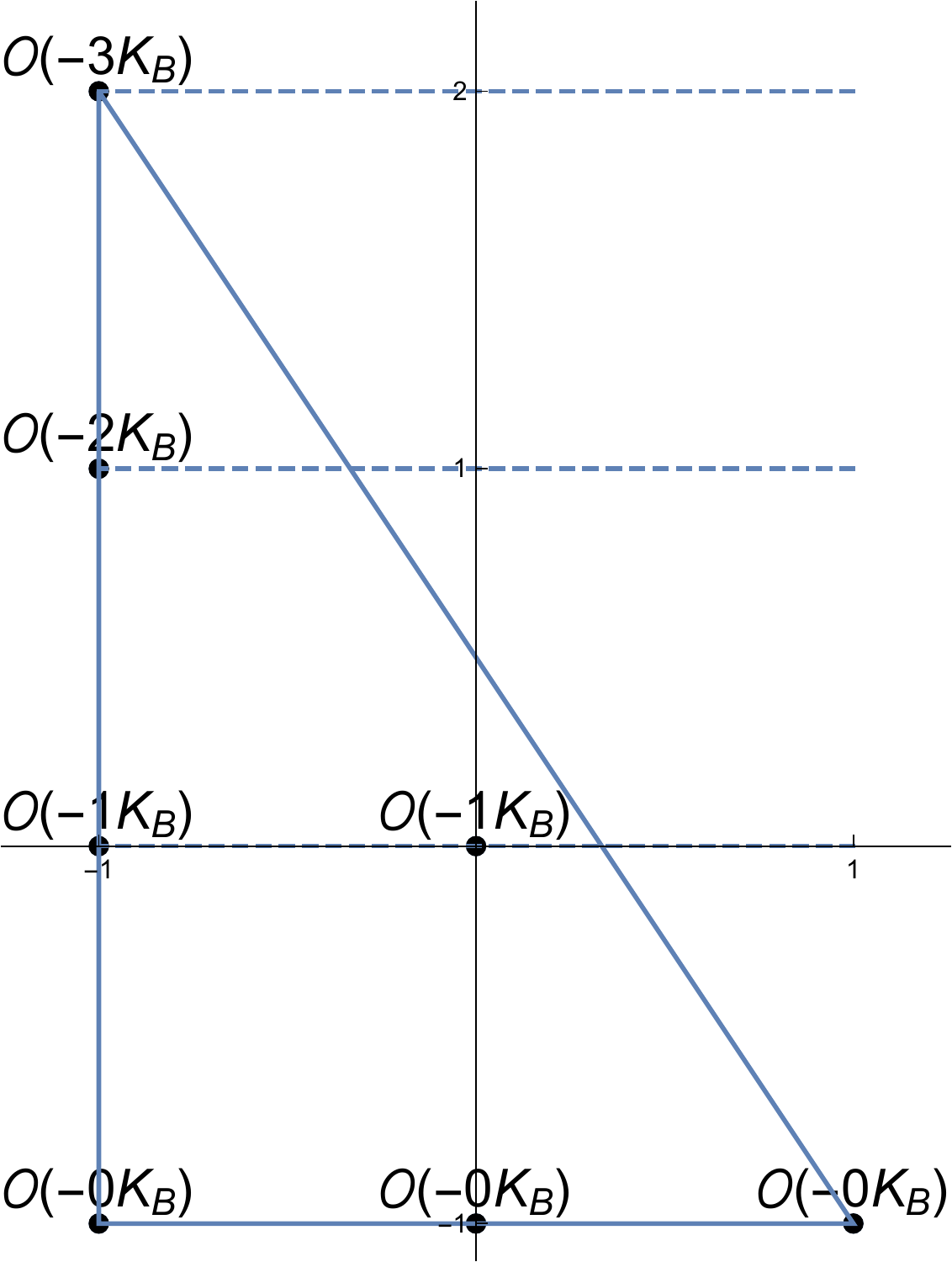} & \includegraphics[width=4.8cm]{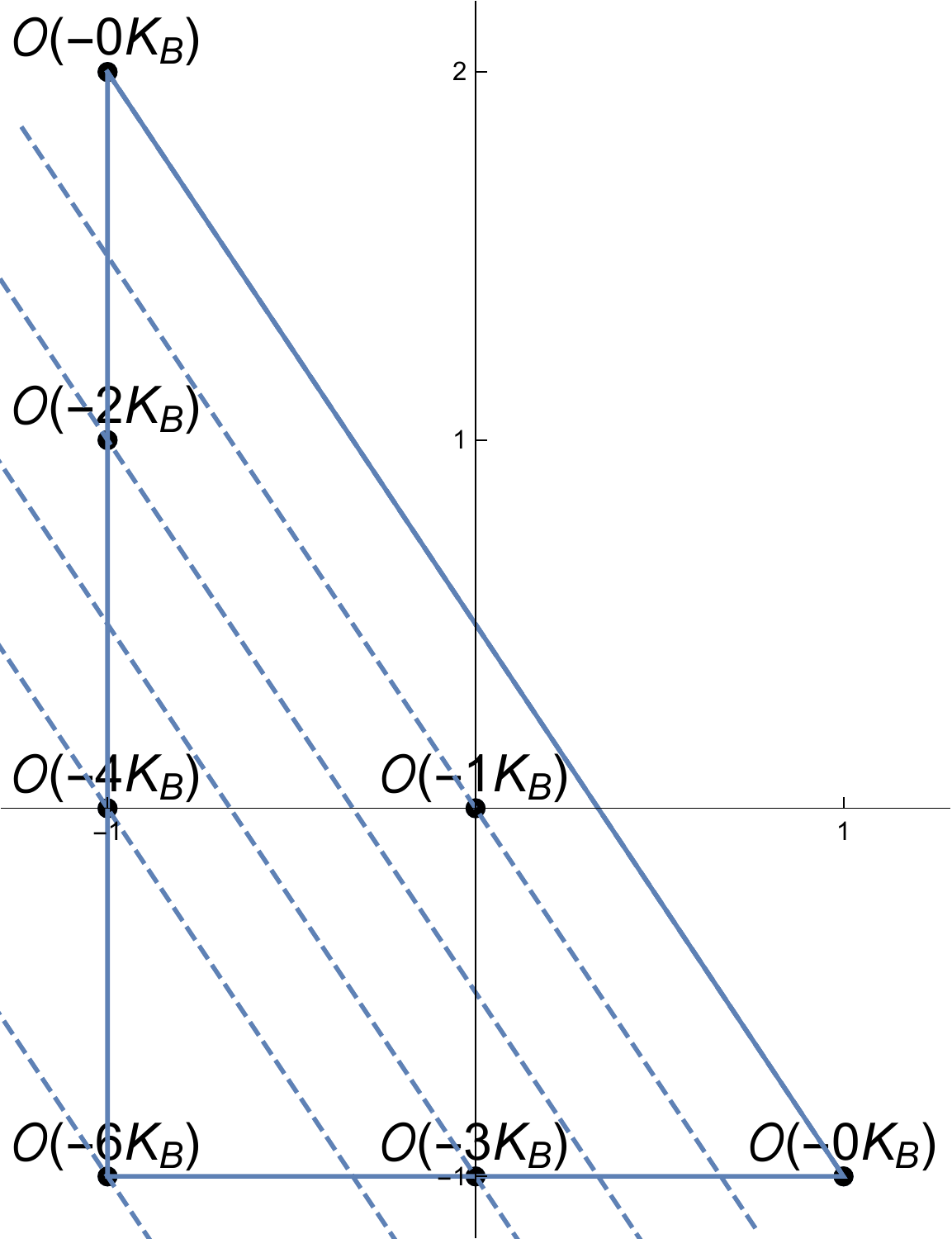}\\
(a)&(b)&(c)
\end{tabular}
\caption{\footnotesize
Different choices of the point $\vF_s$ used to specify a stacking
construction are associated with different
  ``twists'' of the $F$-fiber bundle over the base $B$.  The different
choices of $\vF_s$ for a given fiber type give rise to monomials in
the dual polytope that are sections of different line bundles over the
base, illustrated here for three different choices of $\vF_s$ as
vertices of the fiber $F_{10}= \P^{2,3,1}$.
In the 
  stacking construction, each lattice point in $\ds_2$ is associated
  with a line bundle $\mathcal{O}(-(\vF_s\cdot\mF_j+1)K_B),
  \mF_j\in\ds_2$.  The dashed lines
  are normal to the corresponding $\vF_s$. The lattice points in
  $\ds_2$ on the same dashed line  are associated with sections of the same  line
  bundle over the base. (cf. the $F_{10}$ data in Table \ref{allowed}
  and Table \ref{models}.)}
\label{sections}
\end{figure}

Explicit
examples of $F$-fibered polytopes over Hirzebruch
surfaces $\F_m$ are given in Table \ref{models}, for each fiber type
$F$. The base $\F_m$ is in each case
chosen such that $-m$ saturates the negative curve bound associated
with the specific vertex $\vF_s$ for a given fiber type (see Appendix
\ref{sec:appendix-fibers-dual} for the possible choices of $\vF_s$ for
each fiber type that allow the most negative self-intersection curves
in the base). For
example, the standard stacked $\P^{2,3,1}$-fibered polytopes considered in
\cite{Huang-Taylor-long} have bases  stacked over the vertex $(-3,-2)$
of the fiber $F_{10}$ in Appendix \ref{sec:appendix-fibers}, and there
exist sections in $\mathcal{O}(-nK_B)$ for $n=1,2,3,4,6$ (see Figure
(c) in Table \ref{sections}), so models in this class correspond naturally to the Tate-form Weierstrass models where $a_n=\sigma_n$, and the negative curve bound is
$-12$.  The model listed in Table \ref{models} is the generic
elliptically fibered CY over $\F_{12}$.

The construction just described above gives  the minimal reflexive $F$-fibered polytope
over $B$ that contains the set in equation (\ref{pile}).
While the $F_{10}$ fiber type with $\vF_s=(-3,-2)$ gives the
most generic elliptic Calabi-Yau over any given  toric
base $B$ through this construction,  using the other fiber types or the
other specified points of $F_{10}$ for stacked
stacking polytopes give models  with enhanced
symmetries (these can include
discrete, abelian, and non-abelian symmetries). Further
tunings of the polytope analogous to Tate-tunings for the standard $\P^{2,3,1}$ polytope
can reduce $\ds$ and enlarge $\dd$, giving a much larger class of
reflexive polytopes for Calabi-Yau threefolds.
The explicit construction of the polytopes corresponding to Tate tuned models
via polytope tunings of the standard $F_{10}$-fibered polytope with
$\vF_s=(-3,-2)$ were discussed in section 4.3.3 and Appendix A in
\cite{Huang-Taylor-long}. We have not
attempted systematic
polytope tunings for the other fiber types, but in principle one can work out tuning tables analogous to the Tate table  for the other fiber types.

\section{Results at large Hodge numbers} 
\label{sec:results}

We have systematically run the algorithm described in
Section
\ref{sec:algorithm} to check for a manifest elliptic or genus one
fibration realized through a reflexive 2D fiber  for each
polytope in the Kreuzer-Skarke database
that gives a Calabi-Yau threefold $X$ with $h^{1, 1} (X)$ or $h^{2, 1}
(X)$
greater or equal to 140.  The number of polytopes 
that give rise to Calabi-Yau threefolds with
$h^{1, 1}\geq 140$ 
is $248305$.
Since the set of reflexive polytopes is mirror symmetric (Hodge
numbers $h^{1, 1}, h^{2, 1}$ are exchanged in going from $\dd
\leftrightarrow\ds$), this is also the number of polytopes with
$h^{2, 1}\geq 140$.
(Note, however, that the mirror of an elliptic Calabi-Yau threefold is
not necessarily elliptic.)
There are $495515$ polytopes
with at least one of the Hodge numbers at least 140, and from these
numbers it follows that the number of polytopes with both Hodge
numbers at least 140 is $1095$.
While as described in Section \ref{sec:algorithm}, we have made the
algorithm reasonably
efficient for larger values of $h^{1, 1}$, our
implementation in this initial investigation was in Mathematica, so a
complete analysis of the database using this code was impractical.  We
anticipate that in the future a complete analysis of the rest of the database
can be carried out
with a more efficient code, but our focus here is on identifying the
largest values of $h^{1, 1}, h^{2, 1}$ that are associated with
polytopes that give Calabi-Yau threefolds with no manifest elliptic
fiber.
In \S\ref{sec:prevalence} we analyze the distribution of fibrations at
small $h^{1,1}$.

\subsection{Calabi-Yau threefolds without manifest
genus one fibers}

\begin{figure}
  \centering
  \includegraphics[width=7cm]{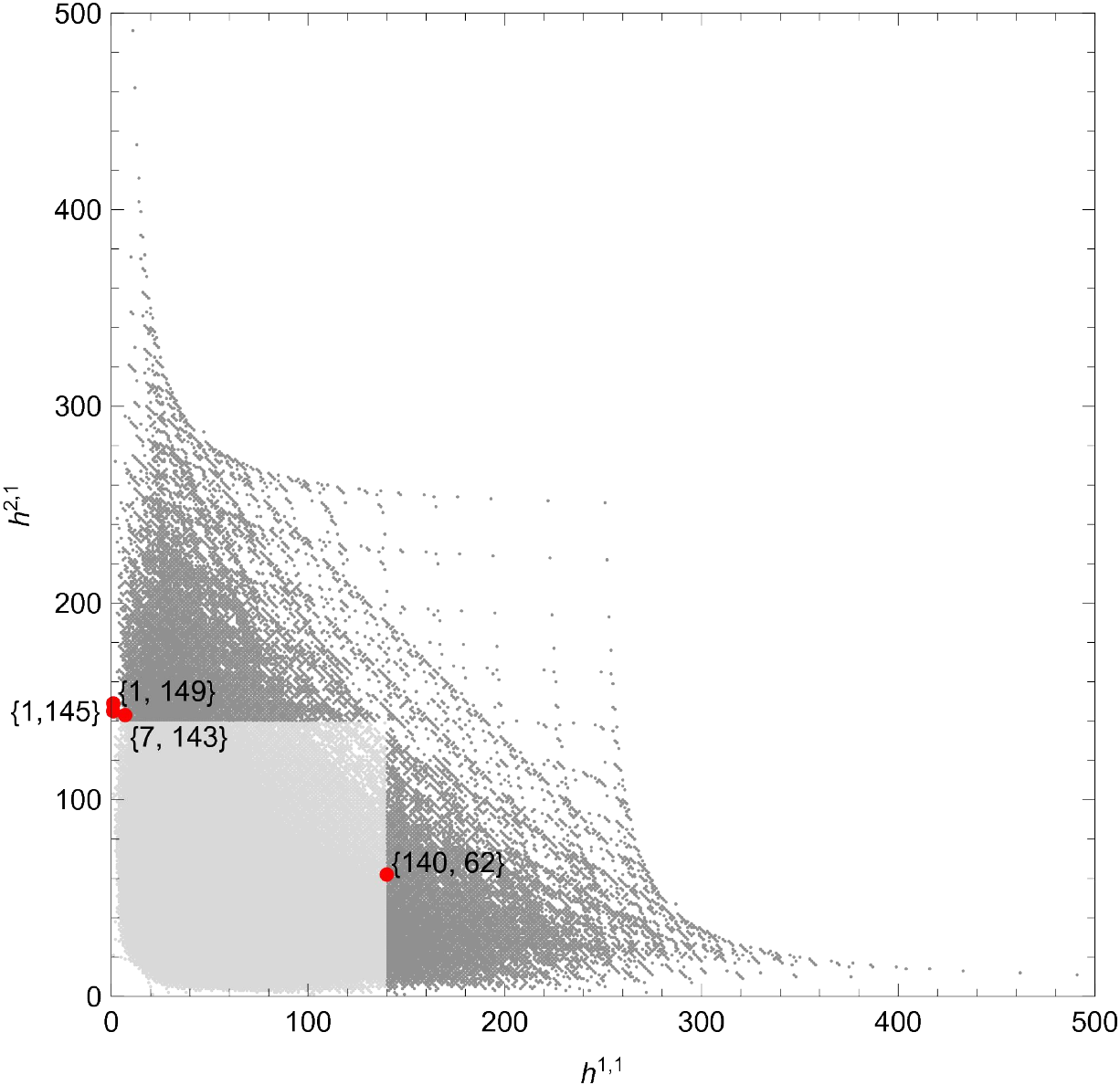}
  \caption{\footnotesize The four Hodge pairs  in the region 
 $h^{1,1}\geq240$ or $h^{2,1}\geq240$
associated with
    polytopes
without reflexive 2D subpolytopes associated with genus one (including
elliptic) fibers.} 
  \label{fig:four}
\end{figure}

Of the 495515 polytopes analyzed at large Hodge numbers, we found that
only four lacked a 2D reflexive polytope fiber, and thus the other
495511 polytopes all lead to Calabi-Yau threefolds with a manifest
genus one fiber.  The Hodge numbers of the four Calabi-Yau threefolds
without a manifest genus one fiber are
\begin{equation}
 (h^{1, 1}, h^{2, 1}) =
(1, 149),  \hspace*{0.1in}
(1, 145), \hspace*{0.1in}
 (7, 143), \hspace*{0.1in} (140, 62) \,.
\label{eq:4-examples}
\end{equation}
(See Figure \ref{fig:four}.) It is of course natural that any Calabi-Yau threefold with $h^{1, 1} =
1$ cannot be elliptically fibered; by the Shioda-Tate-Wazir formula
\cite{stw},
any elliptically fibered Calabi-Yau threefold must have at least
$h^{1, 1} = 2$, with one contribution from the fiber and at least one
more from 
$h^{1, 1}$ of the base, which must satisfy
$h^{1, 1} (B) \geq 1$.
We also expect that any genus one fibered CY3 will have at least a
multi-section \cite{Braun:2014oya, Morrison-WT-sections}, so $h^{1,
  1}\geq 2$ in these cases for similar reasons.

The examples (1, 145) and (1, 149) are the only Hodge numbers from
polytopes in the Kreuzer-Skarke database with $h^{1, 1} = 1, h^{2, 1}
\geq 140$.  Note that the quintic, with Hodge numbers (1, 101), is
another example of a Calabi-Yau threefold with $h^{1, 1}= 1$ that
has no elliptic or genus one fibration.

We list here the
 polytope  structure of the two examples from
 (\ref{eq:4-examples}) that have $h^{1, 1} > 1$, in the form given in
 the Kreuzer-Skarke database.  M refers to the numbers of lattice
 points and vertices of the dual polytope $\ds$, while N refers
 to the numbers of lattice points and vertices of the polytope $\dd$, and H refers to the Hodge numbers $h^{1,1}$ and $h^{2,1}$.
The vectors listed are the vertices of the polytope in the $N$
lattice.
The numbers in parentheses for each polytope refer to the position in
 the list of polytopes
in the Kreuzer-Skarke database that give CY3s
with those specific Hodge numbers.
\begin{itemize}
%
%
%

\item M:196 5 N:10 5 H:7,143 (1$^\text{st}$/54)

Vertices of $\dd$:
$\set{(-1,4,-1,-2), (-1,-1,1,1), (1,-1,0,0), (-1,-1,0,1), (-1,-1,0,3)}$

\item  M:88 8 N:193 9 H:140,62 (6$^\text{th}$/255)

Vertices of $\dd$:
\set{$(-1,2,-1,4), (-1,0,4,-1), (1,-1,-1,-1), (-1,-1,-1,19), $ 

\hspace*{0.6in}$(-1,-1,5,1),(-1,1,0,-1),(-1,1,-1,-1), (-1,-1,-1,-1), (-1,-1,5,-1)$}

\end{itemize}

Note that we have not proven that these Calabi-Yau threefolds do not
have elliptic or genus one fibers, we have just found that such fibers
do not appear in a manifest form from the structure of the polytope.
We leave for further work the question of analyzing non-toric elliptic
or genus one fibration structure of these examples, or others with
smaller Hodge numbers that also lack a manifest genus one fiber; such
an analysis might be carried out using methods similar to those of
\cite{aggl-3}.

\subsection{Calabi-Yau threefolds without manifest elliptic fibers}

Of the 495515 polytopes analyzed, only 384 had fibers of only types
$F_1, F_2, F_4$.  These cases are associated with genus one fibered
Calabi-Yau threefolds that have no manifest toric section, and
therefore are not necessarily elliptically fibered.
Note that we have not proven that these Calabi-Yau threefolds do not
have elliptic fibers; in fact, many toric hypersurface
Calabi-Yau threefolds have been found to have non-toric fibrations
\cite{BGK-geometric}.  It would be interesting to study these examples
further for the presence of non-toric sections.

The largest values of $h^{2, 1}$ and  $h^{1, 1}$  for these genus one
fibered Calabi-Yau threefolds without a manifest toric section are
realized by the examples:

\begin{itemize}
\item M:311 5 N:15 5 H:11, 227 (1$^\text{st}$/19)

Vertices of $\dd$:
$\set{(-1,0,4,-3),(-1,2,-1,0),(1,-1,-1,1),(-1,0,-1,1),(-1,0,-1,3)}$

\item  M:(80; 81; 81; 82) 8 N:(263; 262; 261; 260) 9 H:194, 56 ((7$^\text{th}$; 8$^\text{th}$;
  9$^\text{th}$; 10$^\text{th}$)/52)

Vertices of $\dd$:
\begin{itemize}
\item 7$^\text{th}$ $\set{(-1, 0, 4, -1), (-1, 
  2, -1, -1), (1, -1, -1, -1), (-1, -1, -1, -1), (-1, -1, 6, -1), \\(-1,
   1, 0, 6), (-1, -1, -1, 28), (-1, 1, -1, 10), (-1, -1, 6, 0)}$,\\

\item 8$^\text{th}$ $\set{(-1, 0, 4, -1), (-1, 
  2, -1, -1), (1, -1, -1, -1), (-1, -1, -1, -1), (-1, -1, 6, -1), \\(-1,
   1, 0, 6), (-1, -1, -1, 28), (-1, 0, -1, 19), (-1, -1, 6, 0)}$,\\

\item 9$^\text{th}$ $\set{(-1, 0, 4, -1), (-1, 
  2, -1, -1), (1, -1, -1, -1), (-1, -1, -1, -1), (-1, -1, 5, -1),\\ (-1,
   1, 0, 6), (-1, -1, -1, 28), (-1, 1, -1, 10), (-1, -1, 5, 4)}$,\\

\item 10$^\text{th}$ $\set{(-1, 0, 4, -1), (-1, 
  2, -1, -1), (1, -1, -1, -1), (-1, -1, -1, -1), (-1, -1, 5, -1), \\(-1,
   1, 0, 6), (-1, -1, -1, 28), (-1, 0, -1, 19), (-1, -1, 5, 4)}$

\end{itemize}

\end{itemize}

The fiber type $F_4$ is the only fiber that arises in these five polytopes. In the first case, with Hodge numbers (11, 227), the base of the
elliptic fibration is the Hirzebruch surface $\F_8$.
Analysis of the F-theory physics of the genus one fibration associated
with this polytope suggests that there should in fact be  an elliptic
fiber with a non-toric global section.\footnote{In the F-theory analysis, we consider the Jacobian fibration associated with the $F_4$ fibration.
This is an elliptic fibration with a section, for which a detailed
analysis shows that there are no further enhanced non-abelian gauge
symmetries. There are, however, 150 nodes in the $I_1$ component of the
discriminant locus in the base. Since the generic elliptic fibration
model over $\F_8$ has Hodge numbers (10, 376),
this analysis suggests that there should be an additional section in
this case, which  should correspond to a non-toric section in the original polytope and in
the Jacobian model would give rise to a U(1) abelian factor where the 150
nodes correspond to
matter fields charged under the U(1); the anomaly cancellation
condition is satisfied for the resulting  Jacobian model,  matching with the
shift in Hodge numbers  $(10, 376)+(1,1-150)=(11, 227)$.}
For further
work, it would be nice to prove this and find the non-toric section explicitly. 
Further analysis of the F-theory physics of the other cases may also be
interesting, as well as the question of whether these threefolds admit
elliptic fibrations that are not manifest in the toric description.

\subsection{Fiber types}

The numbers of distinct polytopes in the regions $h^{1,1}, h^{2,1} \geq 140$ that have each fiber type
(not counting multiplicities)
are 

\begin{center}
\begin{tabular}{cccccccc}
$F_1$ &$F_2$  &$F_ 3$ & $F_4$ &$F_5$  &$F_6$  &$F_7$  & $F_8$ \\\hline
612&1&1279&40218&32&19907&20&8579\\\\
 $F_9$ &$F_{10}$    &$F_{11}$  &$F_{12}$  & $F_{13}$ & $F_{14}$ &$F_{15}$  &$F_{16}$\\\hline
2067&487387&24811&850&27631&2438&273&58
\end{tabular}
\end{center}
In Appendix \ref{sec:appendix-results},
we have included a set of figures that
show the distribution of polytopes  containing each fiber type,
according to the Hodge numbers of the associated Calabi-Yau threefolds.
We have shaded the data points of Hodge
pairs varying from light to dark with increasing multiplicities; two factors
contribute to the multiplicity in these figures: the multiplicity of
the polytopes associated with
the same Hodge pair and the multiplicity of
fibers of the same type for a given polytope (note that the latter multiplicity
is not included in the numbers in the table above). We discuss multiple fibrations in the next
subsection.

We can see some interesting patterns in the distribution of polytopes
with different fiber types.  As discussed in \S\ref{ncb}, at least for
polytopes with the stacked fibration form, the only fiber type that
can arise over a base with a curve of self-intersection less than $-8$
is the $\P^{2,3,1}$ ($F_{10}$) fiber (see Table \ref{allowed}).  From
the graphs in Appendix \ref{sec:appendix-results}, it is clear that
this fiber dominates at large Hodge numbers.  The other fiber types
that can arise over a base with a curve of self-intersection less than
$-6$ are $F_4, F_{13}$ (with two possible specified vertices) and $F_6,
F_8, F_{11}$ (with only one specified vertex). The Hodge numbers of
Calabi-Yau threefolds coming from polytopes with fiber types $F_4,
F_6, F_8$ extend to $h^{1,1}=263$, and $F_{11}$ extends to
$h^{1,1}=377$; in fact, the right most data point of the fiber types
$F_4, F_6, F_8, F_9, F_{12}, F_{15}$ is the same: $\set{263,23}$, and
the right most data point of the fiber types $F_{11}$ and $F_{14}$ is
the same: $\set{377,11}$. The fiber $F_{13}$ also continues out to the
largest values of $h^{1, 1}$ as $F_{10}$ does.
Since the largest value of $h^{1,1}$ for a generic elliptic fibration
over a toric base $B$ containing  no curves of self-intersection $< -8$ is 224 \cite{toric, Hodge, Huang-Taylor-long},
these large values of $h^{1,1}$ for fibers other than $F_{10}$ must
involve tuning of relatively large gauge groups.

For $h^{1, 1}> 377$
the only fibers that arise are $F_{10}$ and $F_{13}$. In fact, the
Calabi-Yau threefold with the largest $h^{1, 1}$, which has Hodge
numbers (491, 11),  has two distinct fibrations: one has the
standard $\P^{2,3,1}$ fiber over the 2D toric base
\{$-12//-11//-12//-12//-12//-12//-12//-12//-12//-12//-12//-12//-12//-12//-12//-11//-12,0$\},
represented by the self-intersection numbers of the toric curves, where
// stands for the sequence $-1,-2,-2,-3,-1,-5,-1,-3,-2,-2,-1$; the
other fibration has the fiber $F_{13}$ over the base
\{$-4,-1,-3,-1,-4,-1,-4,-1,-4,0,2$\}.  We leave a more detailed
analysis of the alternative fibration of this Calabi-Yau threefold for
future work.

On the other hand, the fiber $F_{2}$, which is most restricted,  arises from only one
$\dd$ polytope, with multiplicity one: M:40 6 N:186 6 H:149,29, which
also has two different $F_{10}$ subpolytopes.

These observations tell us that, as we might expect, $h^{1,1}$
extends further for the  fiber
subpolytopes
that admit more
negative curves in the base.  Almost half of the fiber types do not
arise for any polytopes at all in the region $h^{2,1}\geq 140$:
$F_2,F_5,F_7,F_{12},F_{14}, F_{15},$ and $F_{16}$.
None of these is allowed over any base with a curve of
self-intersection less than $-6$ (at least in the stacking
construction of \S\ref{ncb}).

\subsection{Multiple fibrations}

Another interesting question is the prevalence of multiple fibrations.
This question was investigated for  complete intersection Calabi-Yau threefolds
in \cite{aggl-2, aggl-3}, where it was shown that many CICY
threefolds have a large number of fibrations.  In the toric
hypersurface context we consider here,
a polytope can have both multiple fibrations by different fiber types
and by the same fiber type.
In this analysis, as in the rest of this paper, we consider only fibrations that are manifest in the
toric description.
We have found that
the total number of (manifest) fibrations in a polytope in
the
two large Hodge number regions ranges from zero to $58$. The  total
numbers of
fibrations and the number of polytopes that have each number of
total fibrations are listed in Table~\ref{t:multiple-fibers}.

\begin{table}
\begin{center}
\begin{tabular}{lccccccc}
 \#  fibrations & 0 & 1 & 2 & 3 & 4 & 5 & 6 \\\hline
\#  polytopes & 4 & 327058 & 113829 & 34657 & 11414 & 4466 & 1955\\
    & (4)    & (327058) & (113829) & (34659) & (11418) & (4465) & (1952) \\\\
\#  fibrations&  7 & 8 & 9 & 10 & 11 & 12 & 13 \\\hline
\#  polytopes& 1003 & 501 & 251 & 150 & 70 & 42 & 32 \\
         & (1003) & (503)    & (251)    & (149)   & (71)    & (42)   & (32)   \\\\
\#   fibrations&  14 & 15 & 16 & 17 & 18 & 20 & 22 \\\hline
\#  polytopes&  31 & 4 & 14 & 6 & 9 & 2 & 6 \\
         & (31)   & (4)      & (14)     & (6)     & (8)     & (2)    & (6)    \\\\
\#    fibrations&  23 & 25 & 26 & 31 & 34 & 37 & 58 \\\hline
\#  polytopes&  2 & 1 & 2 & 1 & 1 & 3 & 1 \\
      &(1)    & (1)      & (2)      & (1)     & (1)     & (2)    &  (0)
\end{tabular}
\end{center}
\caption[x]{\footnotesize  Table of the number of polytopes in the
  large Hodge number regions $h^{1,1}, h^{2,1} \geq 140$ that have
 a given number of distinct (manifest) fibrations.  Numbers in
 parentheses are after modding out by automorphism symmetries (see
 text, Appendix \ref{automorphisms}).}
\label{t:multiple-fibers}
\end{table}

In some cases the number of fibrations is enhanced by the existence of
automorphism symmetries of the polytope.  While a generic polytope has
no symmetries, some polytopes with large numbers of fibrations 
also have many symmetries.  In such cases the number of inequivalent
fibrations can be smaller than the total number of
fibrations.
This issue is also addressed in \cite{Braun:2011ux, aggl-3}.
There are 16 polytopes in the region $h^{1,1}\geq 140$ or
$h^{2,1}\geq 140$ with a non-trivial action of the automorphism symmetry
 on the fibers. We list these 16 polytopes in Appendix
\ref{automorphisms1}. For example, the polytope giving a Calabi-Yau
with Hodge numbers (149, 1) has an automorphism symmetry of order 24,
associated with an arbitrary permutation on 4 of the 5 vertices of the
polytope.  This automorphism symmetry group is described in detail in
Appendix \ref{automorphisms2}; the number of distinct classes of
fibrations modulo automorphisms in this case is reduced to only $8$
instead of 58.

The
polytopes that we have found with a large total number
of (manifest) fibrations
are generally in the large $h^{1,1}$ region; in fact, polytopes in the large $h^{2,1}$ region have at most three fibrations:
\begin{center}
\begin{tabular}{l|cccc}
\# total fibrations&  0 & 1 & 2 & 3 \\\hline
\# polytopes with large $h^{2,1}$ &  3 & 240501 & 7775 & 26
\end{tabular}
\end{center}
The four polytopes with the  two largest numbers of total fibrations
(58, 37 without modding out by automorphisms) are respectively
\begin{equation}
\nonumber
 \set{\set{7, 5, 201, 5, 149, 1, 296}, \set{0, 0, 12, 12, 0, 0, 0, 0, 0, 12, 0, 0, 15, 0, 3, 4}}
\end{equation}
and
\begin{eqnarray}
\nonumber
&&\set{\set{7, 5, 196, 5, 145, 1, 288}, \set{0, 0, 0, 6, 0, 6, 0, 0, 0, 12, 0, 0, 9, 3, 0, 1}}\\\nonumber
&&\set{\set{8, 6, 195, 7, 144, 2, 284}, \set{0, 0, 0, 6, 0, 6, 0, 0, 0, 12, 0, 0, 9, 3, 0, 1}},\\\nonumber
  &&   \set{\set{9, 7, 192, 10, 144, 4, 280}, \set{0, 0, 0, 0, 0, 9, 0, 0, 0, 15, 3, 0, 6, 3, 0, 1}},
\end{eqnarray}
where the numbers are in the format
\begin{center}
\{\{\# lattice points of $\ds$, \# vertices of $\ds$, \# lattice points of $\dd$, \# vertices of $\dd$,\\ $h^{1,1}$, $h^{2,1}$, Euler Number\},\{\#$F_1$,\#$F_2$,$\ldots$,\#$F_{16}$\}\}.
\end{center}
Note that the first two polytopes are, respectively, the mirrors
of the first two polytopes (with $h^{1,1} = 1$) without any fibrations in equation
(\ref{eq:4-examples}).

We also note that in general, the polytopes with larger numbers of
total manifest fibrations fall within a specific range  of values of $h^{1,1}$ and
$h^{2,1}$ (at least in the ranges we have studied here). The ranges of
$h^{1,1}$ and $h^{2,1}$ of the polytopes that have 8 or more
fibrations (without considering automorphisms) are
listed in Table~\ref{t:ranges}.
It may be interesting to note that in a somewhat different context, it
was found in \cite{Wang-WT-MC-2} that a large multiplicity of
elliptically fibered fourfolds arises at a similar locus in the space
of Hodge numbers, at intermediate values of $h^{1,1}$ and small values
of $h^{3,1}$ (which counts the number of complex structure moduli, as
does $h^{2,1}$ for Calabi-Yau threefolds).  It would be interesting to
understand whether these observations stem from a common origin.

\begin{table}
\begin{center}
\begin{tabular}{c|ccccc|}
\hline
\multicolumn{1}{|c|}{\# total fibrations $\geq$} & 8  & 9  & 10 & 11 & 12 \\ \hline
\multicolumn{1}{|c|}{$h^{1,1}$ range}                  &    [140,272]&[140,243]&[140,243]&[140,214]&[140,208]\\ \hline
\multicolumn{1}{|c|}{$h^{2,1}$ range}                  &   [1, 19]& [1, 19]& [1, 16]& [1, 16]& [1, 12]   \\ \hline\\\cline{2-6}
                                                       & 13 & 14 & 15 & 16 & 17 \\ \cline{2-6} 
                                                       &    [140, 208]& [140, 208]& [140, 208]& [140, 208]& [141, 173]    \\ \cline{2-6} 
                                                       &   [1, 11]& [1, 9]& [1, 8]& [1, 8]& [1, 7]    \\ \cline{2-6} \\\cline{2-6} 
                                                       & 18 & 20 & 22 & 23 & 25 \\ \cline{2-6} 
                                                       &    [141, 173]& [141, 173]& [141, 173]& [141, 165]& [141, 154]    \\ \cline{2-6} 
                                                       &    [1, 7]& [1, 6]& [1, 6]& [1, 5]& [1, 5]    \\ \cline{2-6} \\\cline{2-6} 
                                                       & 26 & 31 & 34 & 37 & 58 \\ \cline{2-6} 
                                                       &    [141, 149]& [141, 149]& [141, 149]& [144, 149]& [149, 149]    \\ \cline{2-6} 
                                                       &   [1, 5]& [1, 5]& [1, 4]& [1, 4]& [1, 1]  \\ \cline{2-6} 
\end{tabular}\end{center}
\caption[x]{\footnotesize  Ranges of Hodge numbers in which the
  polytopes with the largest numbers of fibrations (not including
  automorphisms) are localized.}
\label{t:ranges}
\end{table}

\begin{figure}
  \begin{center}
  \includegraphics[width=10cm]{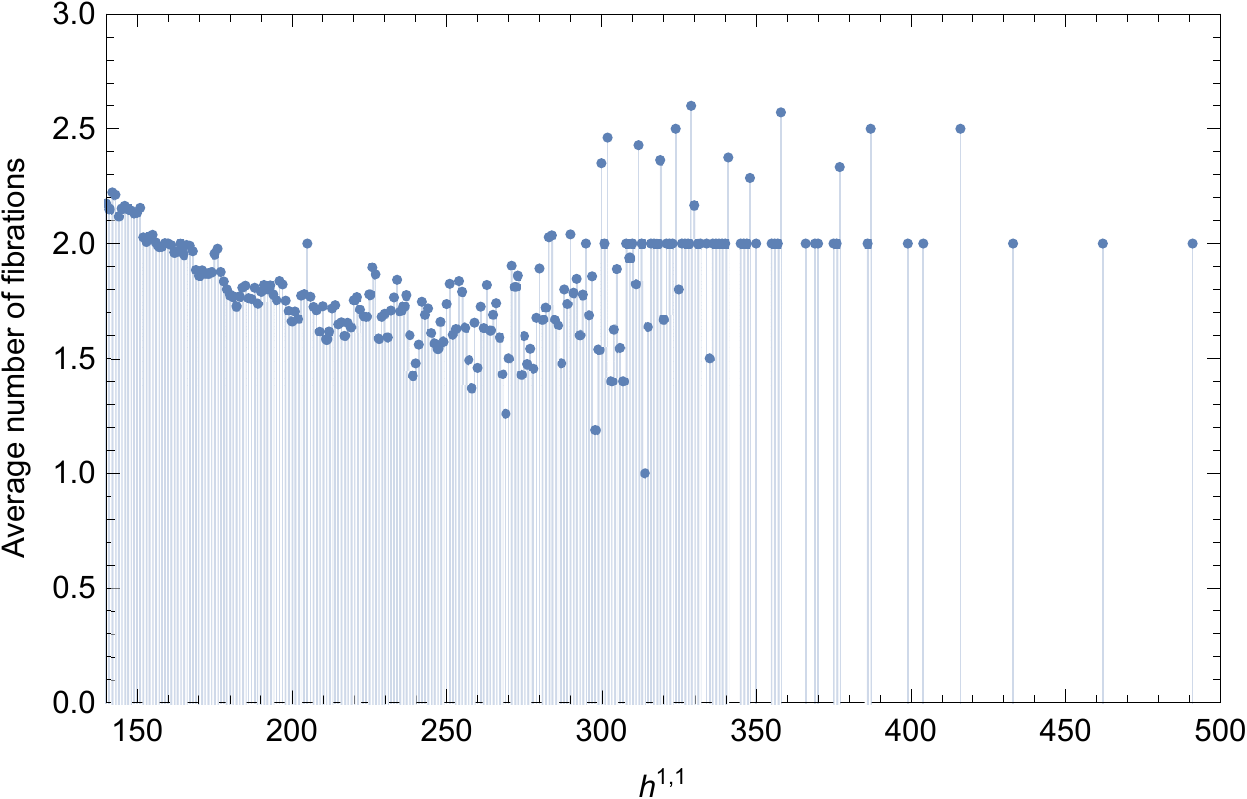}
\end{center}
\caption[x]{\footnotesize  Average number of fibrations for polytopes
  associated with Calabi-Yau threefolds with $h^{1,1} \geq 140$.}
\label{f:average-fibrations}
\end{figure}

It is also interesting to note that while  every Calabi-Yau threefold with $h^{1, 1}>335$ or $h^{2, 1}>256$ has more than
one fibration, the polytopes associated with the largest values of
$h^{1,1}$ have precisely two manifest fibrations, and the average
number of fibrations at large $h^{1,1}$ is close to 2.  In
Figure~\ref{f:average-fibrations}, we show the average number of
fibrations for the polytopes associated with Calabi-Yau threefolds of
Hodge numbers $h^{1,1} \geq 140$.

The maximal number of fibrations for each specific
fiber type in a polytope is 
\begin{center}
\begin{tabular}{cccccccccccccccc}
$F_1$ &$F_2$  &$F_ 3$ & $F_4$ &$F_5$  &$F_6$  &$F_7$  & $F_8$ & $F_9$ &$F_{10}$    &$F_{11}$  &$F_{12}$  & $F_{13}$ & $F_{14}$ &$F_{15}$  &$F_{16}$\\\hline
4& 1& 12& 12& 2& 9& 1& 4& 4& 15& 4& 2& 15& 6& 3& 4\\
(4)& (1)& (6)& (8)& (2)& (9)& (1)& (4)& (4)& (15)& (4)& (2)& (9)& (6)&(3)& (1)
\end{tabular}
\end{center}
Numbers in parentheses are after modding out by automorphism
symmetries; for example, the maximal number of $F_{16}$ fibers,
which comes from the polytope associated with the Hodge pair (149,1), reduces from four
to one (see the last row of the table in Appendix
\ref{automorphisms1}).
 
If we count the distinct fiber types in a polytope, we find that the
maximum number of fiber types that a polytope in the large Hodge
number regions can have is eight. The eight polytopes that have the
maximum number of eight distinct fiber types are 
\begin{eqnarray}
\nonumber
&&\{\{11,6,199,6,151,7,288\},\{0,0,2,1,0,0,0,2,0,3,2,0,2,1,1,0\}\},\\\nonumber&&\{\{12,7,193,8,146,8,276\},\{0,0,2,1,0,0,0,2,0,3,2,0,2,1,1,0\}\},\\\nonumber&&\{\{12,8,201,11,153,6,294\},\{0,0,2,0,0,2,0,0,0,3,2,2,1,1,0,1\}\},\\\nonumber&&\{\{13,8,198,10,151,7,288\},\{0,0,1,1,0,0,0,1,0,2,2,1,1,1,0,0\}\},\\\nonumber&&\{\{15,8,192,12,143,11,264\},\{0,0,2,2,0,1,0,2,0,3,1,0,1,0,1,0\}\},\\\nonumber&&\{\{14,9,184,12,140,8,264\},\{0,0,0,1,0,1,0,1,1,2,1,2,1,0,0,0\}\},\\\nonumber&&\{\{14,9,192,12,146,8,276\},\{0,0,1,1,0,0,0,1,0,2,2,1,1,1,0,0\}\},\\\nonumber&&\{\{16,9,191,13,143,11,264\},\{0,0,1,1,0,1,0,1,0,2,1,1,1,0,0,0\}\}.
\end{eqnarray}
In Table~\ref{t:distinct}, we show the distribution of all polytopes,
polytopes with large $h^{1,1}$, and polytopes with large $h^{2,1}$
according to the number of distinct fiber types.  There are at most
three distinct fiber types in the polytopes in $h^{2,1}\geq
140$. While all fiber types occur in the large $h^{1,1}$ region, the
only fiber types that occur in the large $h^{2,1}$ region are $F_{1},
F_{3}, F_{4}, F_{6}, F_{8}, F_{10}, F_{11}, $ and $F_{13}.$

\begin{table}
\begin{center}
\begin{tabular}{|c|r|r|r|}
\hline
\begin{tabular}[c]{@{}c@{}}\# distinct \\ fiber types\end{tabular} & \# polytopes & \begin{tabular}[c]{@{}c@{}}\# polytopes with \\ $h^{1,1}\geq 140$\end{tabular} & \begin{tabular}[c]{@{}c@{}}\# polytopes with \\ $h^{2,1}\geq 140$\end{tabular} \\ \hline
0                                                                  &   4           &                                                                       1 &                                                                               3 \\ \hline
1                                                                  &      393788        &                                       153601 &                                                                               229443 \\ \hline
2                                                                  &     86008         &                         78995 &                                                                               6460 \\ \hline
3                                                                  &       13354       &          13347 &                                                                               7 \\ \hline
4                                                                  &        1755      &                     1755 &                                                                               - \\ \hline
5                                                                  &     469         &           469 &                                                                               - \\ \hline
6                                                                  &       112       &                112 &                                                                               - \\ \hline
7                                                                  &       17       &                   17 &                                                                               - \\ \hline
8                                                                  &       8       &                             8 &                                                                               - \\ \hline
\end{tabular}
\end{center}
\caption[x]{\footnotesize  Distribution of polytopes by number of
  distinct fiber types}
\label{t:distinct}
\end{table}

Finally, it is interesting to note that only the plot of $F_{10}$ in
Appendix \ref{sec:appendix-results} seems to exhibit  mirror
symmetry to any noticeable extent.  We do not expect elliptic
fibrations to respect mirror symmetry, so this may simply arise from a
combination of the observation that the total set of hypersurface
Calabi-Yau Hodge numbers in the Kreuzer-Skarke database is mirror
symmetric and the observation that in the large Hodge number regions
that we have considered most of the Calabi-Yau threefolds admit
elliptic fibrations described by a $F_{10}$ fibration of the
associated polytope.

\subsection{Standard vs. non-standard $\P^{2,3,1}$-fibered polytopes}

\begin{figure}
  \centering
  \includegraphics[width=7cm]{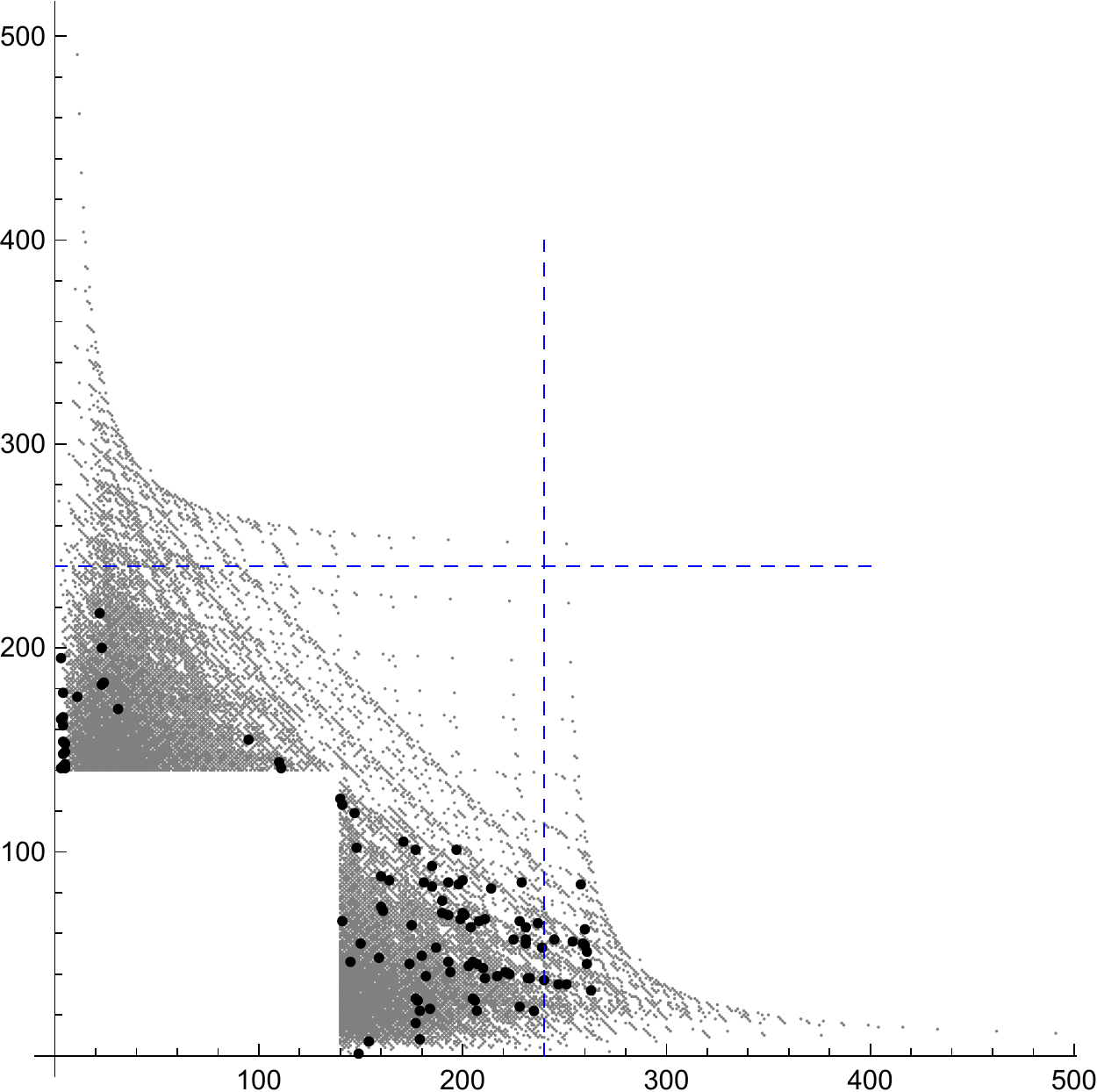}
  \caption{\footnotesize Hodge pairs with only non-standard $F_{10}$-fibered polytopes. The grey dots correspond to all Hodge pairs with $F_{10}$ fibers. The black dots correspond to Hodge pairs with only non-standard $F_{10}$-fibered polytopes.  The vertical and horizontal dashed line correspond to $h^{1,1}=240$ and $h^{2,1}=240$, respectively.} 
  \label{fig:nonstand}
\end{figure}

In \cite{Huang-Taylor-long}, we compared elliptic and toric
hypersurface Calabi-Yau threefolds with Hodge numbers $h^{1,1}\geq
240$ or $h^{2,1}\geq 240$. We found that in the large $h^{1,1}$
region, there were eight Hodge pairs in the KS database that were not
realized by a simple
 Tate-tuned model, and do not correspond to a ``standard
 stacking''
$\P^{2,3,1}$-fibered polytope.
We found, however, that these eight outlying polytopes have a
description in terms of a $\P^{2,3,1}$ fiber structure that
is not of the standard ($v^{(F)}_s=(-3,-2)$)  stacking form, and furthermore it can be seen
do not
respect the stacking framework of \S\ref{ncb}. 
The
Weierstrass models of these Calabi-Yau threefolds
all have the novel feature that they can have gauge groups tuned over
non-toric curves, which can be of higher genus, in the base.
As discussed in \cite{Huang-Taylor-long},
the definition of a standard $\P^{2,3,1}$-fibered polytope
$\dd$ (where the base is stacked over the vertex $(-3,-2)$ of the $F_{10}$ fiber) turns out to be equivalent to the condition that the corresponding $\ds$ has a single lattice point for
each of the choices $\mF_2=(1,-1)$ and $\mF_3=(-1,2)$ in equation
(\ref{dsform}) (where we have numbered  the vertex with the largest
multiple of $- K_B $ as $\mF_1 = (- 1, -1)$), and there is furthermore a
coordinate system in which
this lattice point has coordinates
$\m=(0,0)$ in both cases. We have scanned through
the $F_{10}$-fibered polytopes and used this feature to compute the
fraction of $F_{10}$-fibered polytopes that have the standard versus
non-standard
 form; the results of this analysis are shown in
Table~\ref{t:standard}.

\begin{table}
\begin{center}
\begin{tabular}{|c|c|c|c|}
\hline
                                                                & total \# fibrations & \begin{tabular}[c]{@{}c@{}}\# fibrations in polytopes \\ with $h^{1,1}\geq 140$\end{tabular} & \begin{tabular}[c]{@{}c@{}}\# fibrations in polytopes \\ with $h^{2,1}\geq 140$\end{tabular} \\ \hline
standard                                                        & 433827               & 242562                                                                                    & 192218                                                                                      \\ \hline
non-standard                                                    & 183818               & 130255                                                                                       &53705                                                                                     \\ \hline
\begin{tabular}[c]{@{}c@{}}non-standard\\ fraction\end{tabular} & 0.297611             & 0.349381                                                                                     & 0.218381                                                                                    \\ \hline
\end{tabular}
\end{center}
\caption[x]{\footnotesize  Fractions of fibrations by the fiber
  $F_{10}$ that take the ``standard stacking'' form versus
  other fibrations.}
\label{t:standard}
\end{table}

Of the 488119 $F_{10}$-fibered polytopes, 98758 have more than one
$F_{10}$ fiber. Most of these polytopes have both standard and
non-standard types of fibrations. There are 103 Hodge pairs that have only the
non-standard fibered polytopes. These may give rise to more interesting
Weierstrass models, like those we have studied with $h^{1,1}\geq
240$ in section 6.2 of \cite{Huang-Taylor-long}. As a crosscheck to
the ``sieving'' results there, we have confirmed that
none of these 103 Hodge pairs are in the region
$h^{2,1}\geq 240$, and the 12 Hodge pairs of these 103 pairs that have $h^{1,1}\geq
240$ are exactly the Hodge pairs associated with non-standard $\P^{2,3,1}$-fibered
polytopes in Table 17 of  \cite{Huang-Taylor-long}, together with the
four Hodge pairs of Bl$_{[0,0,1]}\P^{2,3,1}$-fibered polytopes;
the latter four polytopes,  in other words, happen to also be $F_{11}$-fibered, and can be
analyzed as blowups of standard $\P^{2,3,1}$ model (U(1) models). We
list the remaining 91 Hodge pairs that only have non-standard
$\P^{2,3,1}$ fiber types below
(see also Figure \ref{fig:nonstand}):
\begin{itemize}
\item $140\leq h^{1,1}<240$\\
$\{\{149,1\},\{154,7\},\{179,8\},\{177,16\},\{179,22\},\{207,22\},\{235,22\},\{184,23\},\{228,24\}$,\\
$\{178,27\},\{206,27\},\{177,28\},\{205,28\},\{211,38\},\{232,38\},\{233,38\},\{182,39\},\{217,39\}$,\\
$\{223,40\},\{194,41\},\{221,41\},\{210,43\},\{203,44\},\{174,45\},\{207,45\},\{145,46\},\{193,46\}$,\\
$\{205,46\},\{159,48\},\{180,49\},\{187,53\},\{239,53\},\{150,55\},\{231,55\},\{225,57\},\{231,57\},$\\
$\{204,63\},\{231,63\},\{175,64\},\{237,65\},\{141,66\},\{208,66\},\{228,66\},\{199,67\},\{211,67\},$\\
$\{193,69\},\{201,69\},\{190,70\},\{200,70\},\{161,71\},\{160,73\},\{190,76\},\{214,82\},\{185,83\},$\\
$\{198,84\},\{181,85\},\{193,85\},\{229,85\},\{164,86\},\{200,86\},\{160,88\},\{185,93\},\{177,101\}$,\\
$\{197,101\},\{148,102\},\{171,105\},\{147,119\},\{141,123\},\{140,126\}\}$
\item $140\leq h^{2,1}<240$\\
$\{\{3,141\},\{3,165\},\{3,195\},\{4,142\},\{4,148\},\{4,154\},\{4,162\},\{4,166\},\{4,178\},\{5,141\}$,\\
$\{5,143\},\{5,149\},\{5,153\},\{11,176\},\{22,217\},\{23,182\},\{23,200\},\{24,183\},\{31,170\}$,\\
$\{95,155\},\{110,144\},\{111,141\}\}$.
\end{itemize}




\section{Fibration prevalence as a function of $h^{1, 1}(X)$}
\label{sec:prevalence}

In this section we consider the fraction of Calabi-Yau threefolds at a
given value of the Picard number $h^{1, 1}(X)$ that admit a genus one
or elliptic fibration.  We begin in \S\ref{sec:analytic-cubic} with a
summary of some analytic arguments for why we expect that an
increasingly small fraction of Calabi-Yau threefolds will fail to
have
such a fibration as $h^{1, 1}$ increases; we then present some
preliminary
numerical results in \S\ref{sec:small-numbers}.

\subsection{Cubic intersection forms and genus one fibrations}
\label{sec:analytic-cubic}

For some years, mathematicians have speculated that the structure of
the triple intersection form on a Calabi-Yau threefold may make the
existence of a genus one or elliptic fibration increasingly likely as
the Picard number $\rho (X) =h^{1, 1} (X)$ increases.  The rationale
for this argument basically boils down to the fact that a cubic in $k$
variables is increasingly likely to have a rational solution as $k$
increases.  In this section we give some simple arguments that explain
why in the absence of unexpected conspiracies this conclusion is
true.  If this result could be made rigorous it would be a significant
step forwards towards proving the finiteness of the number of distinct
topological types of Calabi-Yau threefolds.

As summarized in \cite{aggl-2}, the following conjecture is due to Koll\'ar
\cite{Kollar}:
\begin{conjecture}
Given  a Calabi-Yau $n$-fold $X$, $X$ is genus one (or elliptically)
fibered iff there exists a divisor $D \in H^{2} (X,\Q)$ that satisfies
$D^n = 0, D^{n -1} \neq 0$, and $D \cdot C \geq 0$ for all algebraic
curves $C \subset X$.
\end{conjecture}
Basically the idea is that $D$ corresponds to the lift $D =\pi^{-1} (D^{(B)})$
of a divisor $D^{(B)}$ on the base of the fibration, where the $(n - 1)$-fold
self-intersection of $D$  gives a  positive multiple of the fiber
$F = \pi^{-1}(p)$, with $p$ a point on the base.
This conjecture was proven  already for $n = 3$ 
by Oguiso and Wilson \cite{Oguiso, Wilson} under the additional
assumption that either $D$ is effective or $D \cdot c_2 (X)\neq 0$.
In the remainder of this section, as elsewhere in the paper, we often simply refer to a Calabi-Yau as genus one
fibered as a condition that includes both elliptically fibered
Calabi-Yau threefolds and more general genus one fibered threefolds.

 In the case $n = 3$, to show that a Calabi-Yau threefold is
genus one fibered, we thus wish to identify an effective divisor $D$ whose triple
intersection with itself vanishes.  The triple intersection form can
be written in a particular basis $D_i$ for $H^2 (X,\Z)$ as
\begin{equation}
\langle   A, B, C \rangle =
\sum_{i, j, k} \kappa_{i jk} a_ib_jc_k \,,
\label{eq:}
\end{equation}
where $A =\sum_i a_i D_i$, etc., and $D_i \cap D_j \cap D_k = \kappa_{i j k}$
The condition that there is a divisor $D = \sum_i d_i D_i$ satisfying
 $D^3 = 0$ is then the condition that the cubic 
intersection form on $D$ vanishes
\begin{equation}
D^3 = \langle D, D, D \rangle =
\sum_{i, j, k}^{}  \kappa_{ijk} d_id_jd_k  = 0\,.
\label{eq:cubic-condition}
\end{equation}
We are thus interested in finding a solution over the rational numbers
of a cubic equation in $k = \rho (X)$ variables.  The curve condition
provides a further constraint that $D$ lies in the positive cone
defined by $D \cdot C \geq 0$ for all algebraic curves $C \subset X$.
Note that identifying a rational solution $D$ to
(\ref{eq:cubic-condition}) immediately leads to a solution over the
integers $\hat{d}_i \in\Z\ \forall i$, simply by multiplying by the LCM of
all the denominators of the rational solution $d_i$.

There are basically two distinct ways in which the conditions
for the existence of a divisor  in the positive cone satisfying $D^3=0$
can
fail.
We consider each in turn.
Note that even when the condition $D^3 = 0$ is satisfied, the
condition for an elliptic fibration can fail if $D^2 = 0$, in which
case $D$ itself corresponds to a K3 fiber; this class of fibrations is
also interesting to consider but seems statistically likely to become
rarer as $\rho$ increases.

\subsubsection{Number theoretic obstructions}

There can be a number theoretic obstruction to the existence of a solution
 to a degree $n$ homogeneous equation over the rationals such as
 (\ref{eq:cubic-condition}).\footnote{Thanks to Noam
  Elkies for explaining to us various aspects of the mathematics in this
  section.}  For example, there cannot be an integer
 solution in the variables $x, y, z, w$ of the equation
\begin{equation}
x^3 + x^2 y + y^3 + 2z^3 + 4w^3  = 0\,.
\label{eq:}
\end{equation}
This can be seen as follows: if all the variables $x, y, z, w$ are
even, we divide by the largest possible power of 2 that leaves them
all as integers.  Then there must be a solution with
at least one variable
odd.  The variable $x$ cannot be odd, since if $y$ is odd or even the
LHS is odd.  Similarly, $y$ cannot be odd.  So $x, y$ must be even in
the minimal solution.  But $z$ cannot be odd or the LHS would be
congruent to 2 mod 4.  And $w$ cannot be odd if the others are even
since then the LHS would be congruent to 4 mod 8.

Such number-theoretic obstructions can only arise for small numbers of
variables $k$.  It was conjectured long ago that for a homogeneous
degree $n$ polynomial the maximum number of variables for which such a
number-theoretic obstruction can arise is $n^2$ \cite{Mordell}.  While
there is a counterexample known for $n = 4$, where there is an
obstruction for a quartic with 17 variables, it was proven in
\cite{Heath-Brown} that every {\it non-singular} cubic form in 10
variables with rational coefficients has a non-trivial rational zero.
And the existence of a rational solution has been proven for general
(singular or non-singular) cubics in 16 or more variables
\cite{Davenport}.  Thus, no number-theoretic obstruction to the
existence of a solution to $D^3 = 0$ can arise when $\rho (X) =h^{1,
  1} (X) > 15$, and there are also quite likely no obstructions for $\rho
  (X) > 9$ though this stronger bound is not proven as far as the
  authors are aware.

\subsubsection{Cone obstructions}

If the coefficients in the cubic conspire in an appropriate way, the
cubic can fail to have any solutions in the K\"ahler cone.  We now
consider this type of obstruction to the existence of a solution.  
For
example, the cubic
\begin{equation}
\sum_{i} d_i^3 + \sum_{i, j}d_i^2 d_j + \sum_{i, j, k}d_id_jd_k = 0
\label{eq:}
\end{equation}
has no nontrivial solutions in the cone $d_i \geq 0$ since all
coefficients are positive. 
The absence of solutions in a given cone becomes increasingly
unlikely, however, as the number of variables increases
(again, in the absence of highly structured cubic coefficients).
A somewhat rough and naive approach to
understanding this is to consider adding the variables one at a time,
assuming that the coefficients are random and independently
distributed numbers.
In
this analysis we do not worry about the existence of rational
solutions; in any given region, the existence of a rational solution
should depend upon the kind of argument described in the previous
subsection.
We assume for simplicity that the cone condition states simply that
$d_i \geq 0\ \forall i$; a more careful analysis would consider cones
of different sizes and angles.
For two variables  $x = d_1, y = d_2$ we have a cubic equation
\begin{equation}
\kappa_{111} x^3 + 3\kappa_{112} x^2 y + 3 \kappa_{122}  xy^2 +
\kappa_{222} y^3 \,.
\label{eq:}
\end{equation}
Now assume that $x$ is some fixed value $x \geq 0$.
This cubic always has at least one real solution $(x, y)$.  
If the coefficients in the cubic are randomly distributed, we expect
roughly a 1/2 chance that $y \geq 0$ for this real solution.  Now add
a third variable.  If the above procedure gives a solution $(x, y, z =
d_3 = 0)$ in the positive cone, we are done.  If not, we plug in some
fixed positive values $x, y \geq 0$ and the condition becomes a cubic
in $z$.  Again, there is statistically roughly a 1/2 chance that a
given real solution for $z$ is positive.  So for 3 variables we expect
at most a probability of roughly $1/4$ that there is no solution in
the desired cone.  Similarly, for $k$ variables, this simple argument
suggests that most a fraction of $1/2^{k-1}$ of random cubics will
lack a solution in the desired cone.

This is an extremely rough argument, and should not be taken
particularly seriously, but hopefully it illustrates the general sense
of how it becomes increasingly difficult to construct a cubic that has
no solutions in $k$ variables within a desired cone.  Interestingly,
the rate of decrease found by this simple analysis matches quite
closely with what we find in a numerical analysis of the
Kreuzer-Skarke data at small $k =\rho (X) =h^{1, 1} (X)$.

\subsection{Numerical results for Calabi-Yau threefolds at
small $h^{1,1} (X)$}
\label{sec:small-numbers}

We have done some preliminary analysis of the distribution of
polytopes without a manifest reflexive 2D fiber for cases giving
Calabi-Yau threefolds with small $h^{1, 1}$.  The results of this are
shown in Table~\ref{t:small-h11}.

It is interesting to note that the fraction of polytopes without a
genus one (or elliptic) fiber that is manifest in the toric geometry
decreases roughly exponentially, approximately as $p ({\rm no  \, fiber})\sim 0.1 \times 2^{5-h^{1, 1}}$ in the range $h^{1, 1}\sim 4$---$7$.
Comparing to the total numbers of polytopes in the KS database that lack a manifested genus one fiber, if
this fraction continues to exhibit this pattern, the total number of
polytopes out of the 400 million in the full KS database would be
something like 14,000.
(Note, however, that the polytope identified in the database that has
no manifest fibration and corresponds to a Calabi-Yau with $h^{1,1} =
140$ would be extremely unlikely if this exponential rate of decrease
in manifest fibrations continues; this suggests that the tail of the
distribution of polytopes lacking a manifest fibration does not
decrease quite so quickly at large values of $h^{1,1}$.
Because the analytic argument of the previous section involves all
fibrations, not just manifest ones, it may be that this asymptotic is
still a good estimate of actual fibrations if most of the polytopes at
large $h^{1,1}$ that lack manifest fibrations actually have other
fibrations that cannot be seen from toric fibers.)

\begin{figure}
  \centering
  \includegraphics[width=15cm]{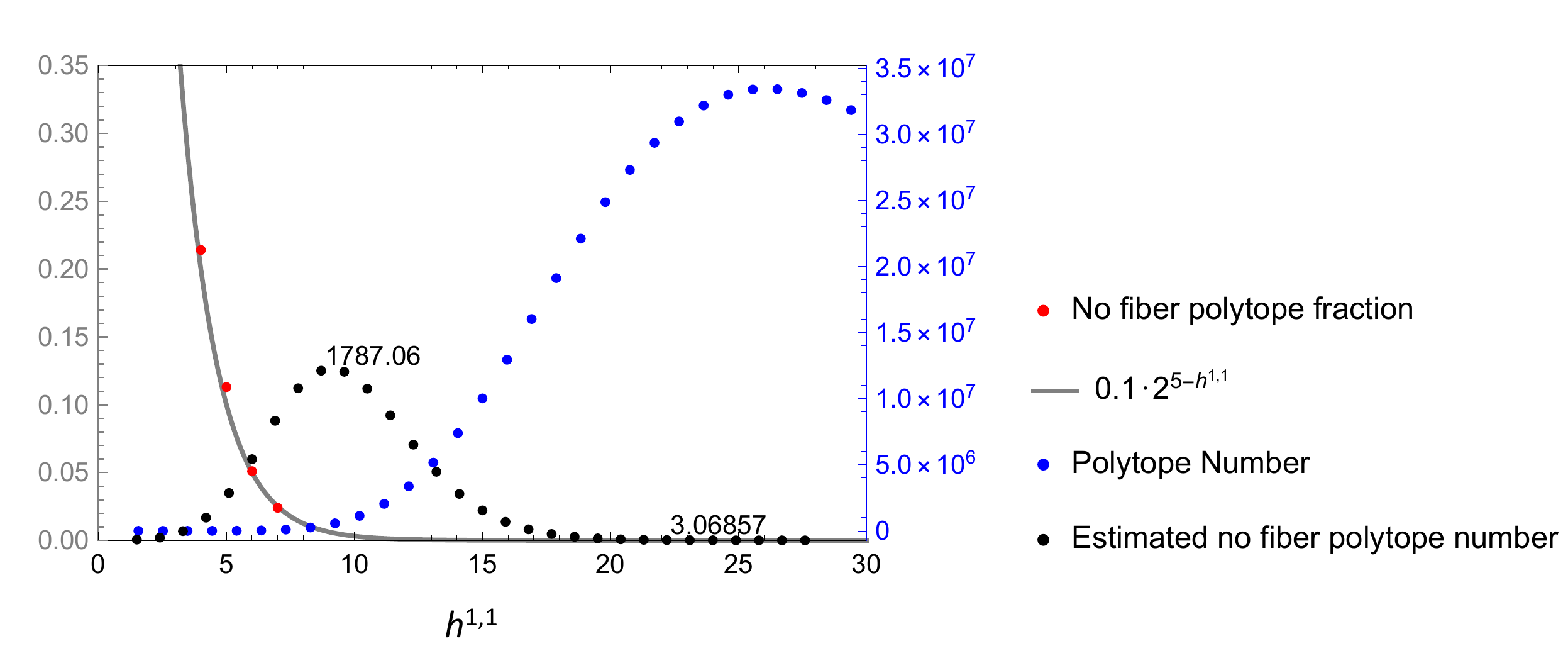}
  \caption{\footnotesize
The fraction of polytopes without a manifest reflexive fiber goes roughly as $0.1
\times 2^{5 - h^{1,1}}$ for small values of $h^{1,1}$.  Continuing
this estimate to higher values of $h^{1,1}$,
    the estimated number of polytopes with no fiber has  a peak value
around $1800$ at $h^{1,1}\sim 9$
 and drops below five around $h^{1,1} \sim 24$. The estimated number of total polytopes with no  manifest fiber is around $14,000$.} 
  \label{fig:est}
\end{figure}

The naive distribution of the estimated number of polytopes from the
simple exponentially decreasing estimate is shown in the black dots in
Figure \ref{fig:est}.
Even with some uncertainty about the exact structure of the tail of
this distribution,
this seems to give good circumstantial
evidence that at least among this family of Calabi-Yau threefolds, the
vast majority are genus one or elliptically fibered, and that the
Calabi-Yau threefolds like the quintic that lack genus one fibration
structure are exceptional rare cases, rather than the general rule.

\begin{table}[]
\centering
\begin{tabular}{|c|cccccc|}
\hline
$h^{1,1}$                                                                             & $2$     & $3$     & $4$     & $5$     & $6$     & $7$     \\ \hline
Total \# polytopes
& $36$    & $244$   & $1197$  & $4990$  & $17101$ & $50376$ \\ \hline
\# without reflexive fiber $\dd_2$
& $23$    & $91$    & $256$   & $562$   & $872$   & $1202$  \\ \hline
\% without reflexive fiber
& $0.639$ & $0.373$ & $0.214$ & $0.113$ & $0.051$ & $0.024$ \\ \hline
\end{tabular}
\caption{\footnotesize The numbers of polytopes without a  2D
  reflexive fiber, corresponding to Calabi-Yau threefolds without a
  manifest genus one fibration, for small values of $h^{1, 1}$}
\label{t:small-h11}
\end{table}

\section{Conclusions} 
\label{sec:conclusions}

The results reported in this paper indicate that most Calabi-Yau
threefolds that are realized as hypersurfaces in toric varieties have
the form of a genus one fibration.  At large Hodge numbers almost all
Calabi-Yau threefolds in the Kreuzer Skarke database satisfy the
stronger condition that they are elliptically fibered.  This
contributes to the growing body of evidence that most Calabi-Yau
threefolds lie in the finite class of elliptic fibrations.
We have shown that all known Calabi-Yau threefolds where
at least one of the Hodge numbers is greater than 150 must have a
genus one fibration, and all CY3's with
$h^{1, 1}\geq 195$ or $h^{2, 1}\geq 228$ have an elliptic fibration.  
We have also shown that the fraction of toric hypersurface Calabi-Yau
threefolds that are not manifestly genus one fibered decreases exponentially
roughly as $0.1 \times 2^{5 - h^{1,1}}$ for small values of
$h^{1,1}$.  These results correspond well with the recent
investigations in \cite{aggl-2, aggl-3, agh-non-simply}, which showed
that over 99\% of all complete intersection Calabi-Yau (CICY)
threefolds have a genus one fibration (and generally many distinct
fibrations), including all CICY threefolds with $h^{1,1} > 4$, and
that similar results hold for the only substantial known class of
non-simply connected Calabi-Yau threefolds.

Taken together, these empirical results, along with the analytic
arguments described in \S\ref{sec:analytic-cubic}, suggest that it becomes increasingly
difficult to form a Calabi-Yau geometry that is not genus one or
elliptically fibered as the Hodge number $h^{1,1}$ increases.
Proving that any
Calabi-Yau with Hodge numbers beyond a certain value must admit an
elliptic fibration is a significant challenge for mathematicians;
progress in this direction might help begin to place some explicit
bounds that would help in proving the finiteness of the complete set
of Calabi-Yau threefolds.

There are a number of ways in which the analysis of this paper could
be extended.  Clearly, it would be desirable to analyze the fibration
structure of the full set of polytopes in the Kreuzer-Skarke database,
which could be done by implementing the
algorithm used in this paper using faster and more powerful
computational tools.
It is also important to note that while the simple criteria we used
here showed already that most known Calabi-Yau threefolds at large
Hodge numbers are elliptic or more generally genus one fibered, the
cases that are not recognized as fibered by these simple criteria may still
have genus one or elliptic fibers.  In particular, while we have identified a couple of Calabi-Yau threefolds with
$h^{1, 1} > 1$ and either $h^{1, 1}$ or $h^{2, 1}$ greater than 140
that do not admit an explicit toric genus one fibration that can be
identified by a 2D reflexive fiber in the 4D polytope,
it seems quite likely that the Calabi-Yau threefolds associated with these
polytopes may have a non-toric  genus one or elliptic fibration
structure. Such fibrations could be identified by a more extensive
analysis along the lines of \cite{aggl-3}.

For Calabi-Yau threefolds that do not admit any genus one or elliptic fibration, it
would be interesting to understand whether there is some underlying
structure to the triple intersection numbers that is related to those
of elliptically fibered Calabi-Yau manifolds, and whether there are
simple general classes of transitions that connect the
non-elliptically fibered threefolds to the elliptically fibered CY3's,
which themselves all form a connected set through transitions
associated with blow-ups of the base and Higgsing/unHiggsing processes
in the corresponding F-theory models.  We leave further investigation
of these questions for future work.

Finally, it of course would be interesting to extend this kind of
analysis to Calabi-Yau fourfolds.
An early analysis of the fibration structure of some known toric
hypersurface Calabi-Yau fourfolds was carried out in
\cite{Rohsiepe:2005qg}.
The analysis of fibration
structures of complete intersection Calabi-Yau fourfolds in
\cite{Gray-hl} suggests that again most known constructions should
lead predominantly to Calabi-Yau fourfolds that are genus one or
elliptically fibered.  The classification of hypersurfaces in
reflexive 5D polytopes has not been completed, although the complete
set of $3.2 \times 10^{11}$ associated weight systems has recently
been constructed \cite{ss}.  In fact,
recent work on classifying toric threefold bases that can support
elliptic Calabi-Yau fourfolds suggests that the number of such
distinct bases already reaches enormous cardinality on the order of
$10^{3000}$ \cite{Halverson-ls, Wang-WT-MC-2}.  Thus, at this point
the known set of elliptic Calabi-Yau fourfolds is much larger than any
known class of Calabi-Yau fourfolds from any other construction.

\acknowledgments{
We would like to thank Lara Anderson,
Andreas Braun,
Noam Elkies,
James Gray,
Sam Johnson,
Nikhil Raghuram, 
David
  Morrison, 
Andrew Turner, 
Yinan Wang, and Timo Wiegand  for helpful discussions. 
This material is based upon work supported by the U.S.\ Department of
Energy, Office of Science, Office of High Energy Physics under
grant Contract Number
DE-SC00012567.
WT would like to thank the Aspen Center for Physics
for hospitality during the completion of this work; the Aspen Center
for Physics is supported by National Science Foundation grant
PHY-1607611.
}

\newpage
\appendix

\section{The 16 reflexive 2D fiber polytopes $\dd_2$}
\label{sec:appendix-fibers}

We list here the 16 reflexive 2D polytopes $\dd_2$.  The dual
polytopes $\ds_2$ are listed in
Appendix~\ref{sec:appendix-fibers-dual}.
With each polytope we also provide the value $I_{\rm max}$ associated
with the maximum possible value of $v \cdot m$ where $v \in\dd_2, m
\in\ds_2$.
As discussed in the main text, the three fibers $F_1 =\P^2, F_2 = \P^1
\times \P^1 =\F_0, F_4 =\F_2$ have no $-1$ curves, associated with
divisors that give global sections; all other fibers have $- 1$ curves
and correspond to elliptic fibers of the Calabi-Yau threefold.

\vspace*{0.2in}
{\centering
\begin{tabular}{cccc}
\includegraphics[height=3.3cm]{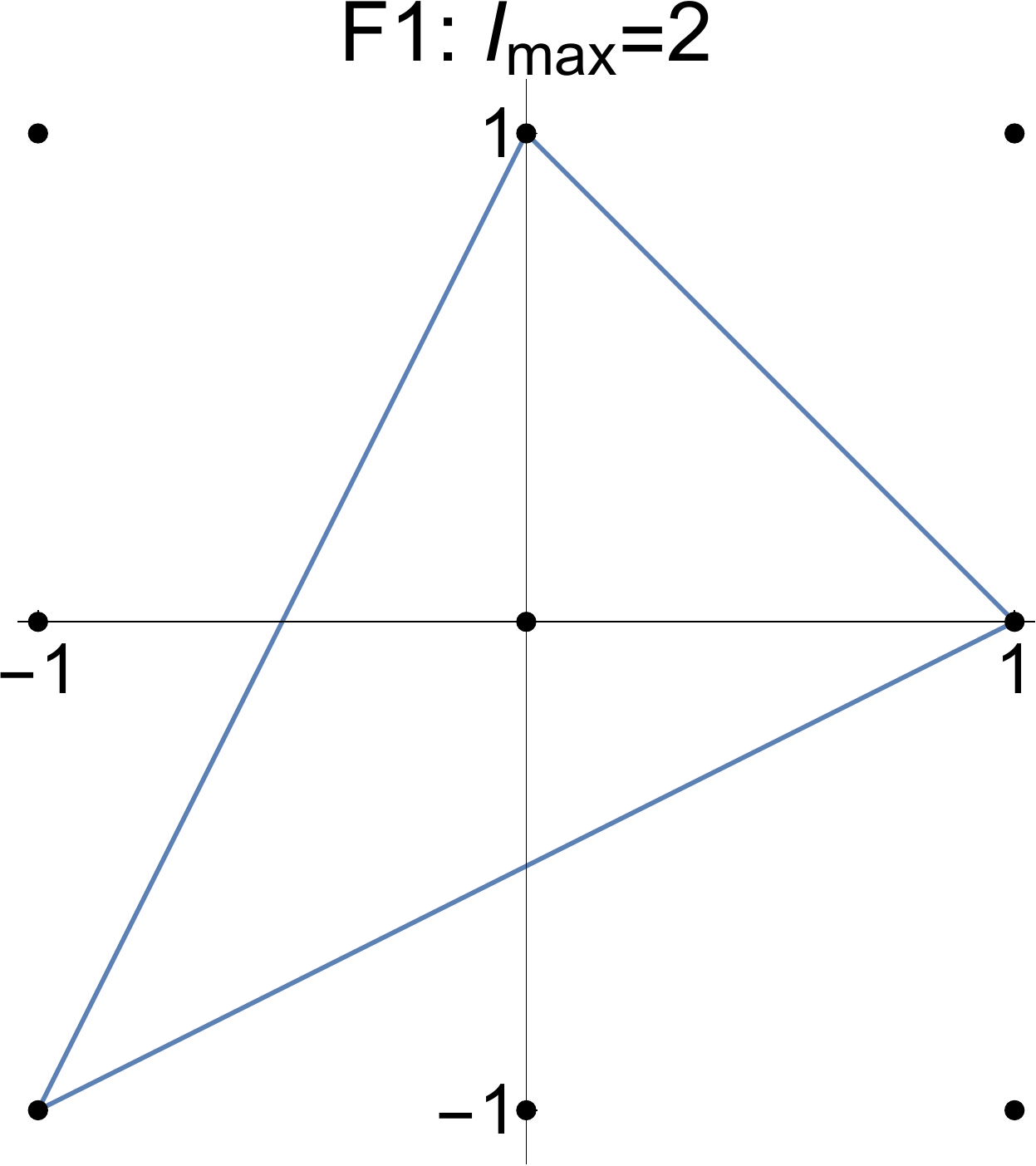}  &\includegraphics[height=3.3cm]{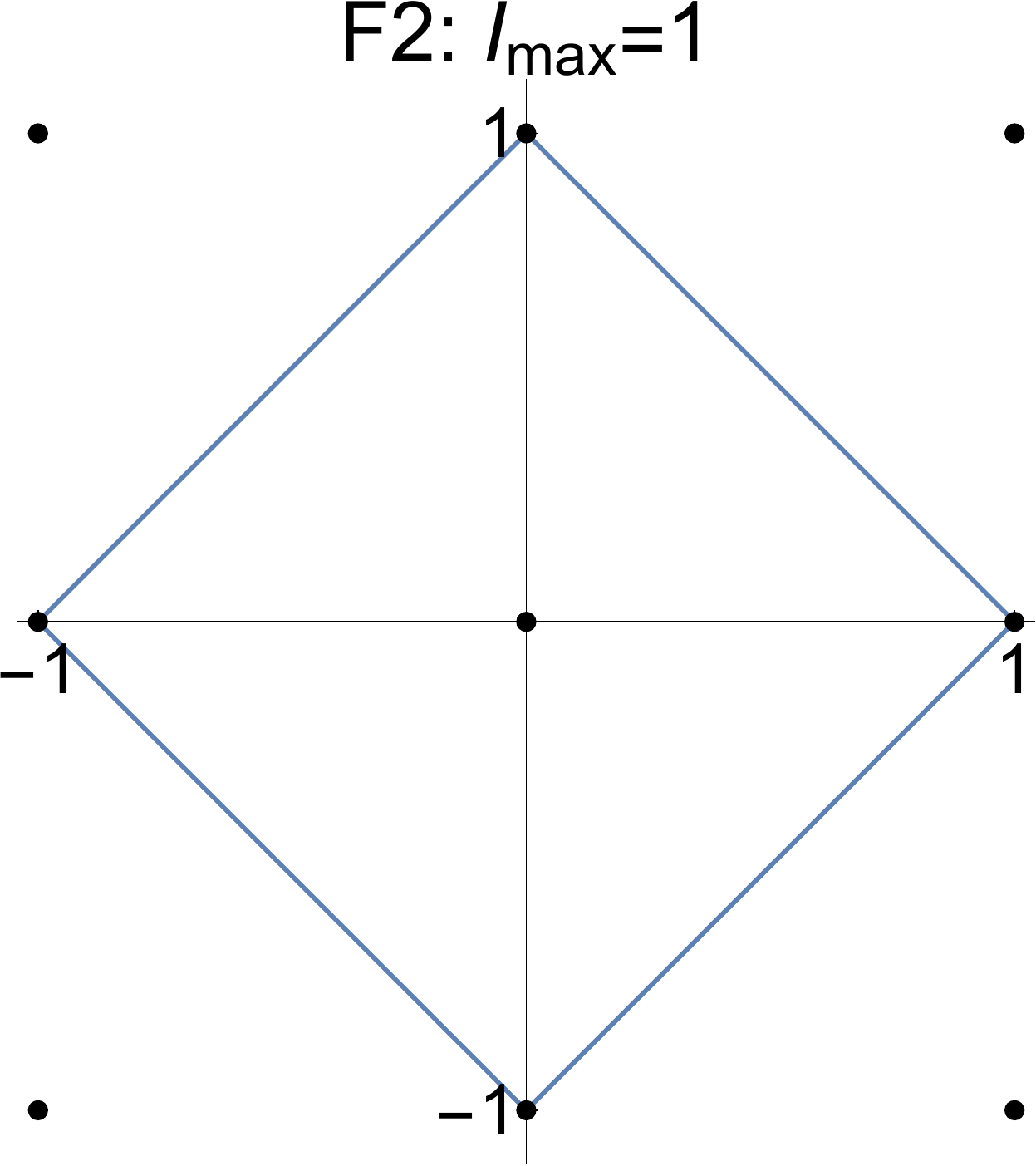}    & \includegraphics[height=3.3cm]{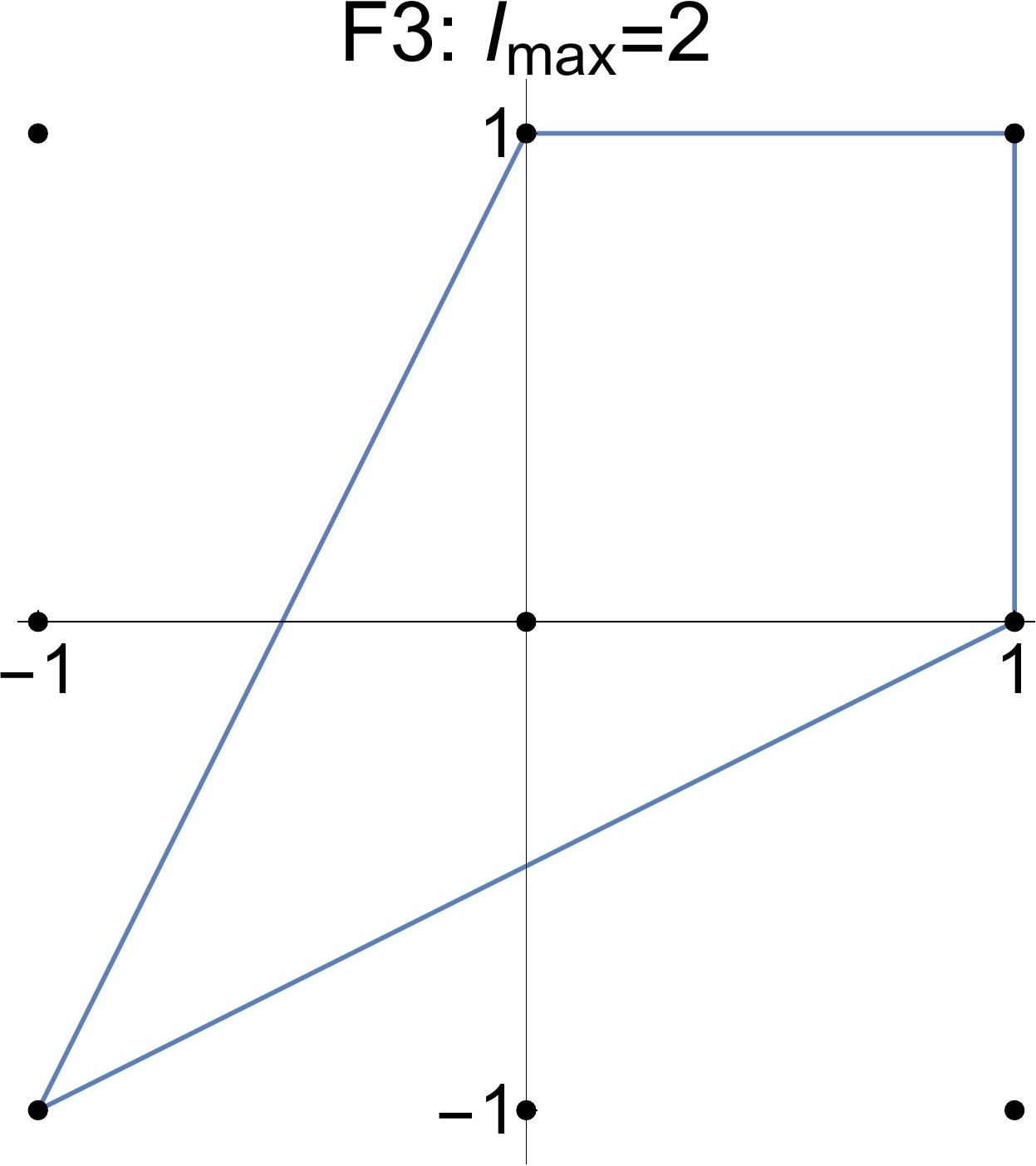}   & \includegraphics[height=3.3cm]{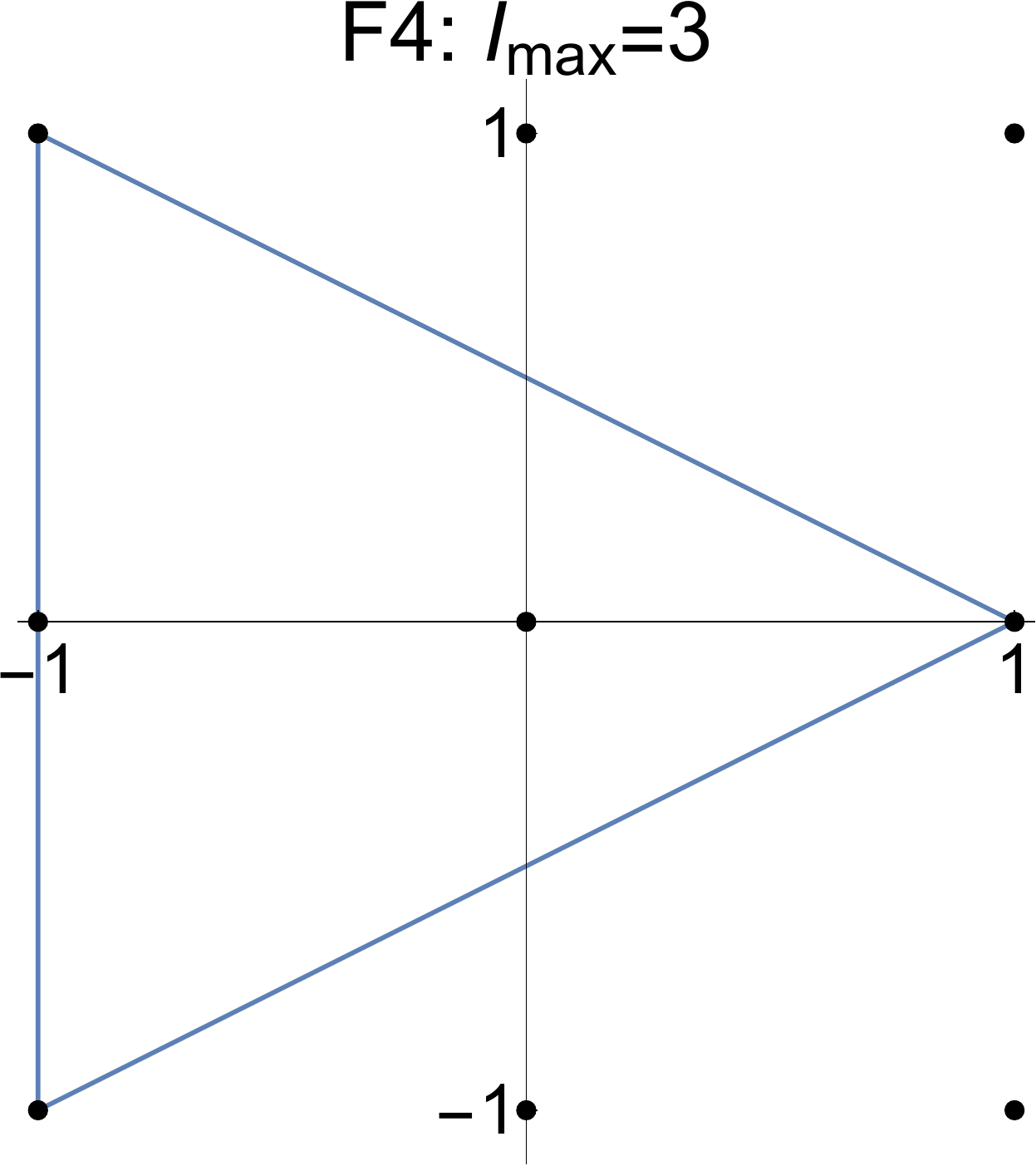}   \\\\
\includegraphics[height=3.3cm]{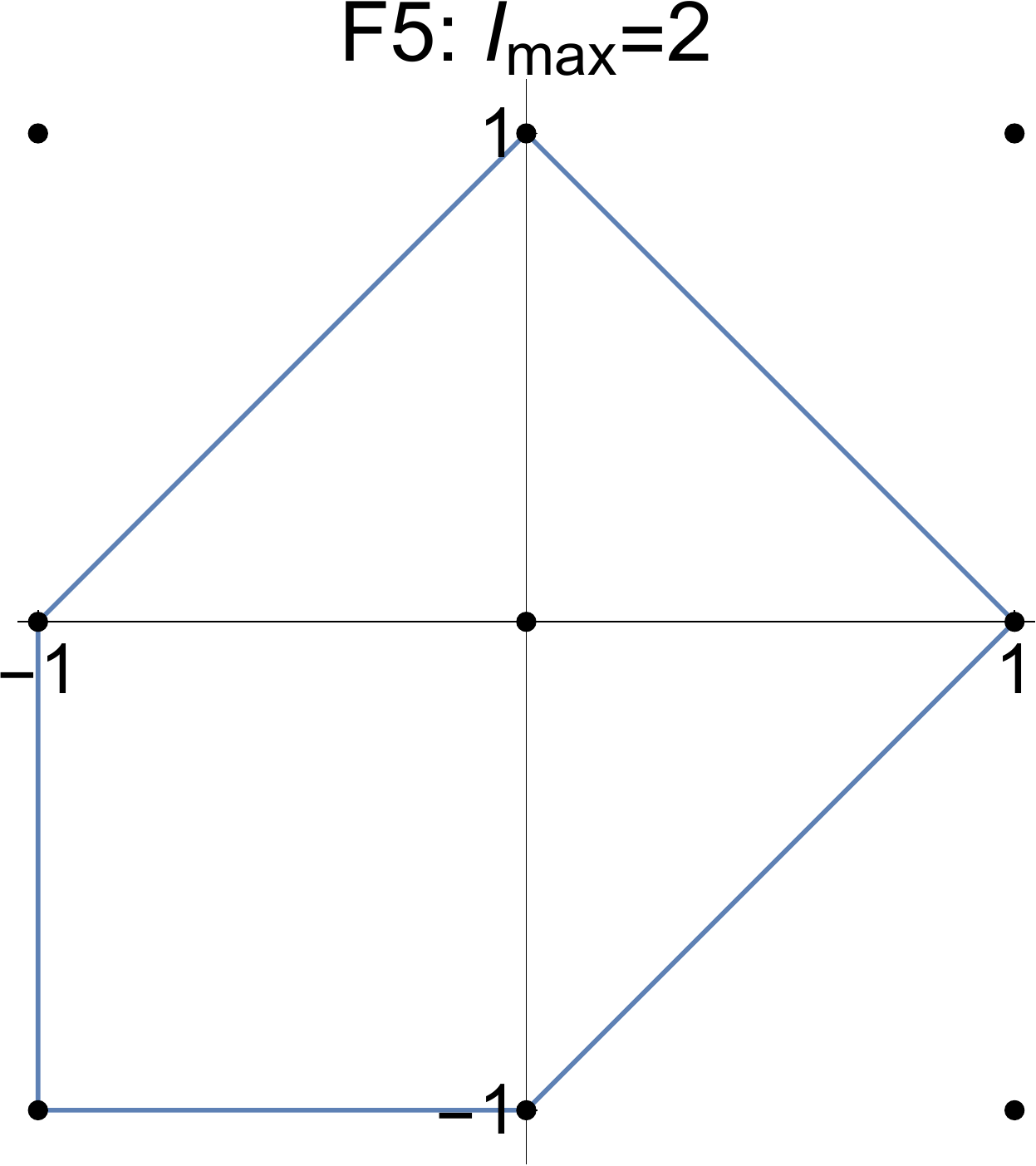}          & \includegraphics[height=3.3cm]{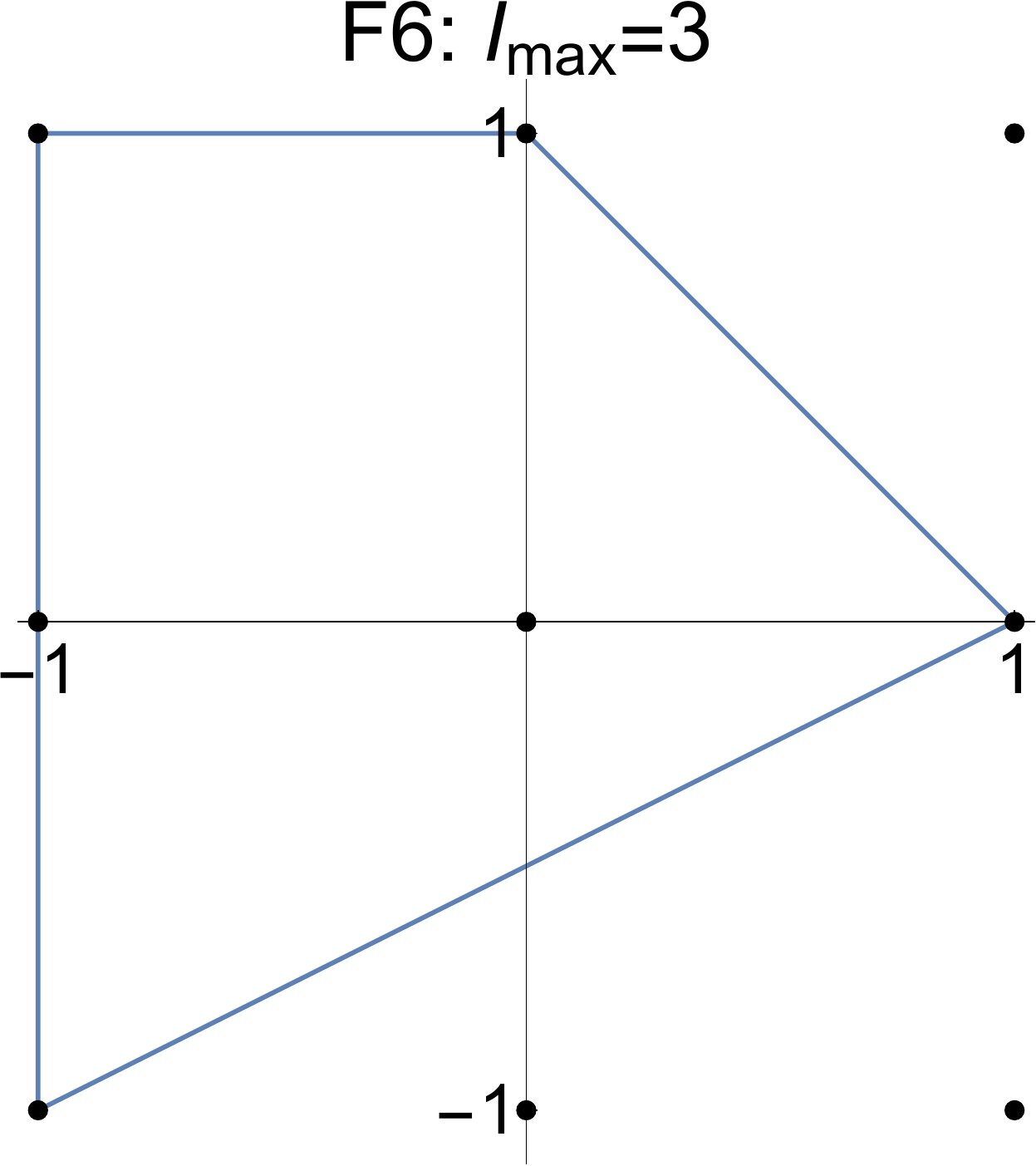}       &\includegraphics[height=3.3cm]{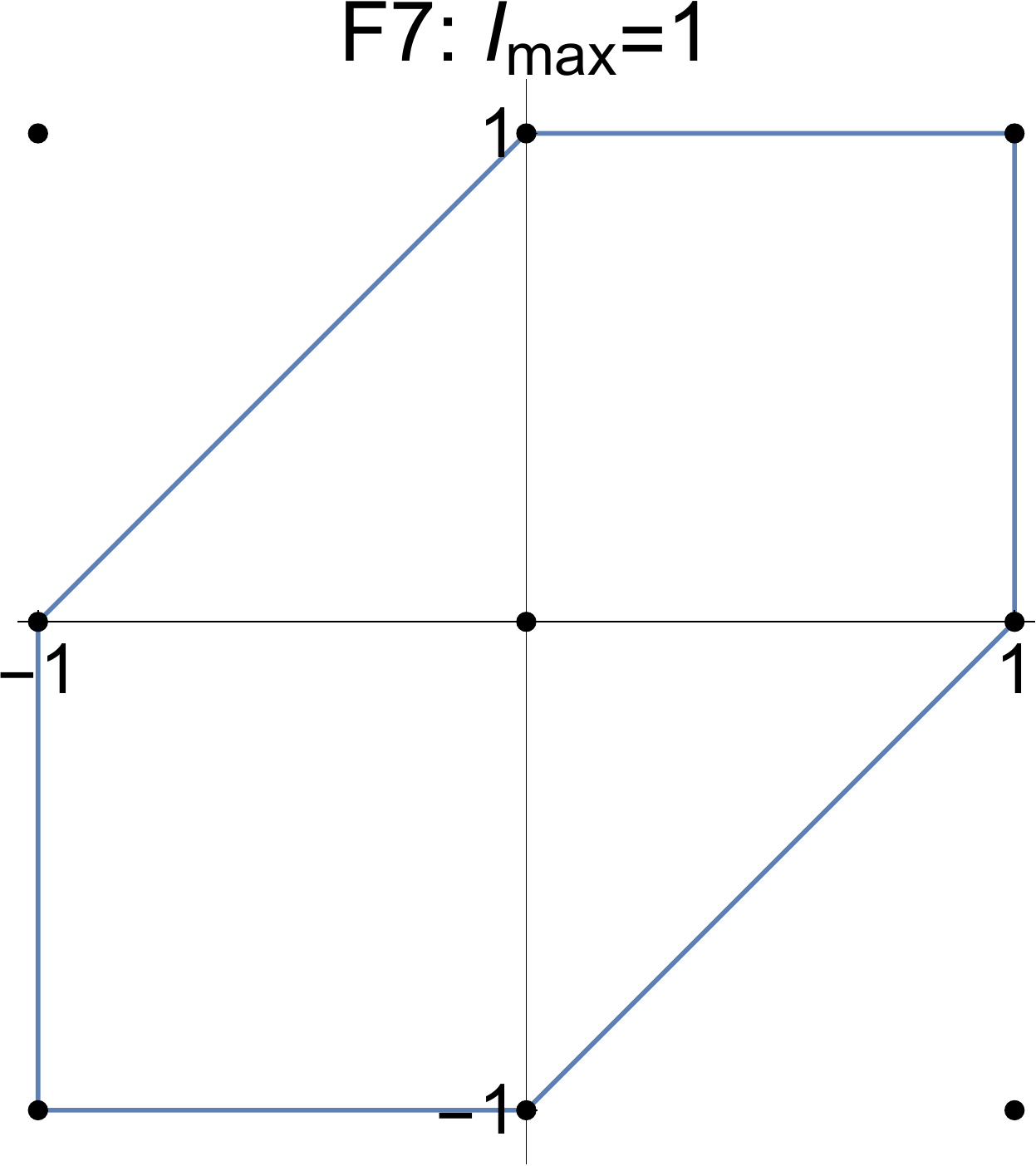}        &\includegraphics[height=3.3cm]{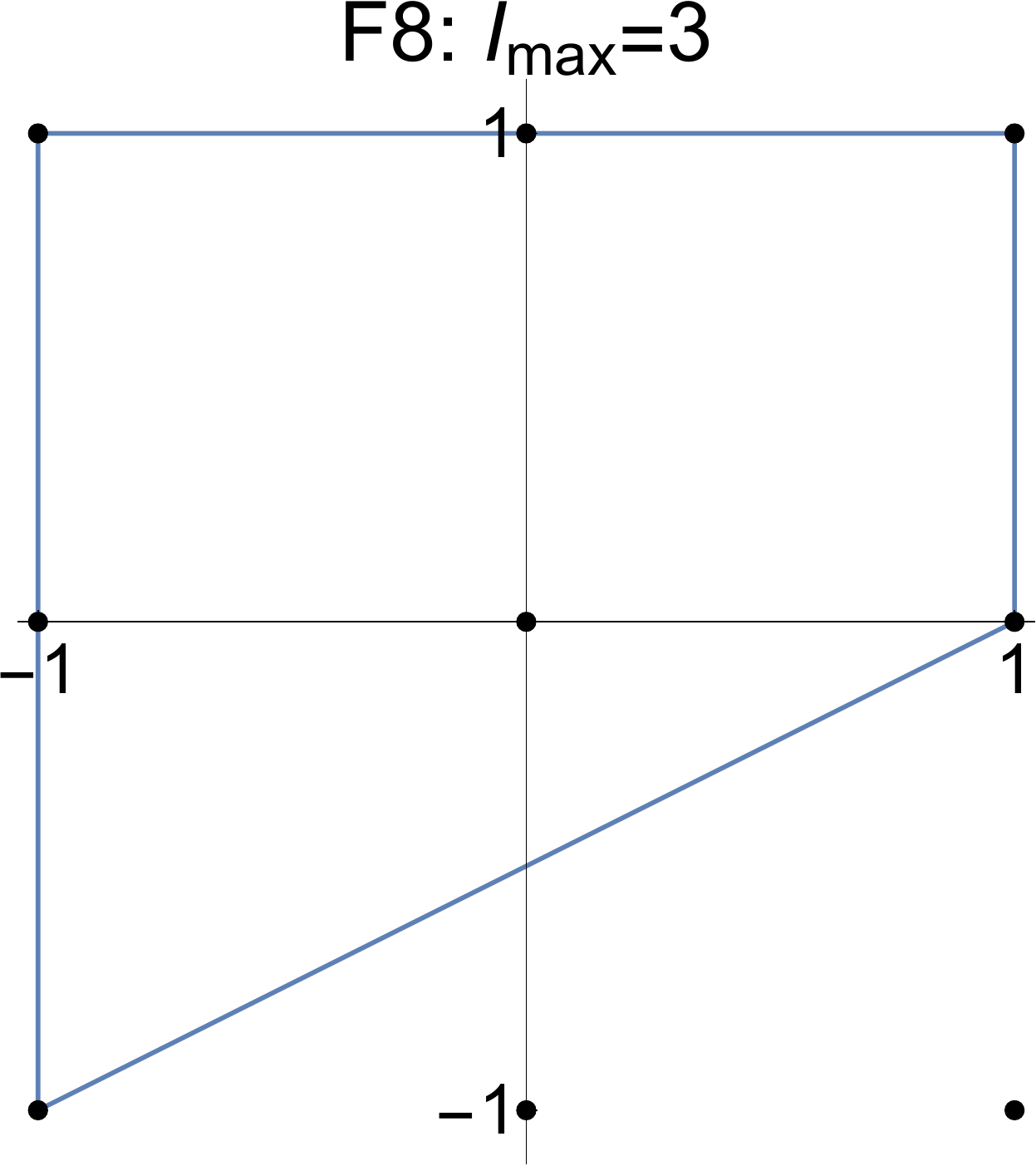}        \\\\
\includegraphics[height=3.3cm]{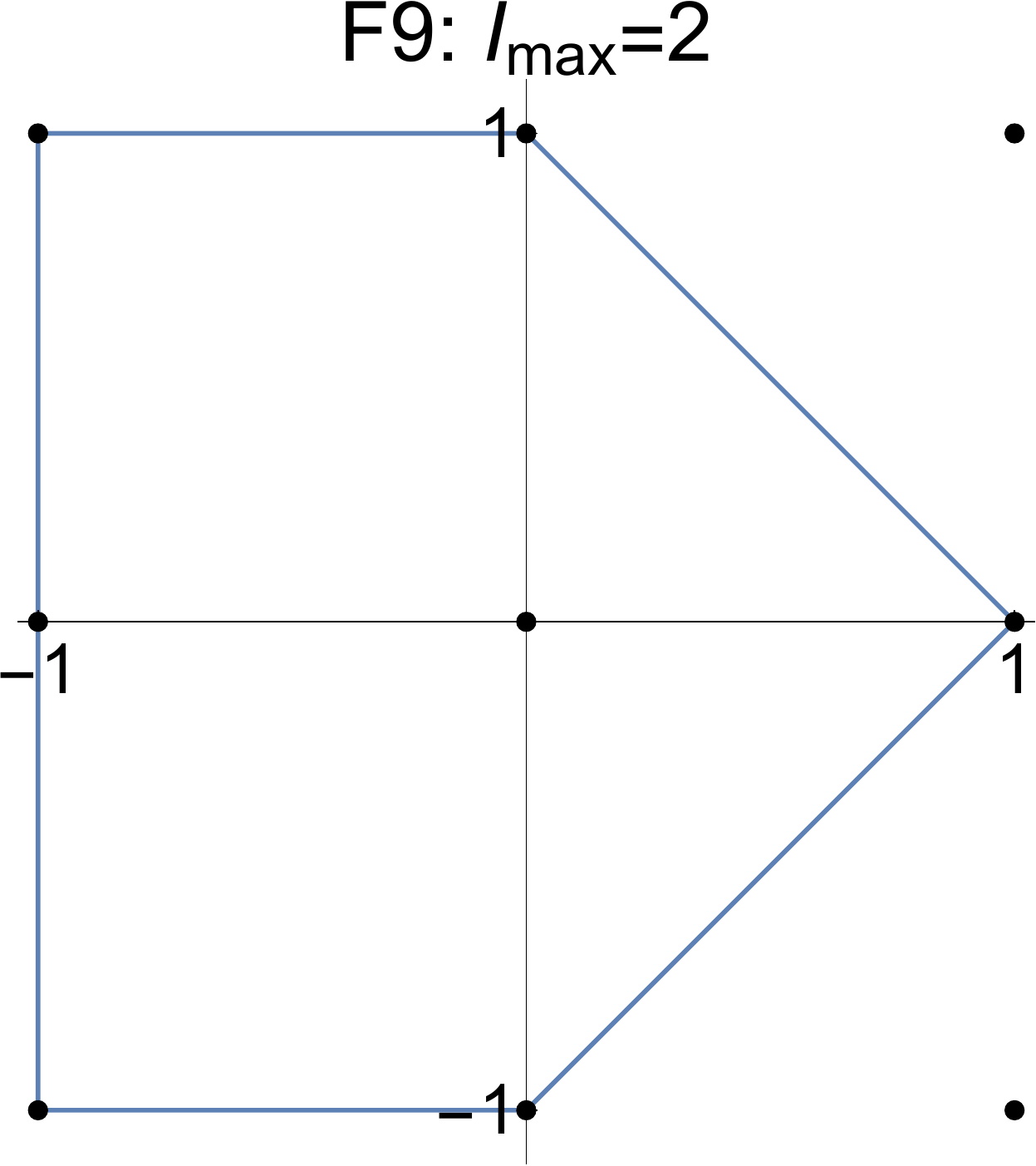}          & \includegraphics[height=3.3cm]{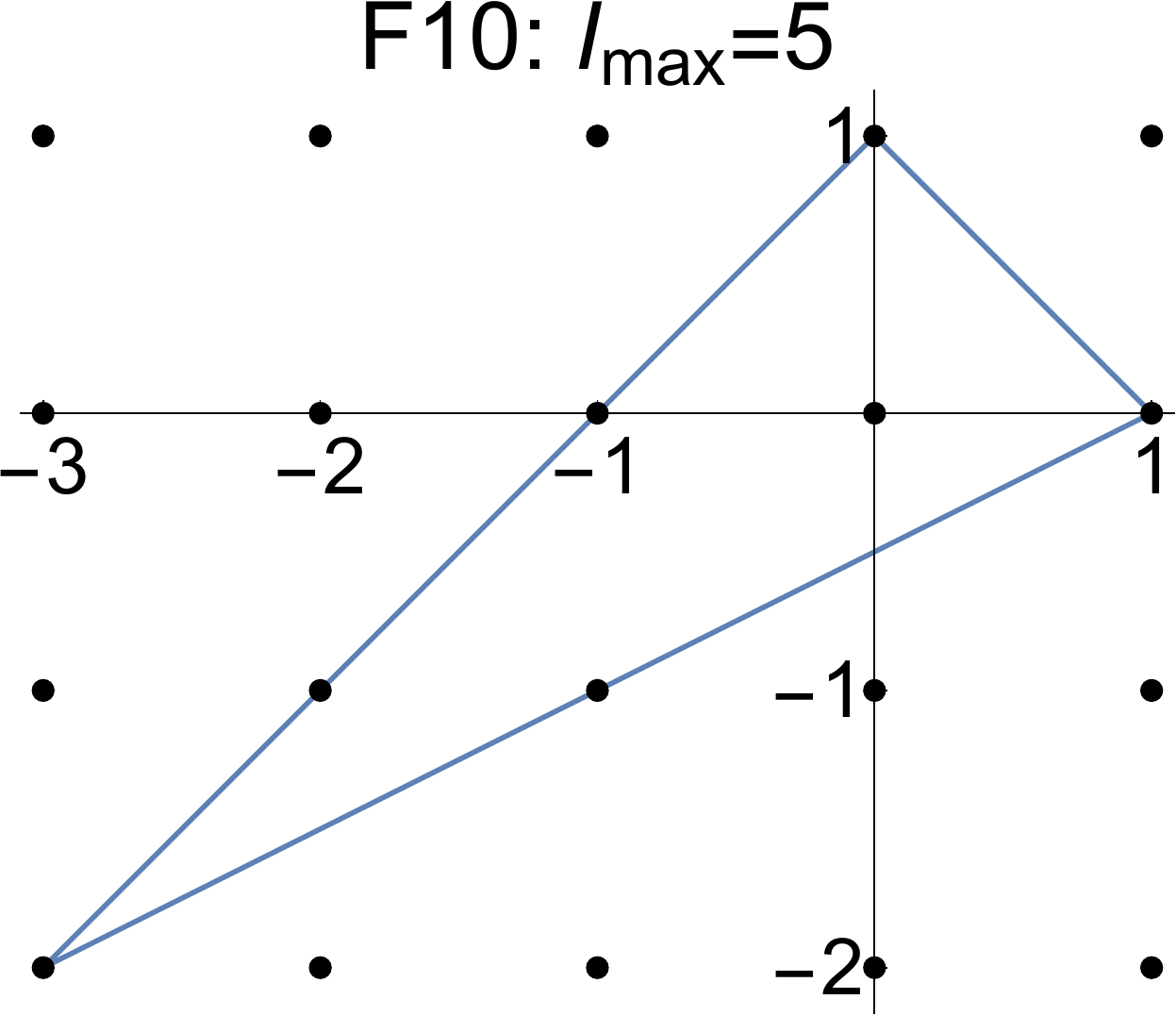}       &\includegraphics[height=3.3cm]{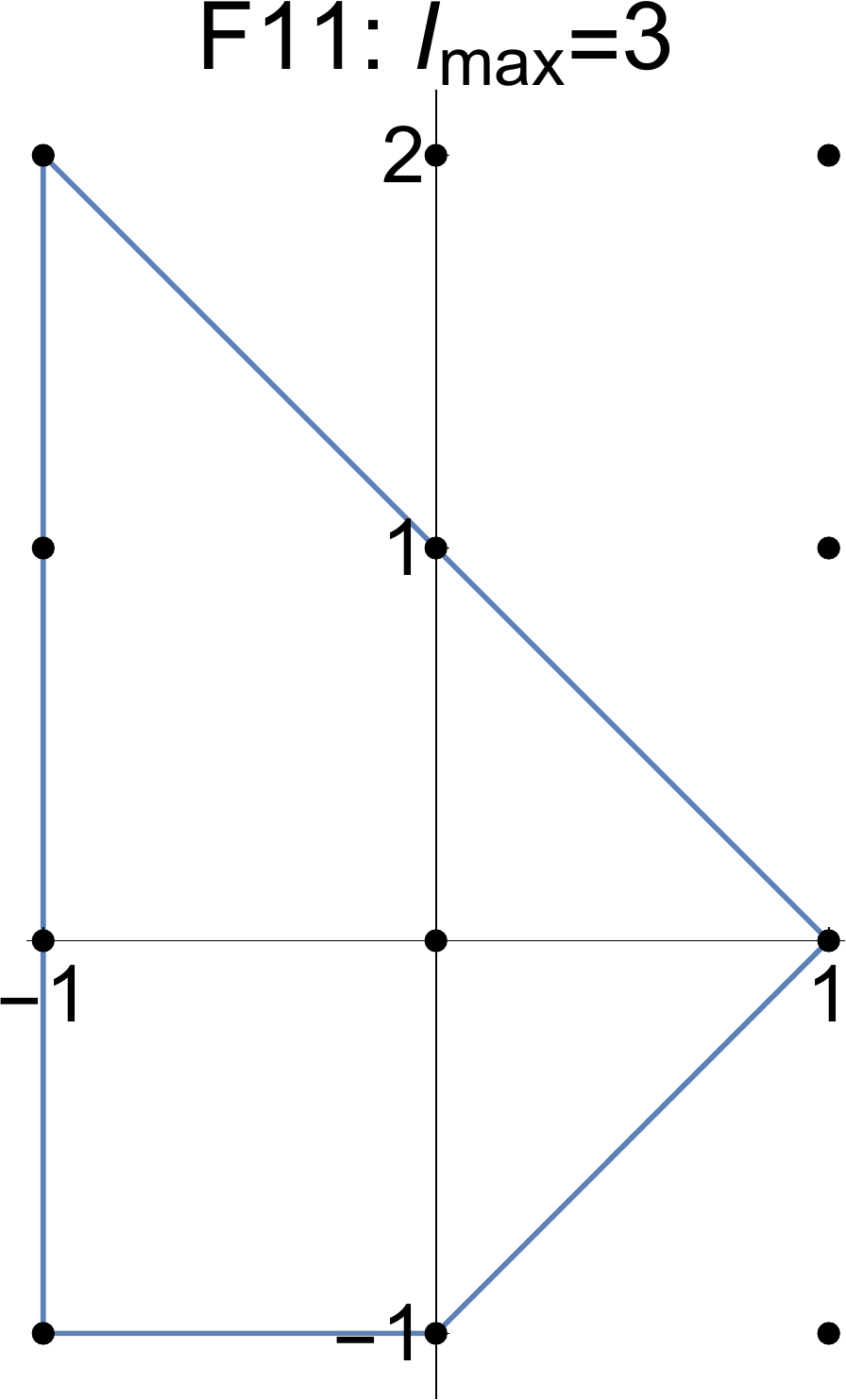}        &\includegraphics[height=3.3cm]{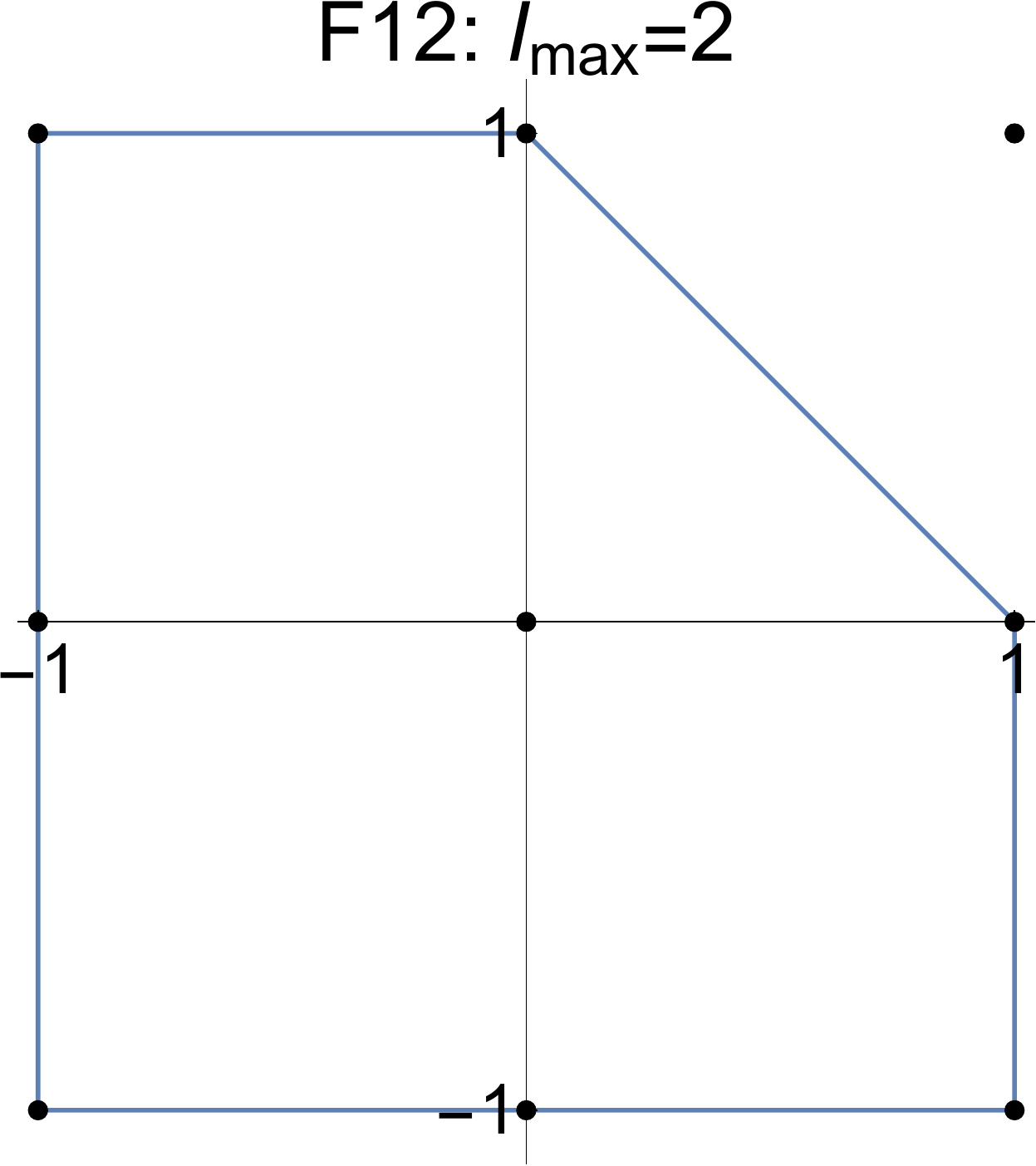}        \\\\
\includegraphics[height=3.3cm]{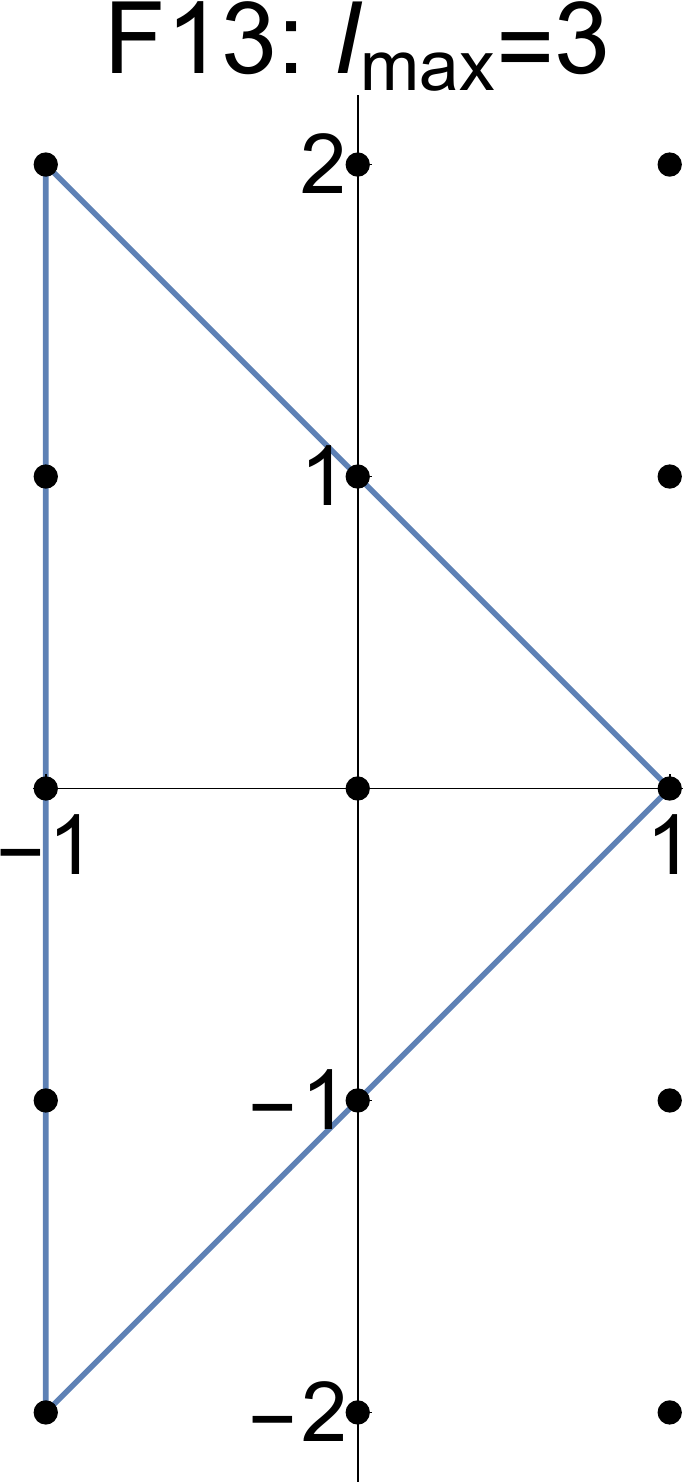}          & \includegraphics[height=3.3cm]{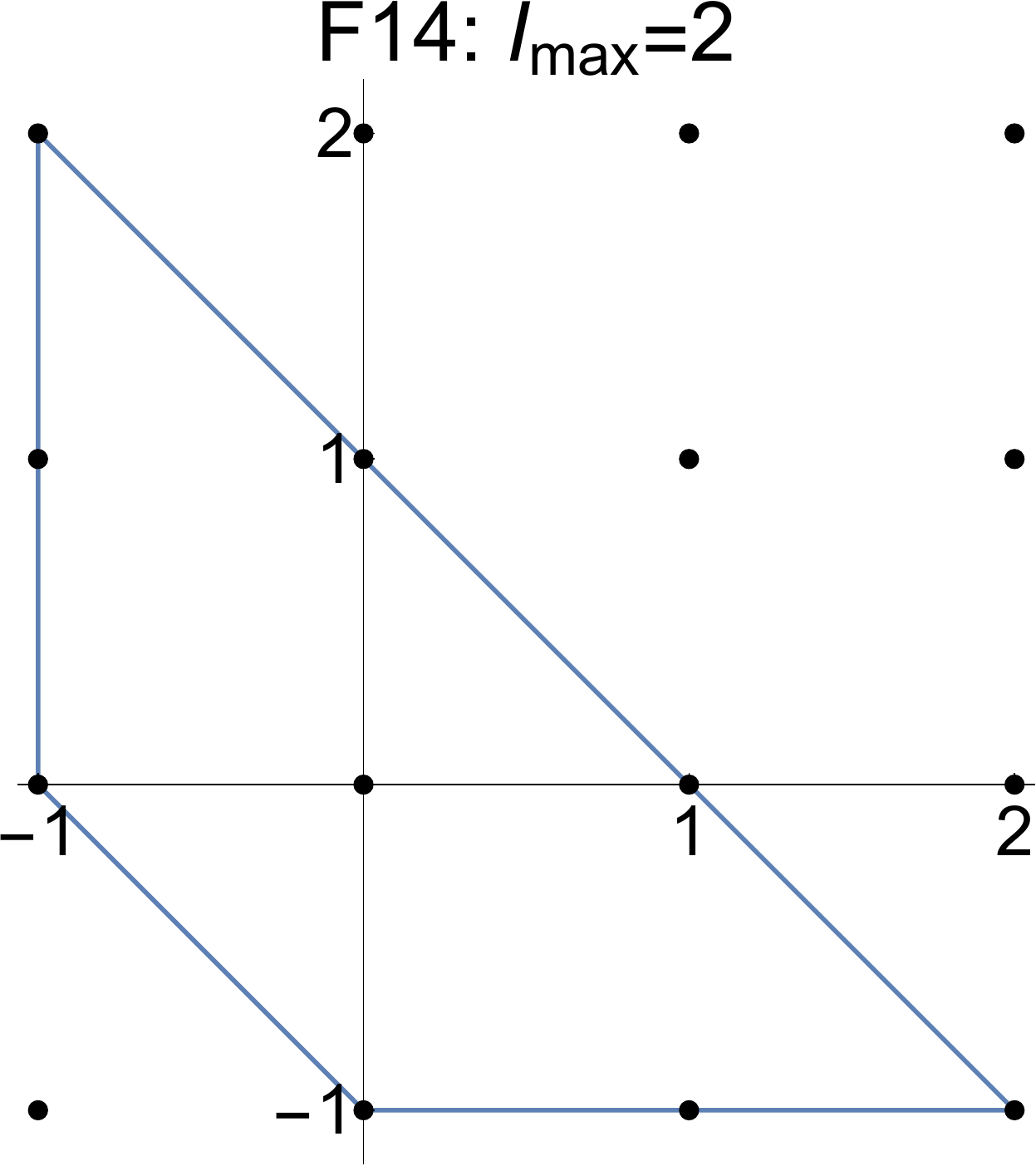}       & \includegraphics[height=3.3cm]{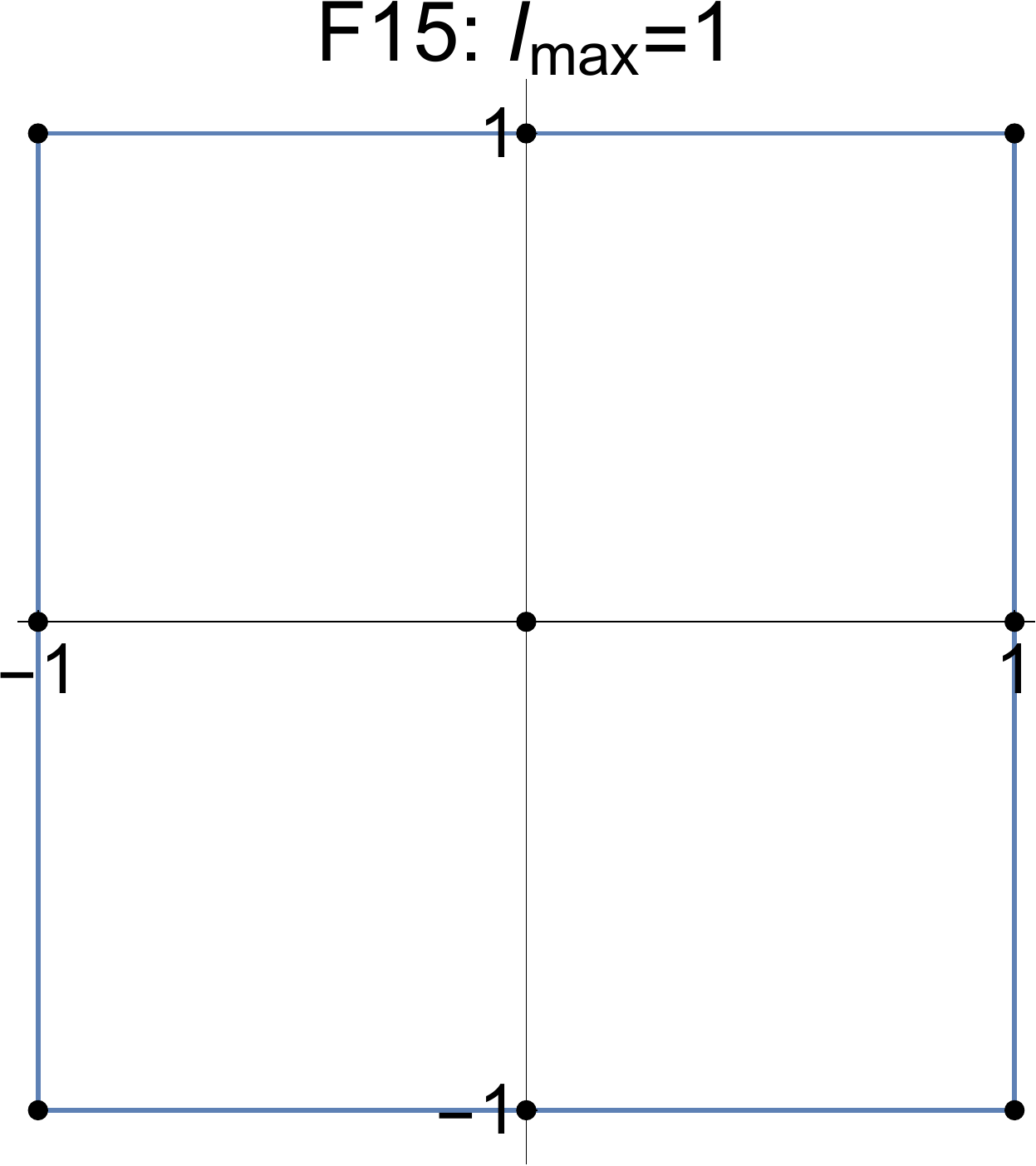}       &\includegraphics[height=3.3cm]{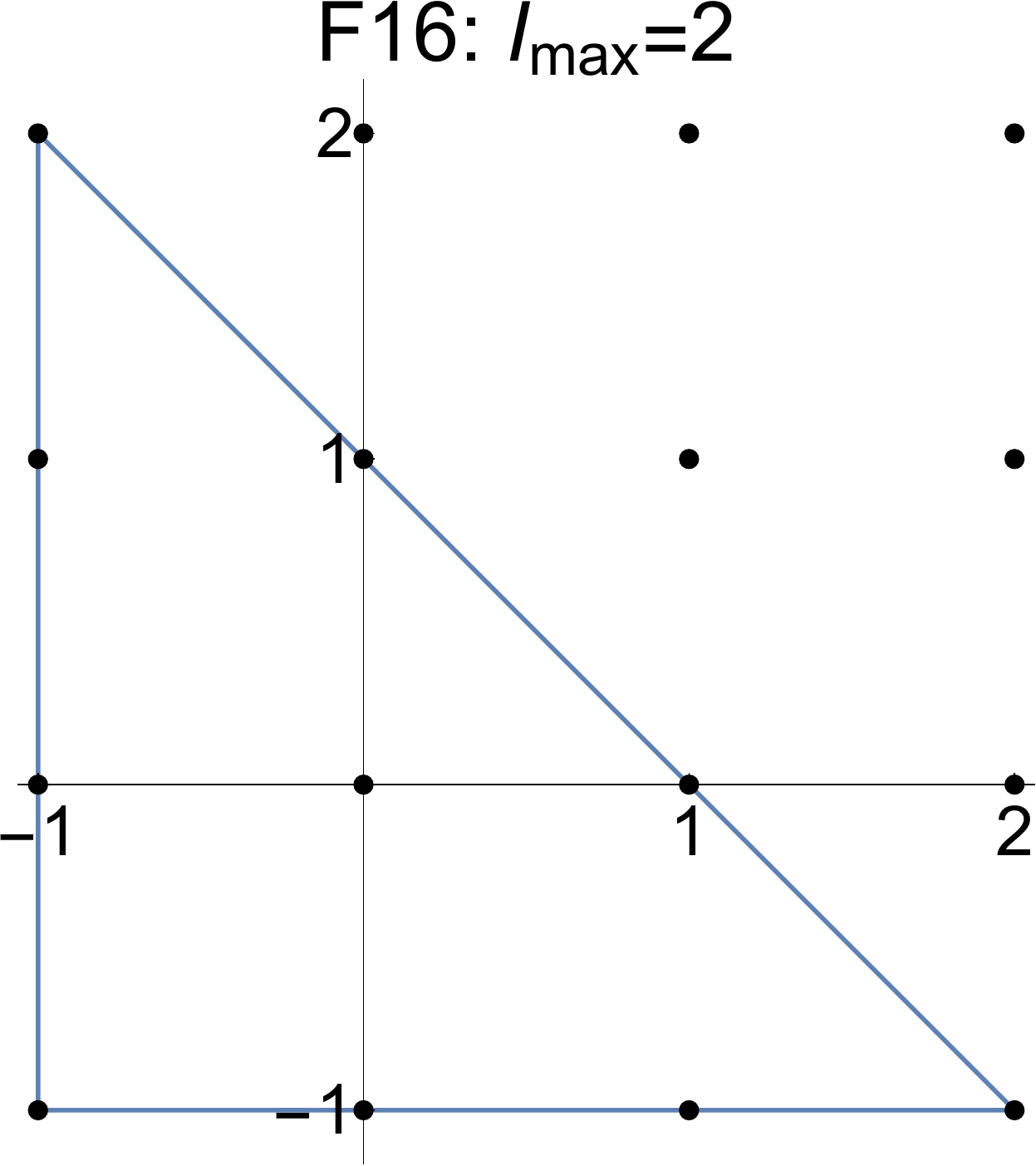}       
\end{tabular}
}
\newpage
\section{The 16 dual polytopes $\ds_2$}
\label{sec:appendix-fibers-dual}

The dual polytopes $\ds_2$ for the 16 reflexive 2D fiber polytopes
listed in the previous Appendix.  For each fiber type $\dd_2$ in Appendix \ref{sec:appendix-fibers}, a
lattice point $\vF \in\dd_2$ is given such that
 a fibration built from the  stacking construction
 (\S\ref{sec:stacked}) over the point $\vF$ allows the most negative
 curve
 self-intersection in the base among all  stackings with that fiber.
\vspace*{0.2in}
 
{\centering
\begin{tabular}{cccc}
\includegraphics[height=3.3cm]{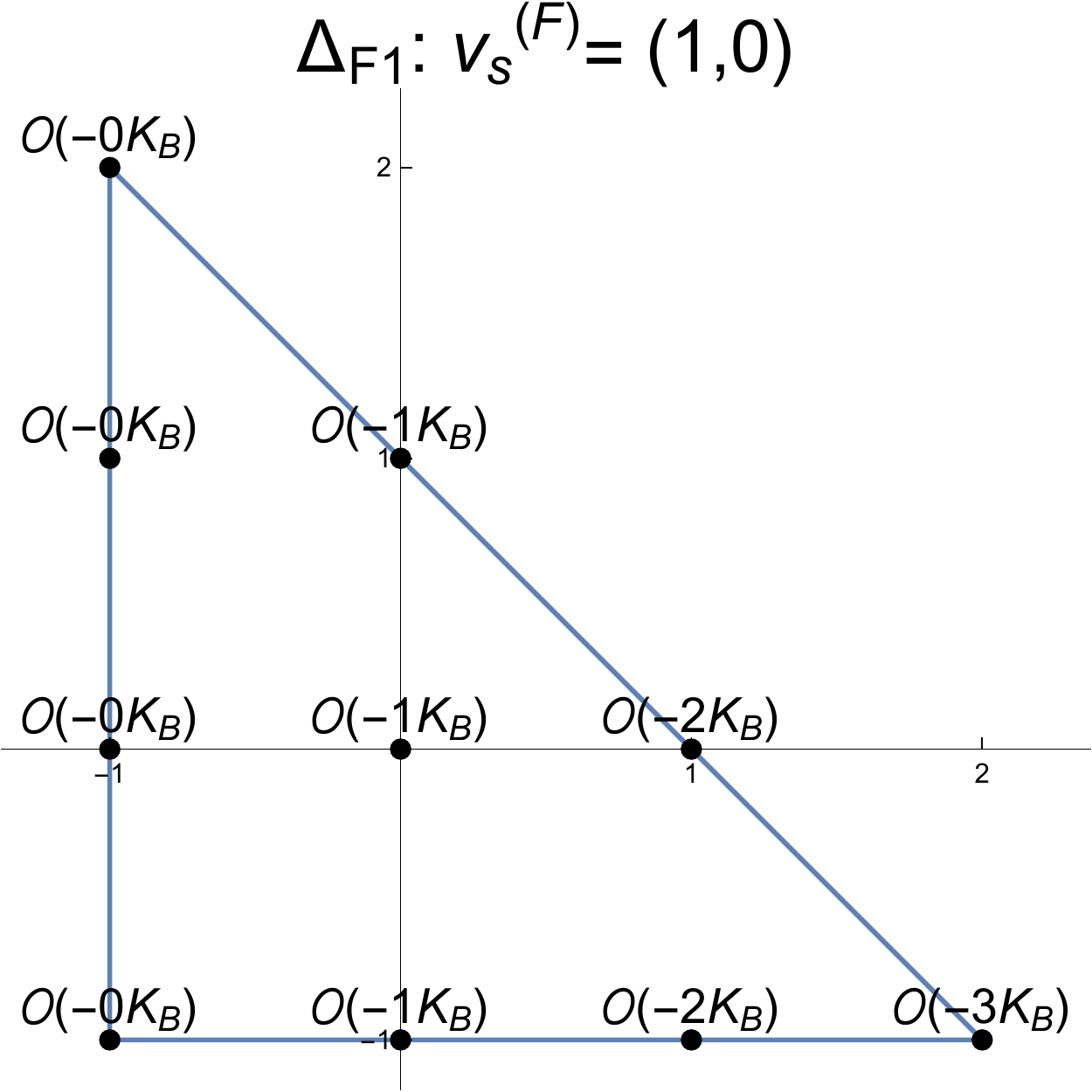}  &\includegraphics[height=3.3cm]{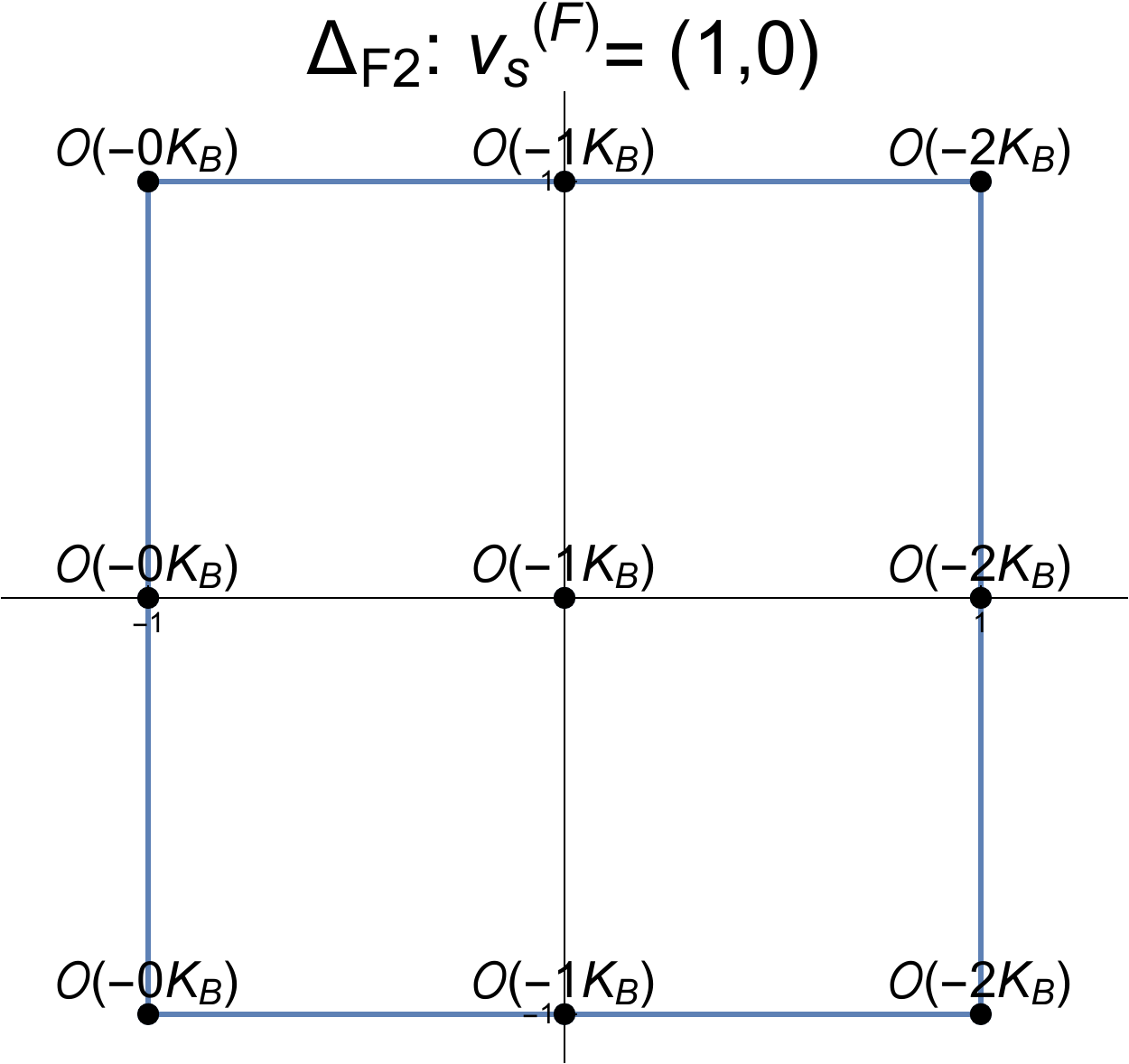}    & \includegraphics[height=3.3cm]{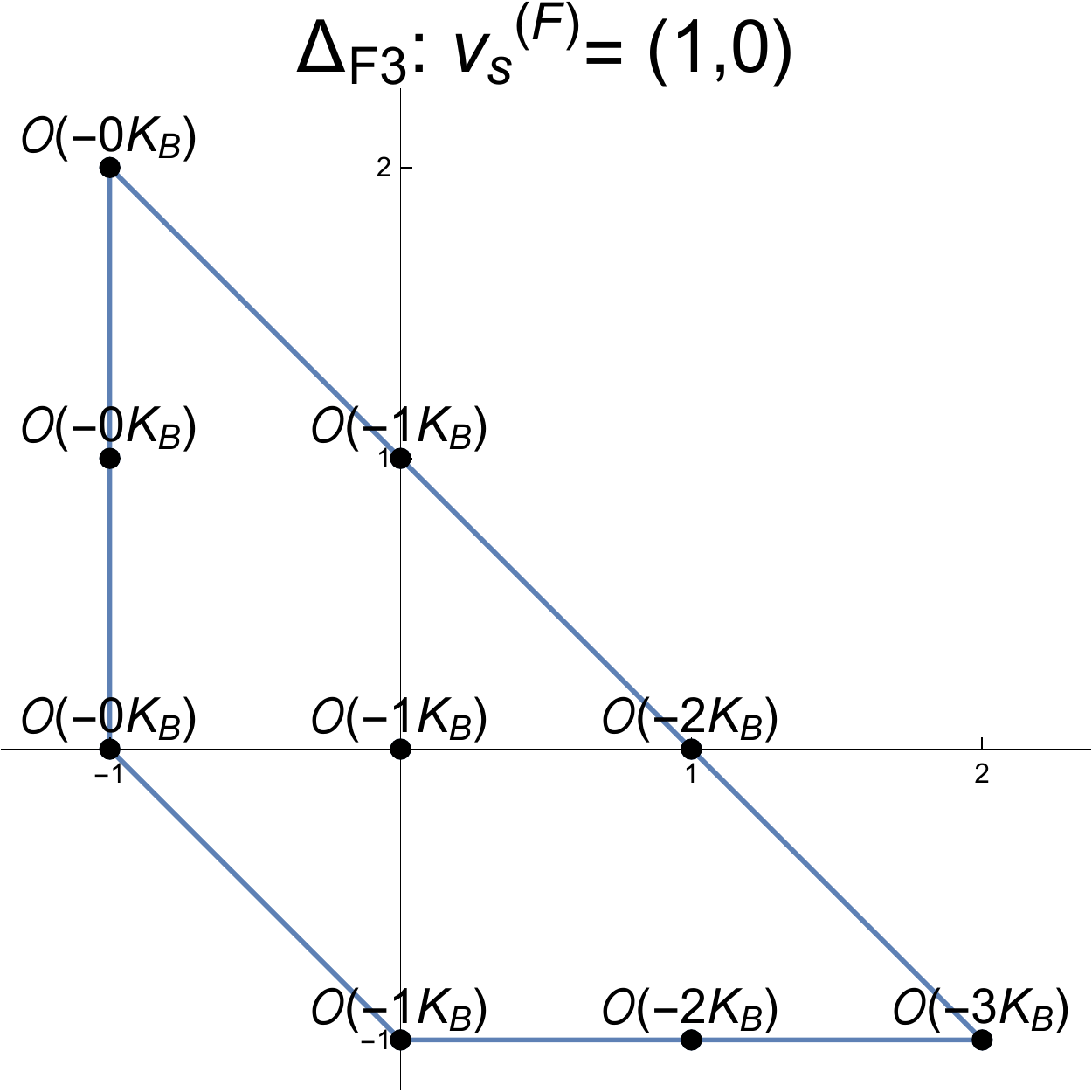}   & \includegraphics[height=3.3cm]{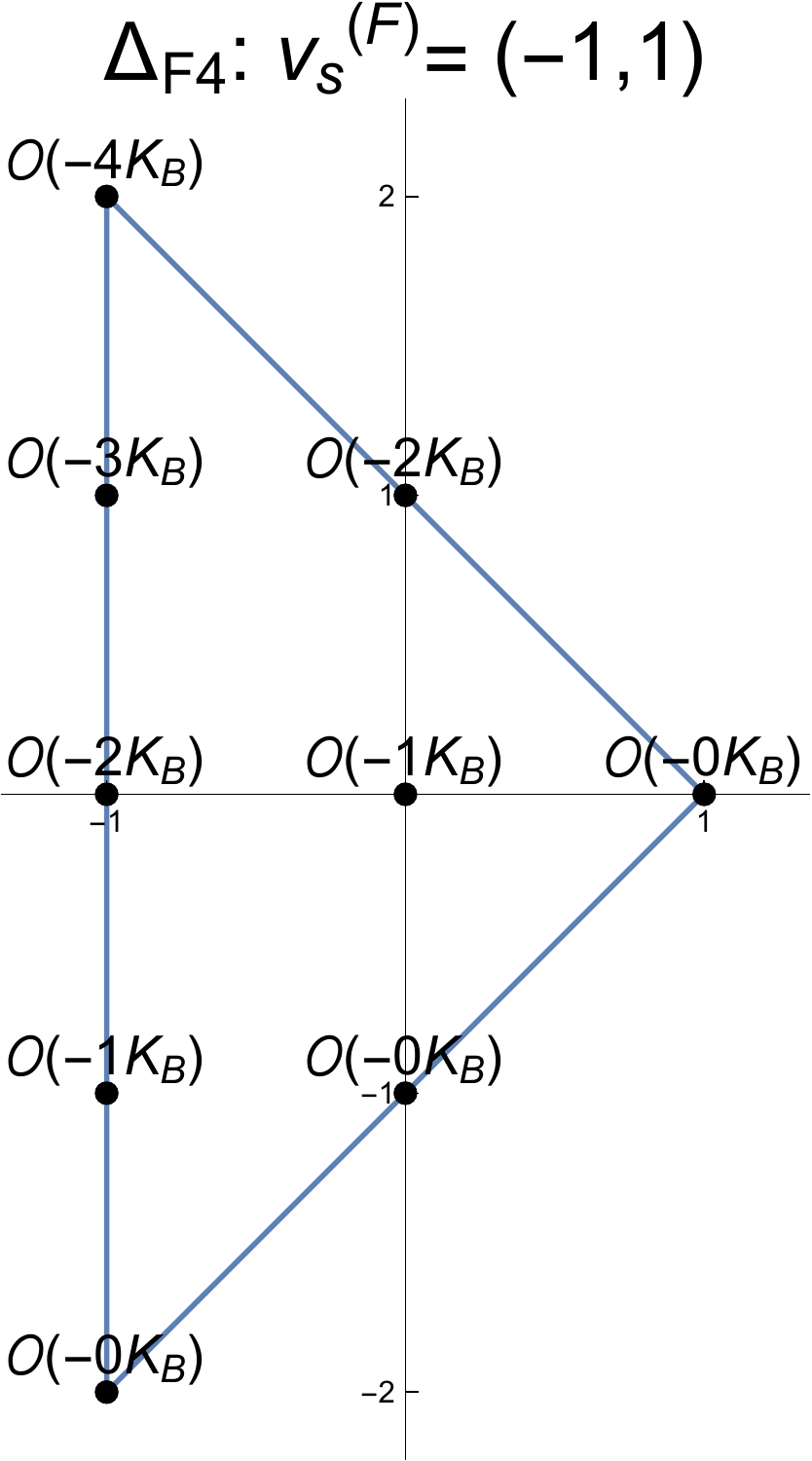}   \\\\
\includegraphics[height=3.3cm]{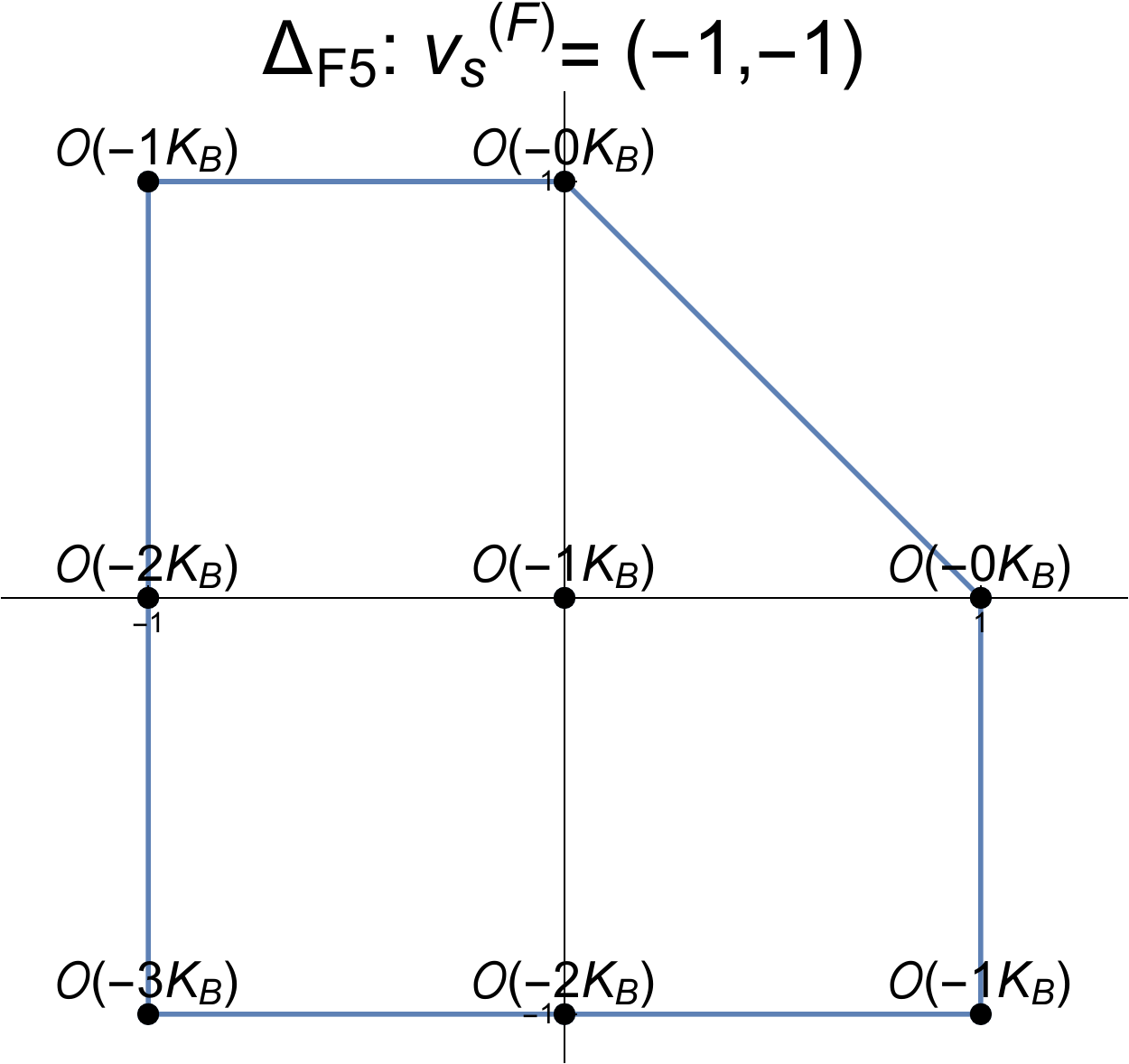}          & \includegraphics[height=3.3cm]{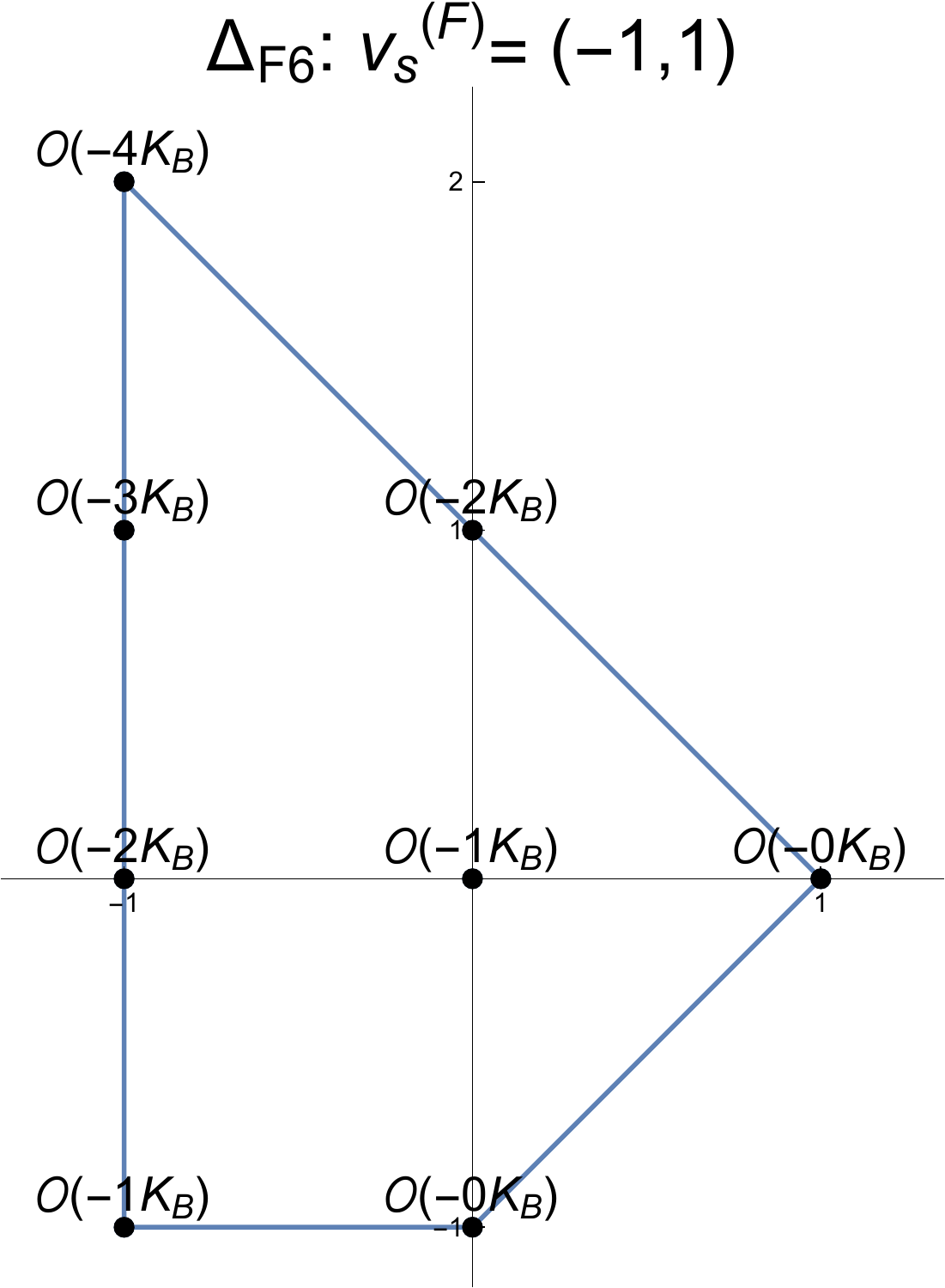}       &\includegraphics[height=3.3cm]{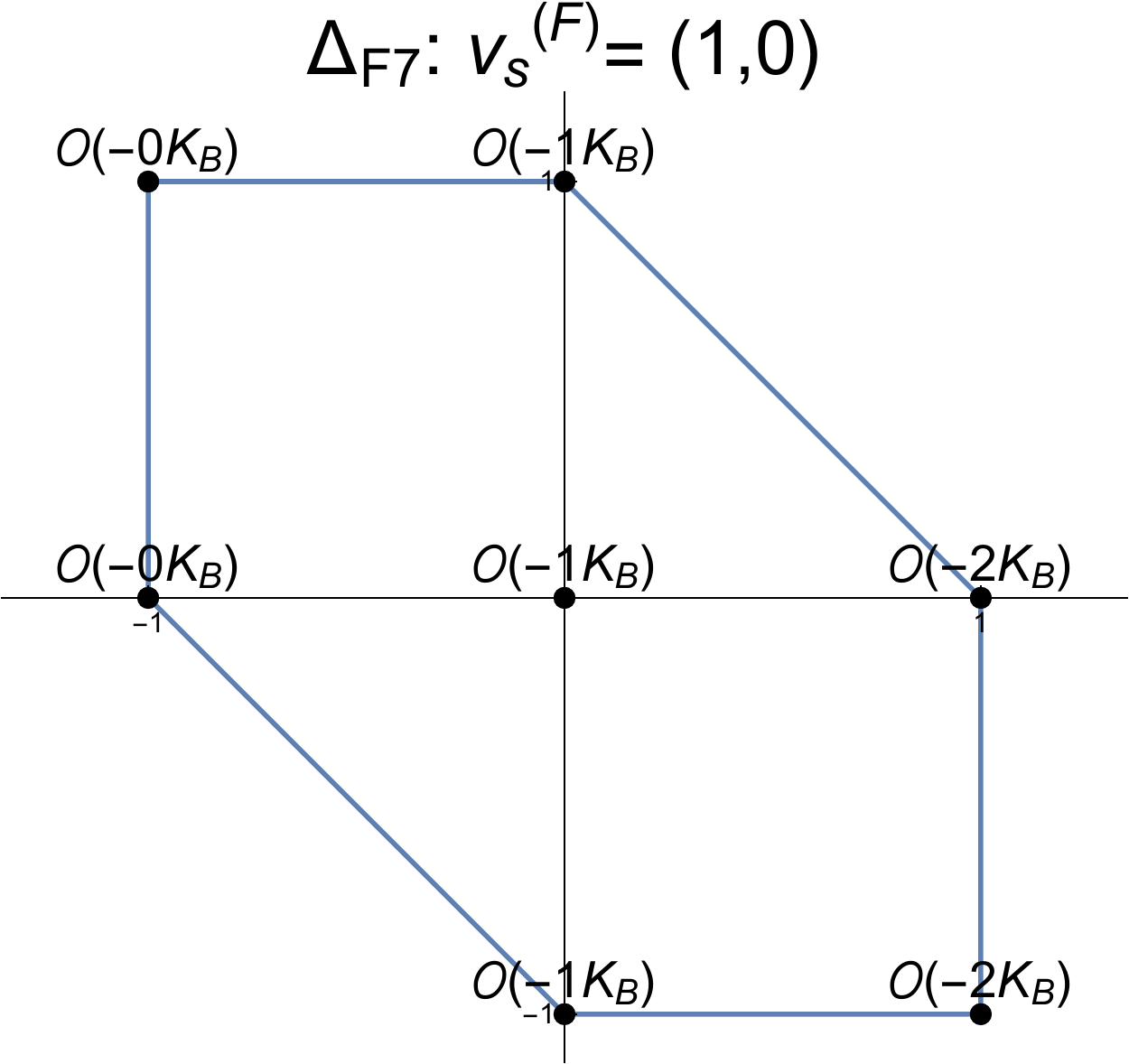}        &\includegraphics[height=3.3cm]{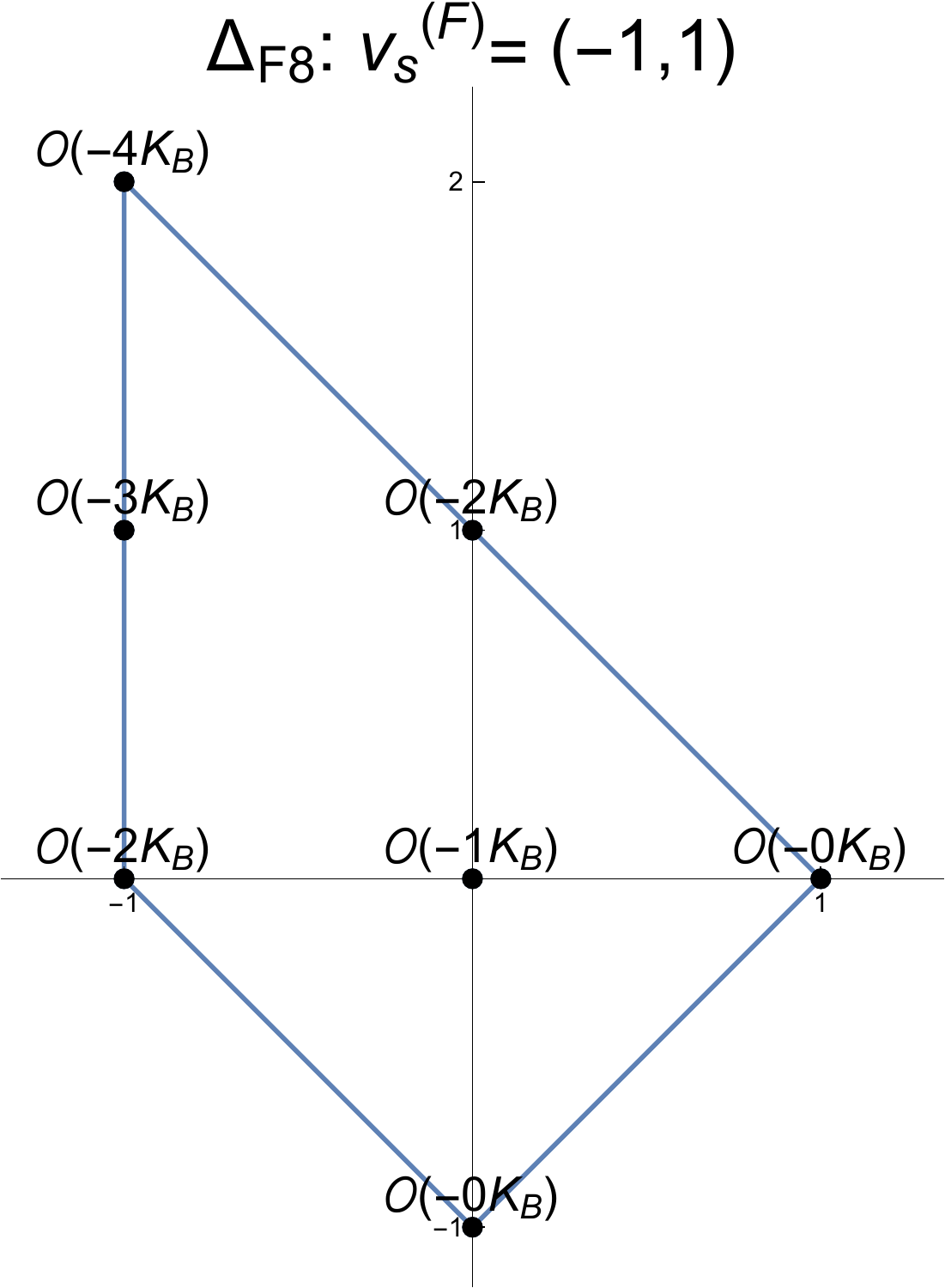}        \\\\
\includegraphics[height=3.3cm]{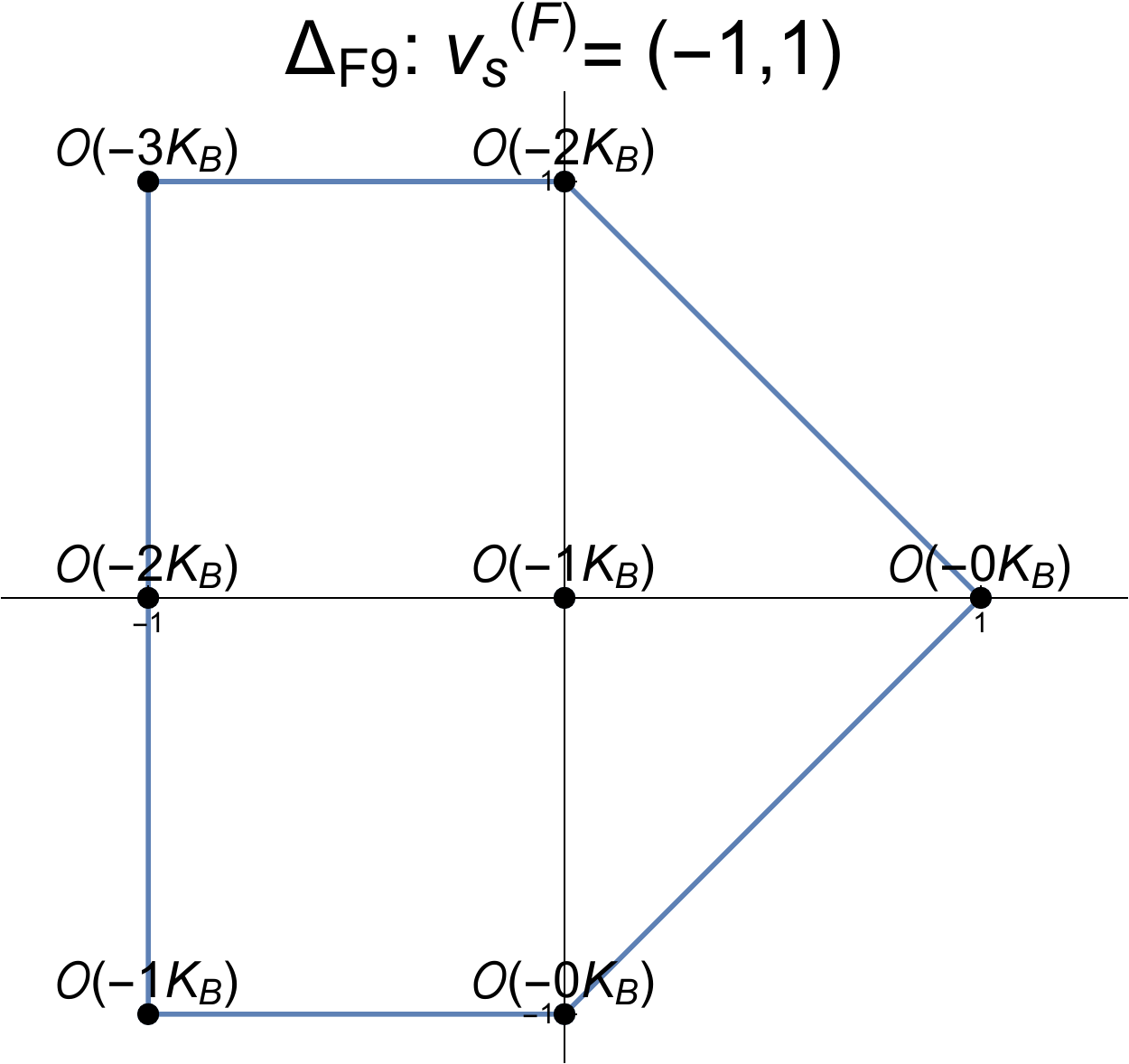}          & \includegraphics[height=3.3cm]{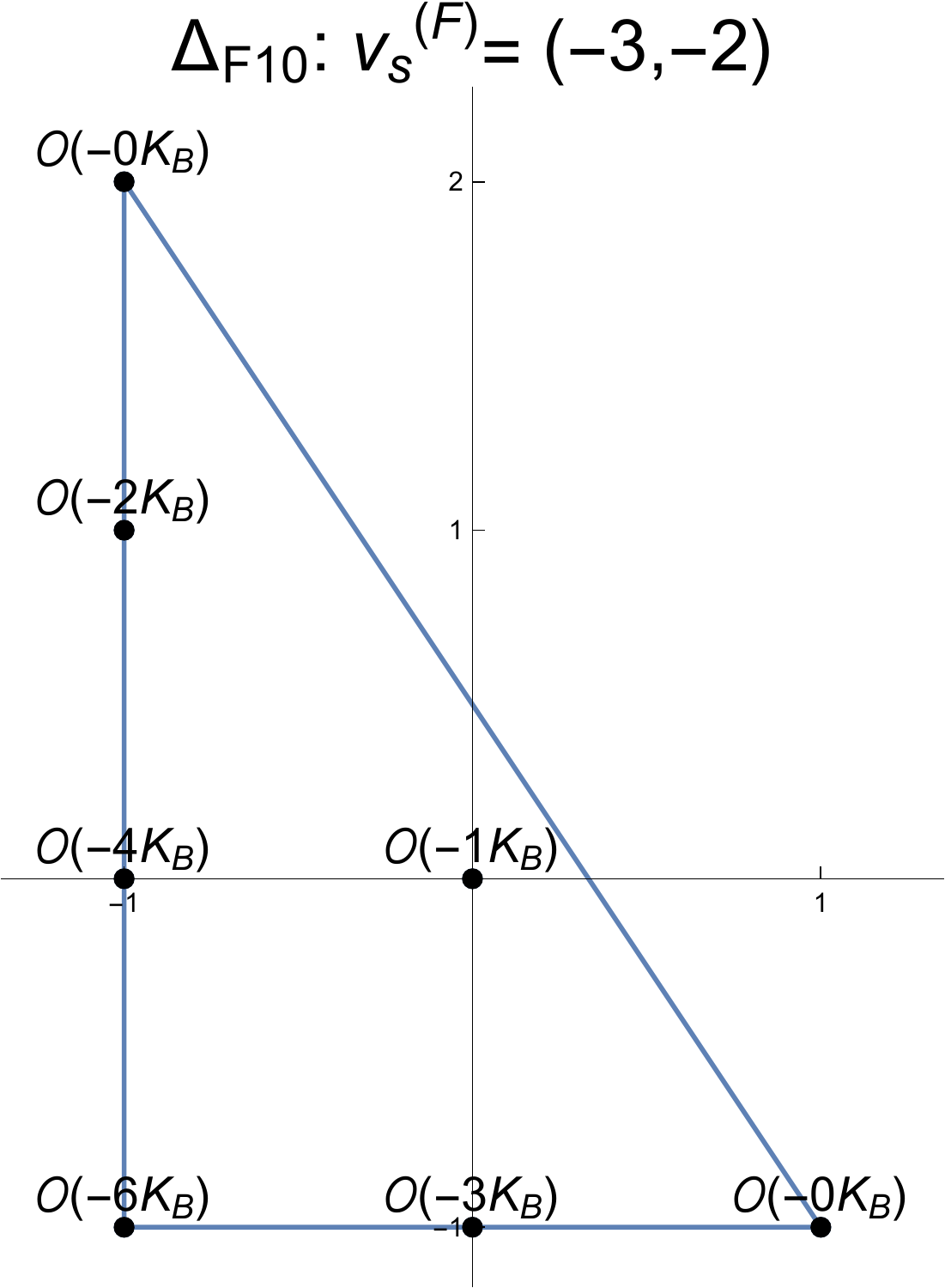}       &\includegraphics[height=3.3cm]{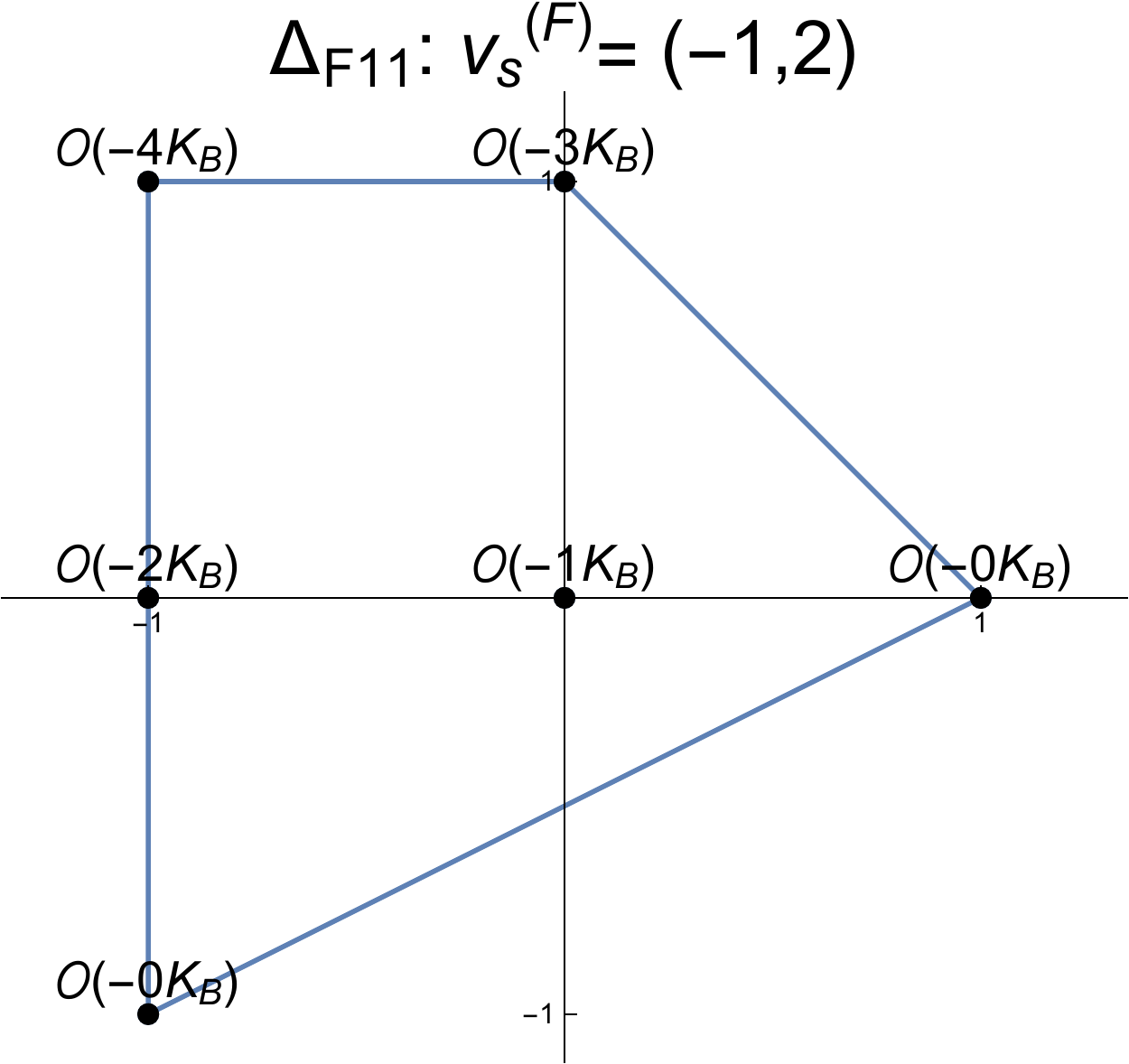}        &\includegraphics[height=3.3cm]{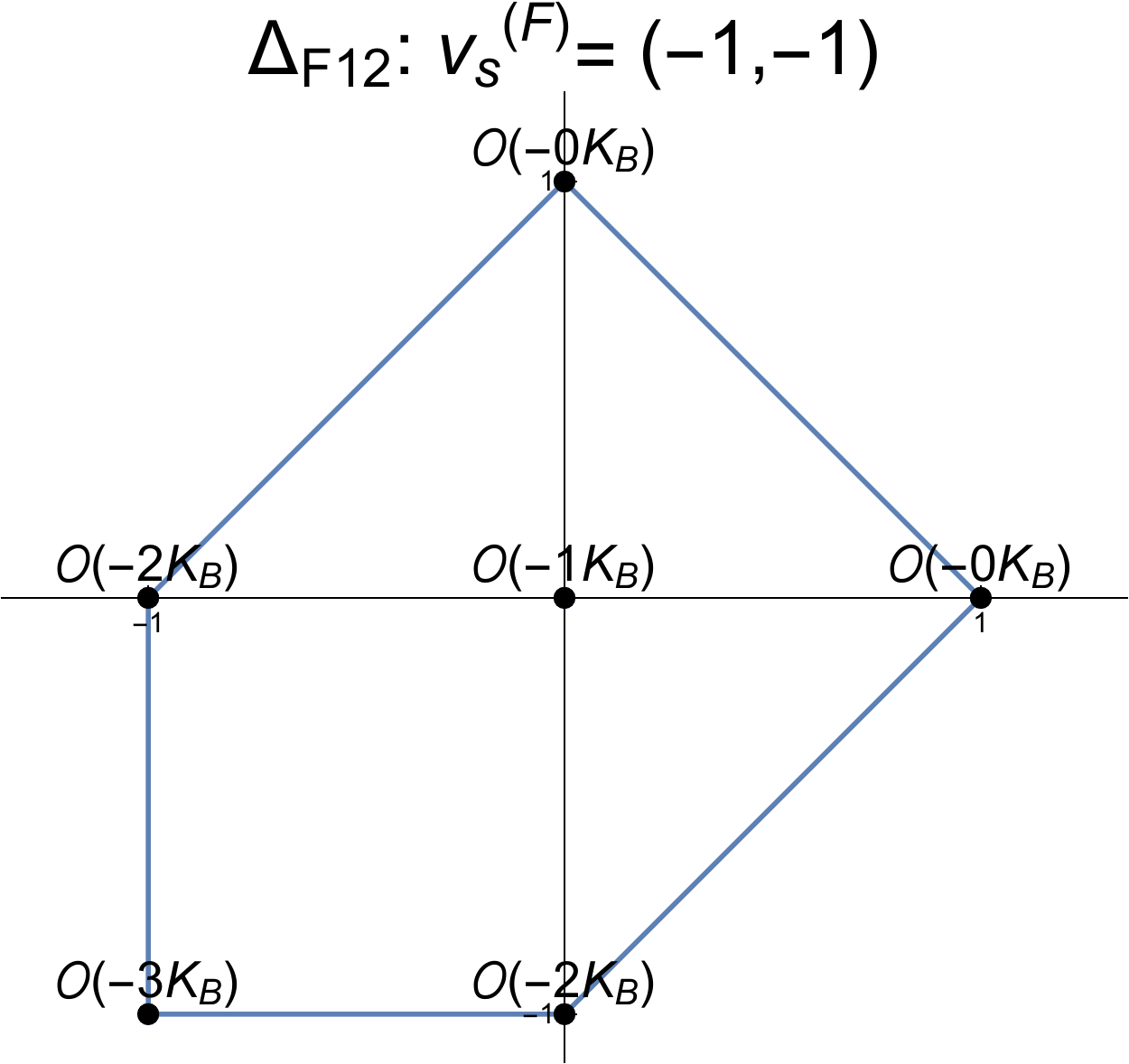}        \\\\
\includegraphics[height=3.3cm]{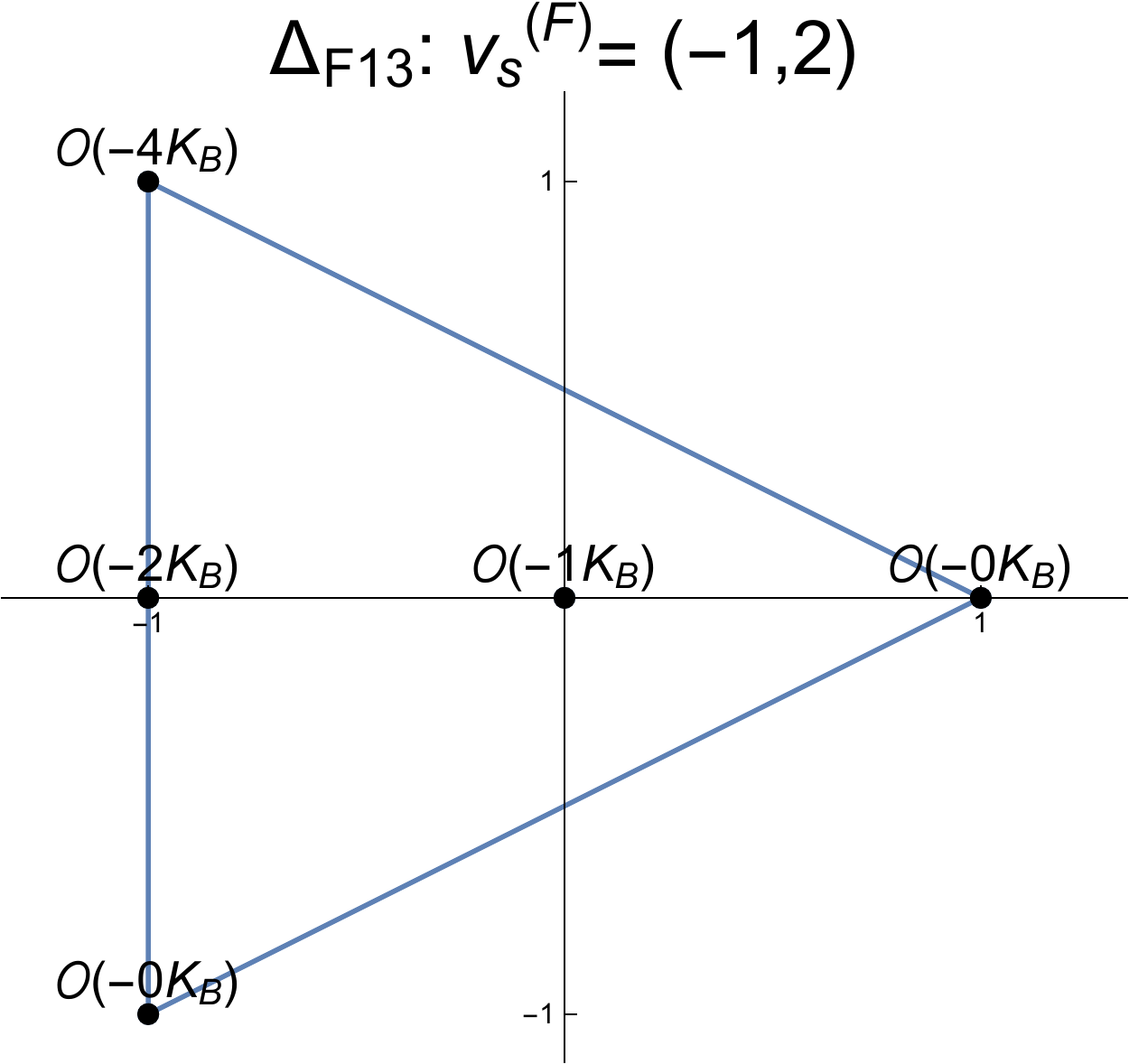}          & \includegraphics[height=3.3cm]{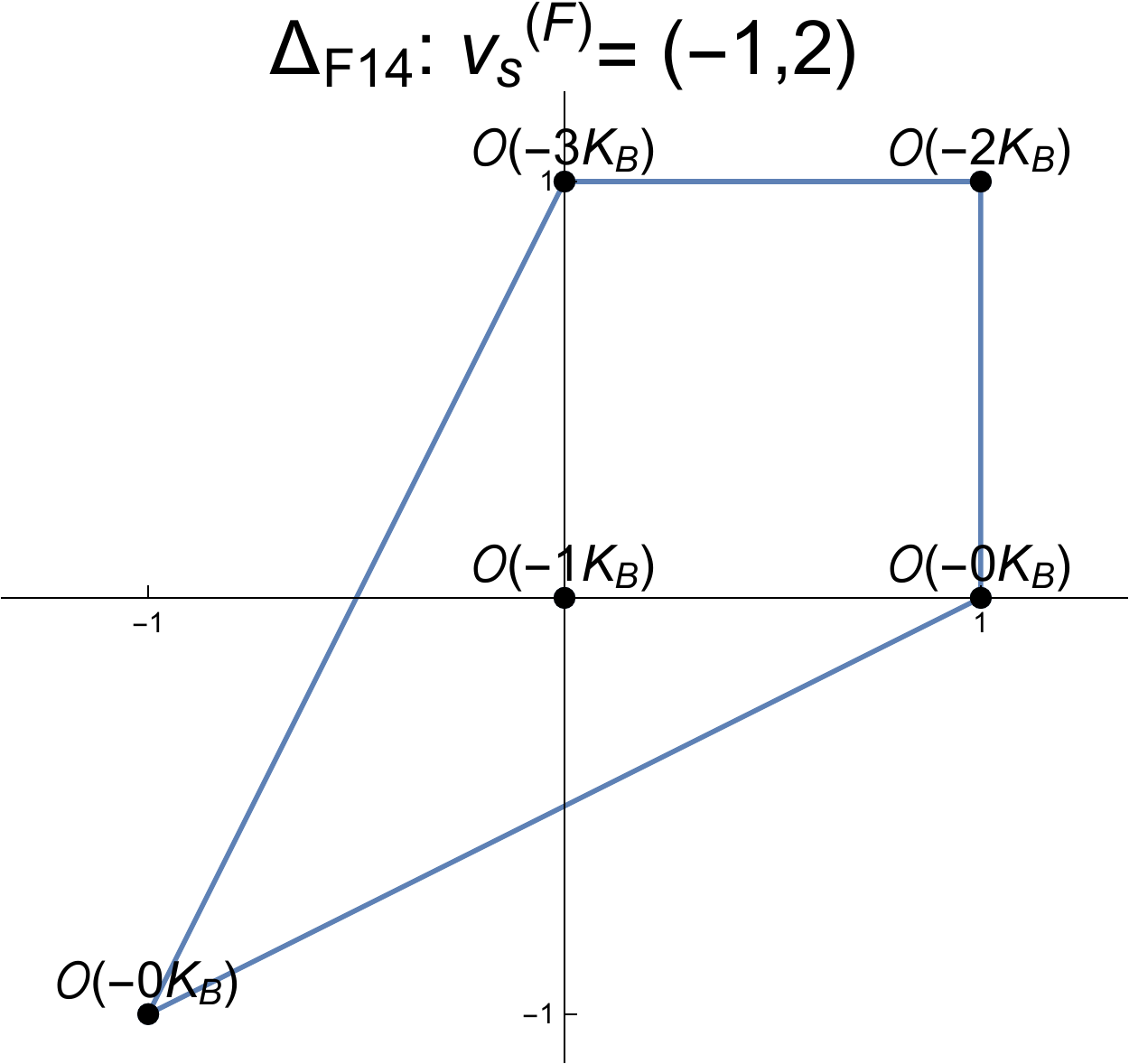}       & \includegraphics[height=3.3cm]{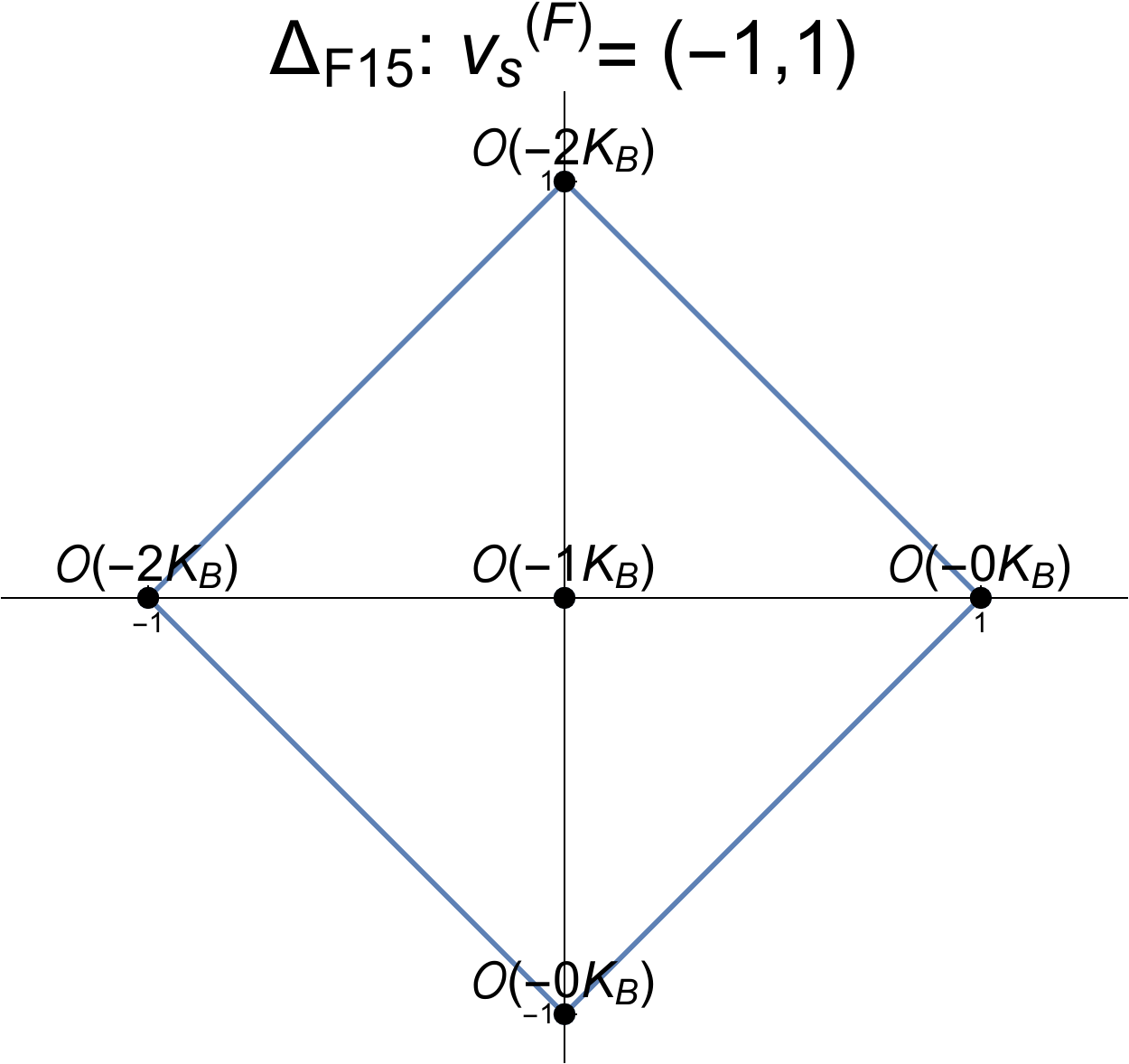}       &\includegraphics[height=3.3cm]{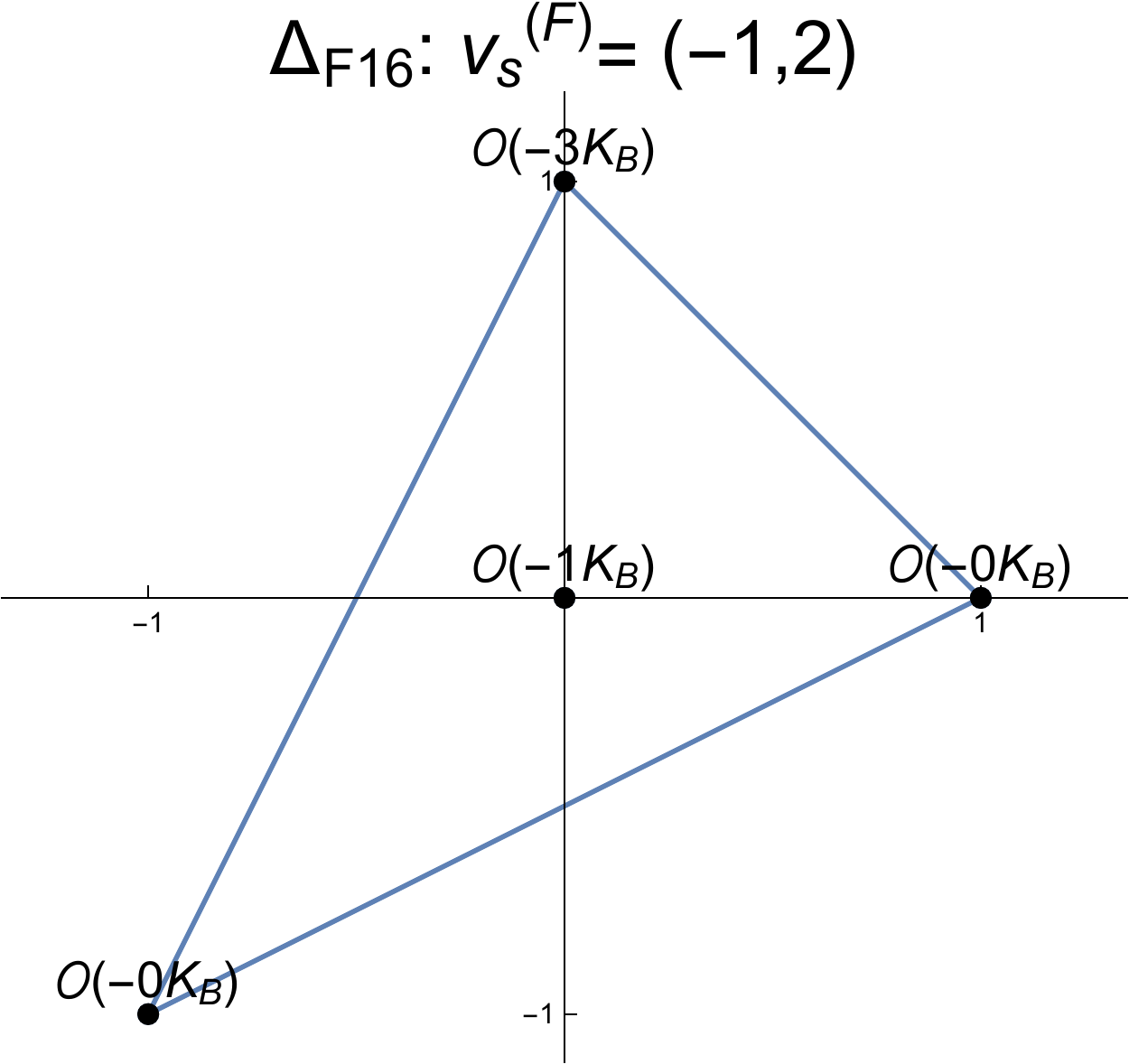}
\end{tabular}
}
\newpage

\section{Distribution of polytopes with each fiber type}
\label{sec:appendix-results}

The figures in this Appendix depict the distribution of Hodge numbers
for the Calabi-Yau threefolds associated with the polytopes that have
each type of reflexive 2D fiber. 
The
largest values of $h^{1,1}$ and $h^{2,1}$
for Calabi-Yau threefolds associated with polytopes having each fiber type
are shown in the figure. In each figure, the density scale at the right indicates the color coding according to
the total number of fibrations at each Hodge number pair, which
results both from the multiplicity of the fibers of a given polytope
and from the multiplicity of the polytopes at each Hodge number pair.

{\centering
\begin{tabular}{cc}
\includegraphics[height=6.5cm]{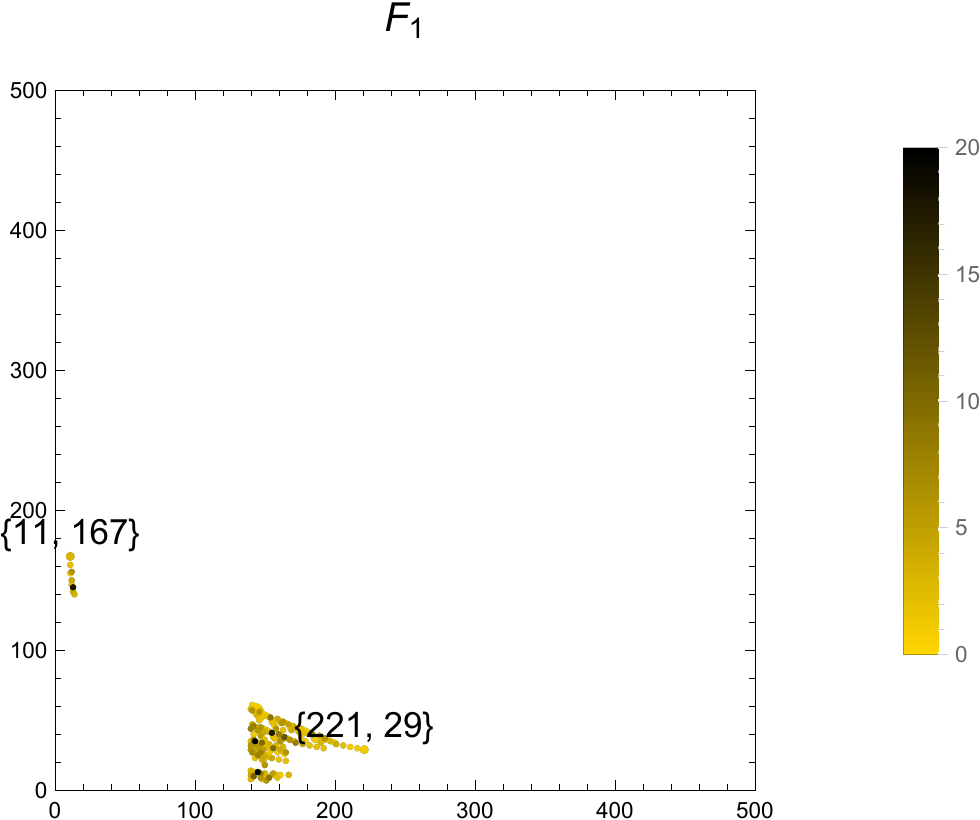}   & \includegraphics[height=6.5cm]{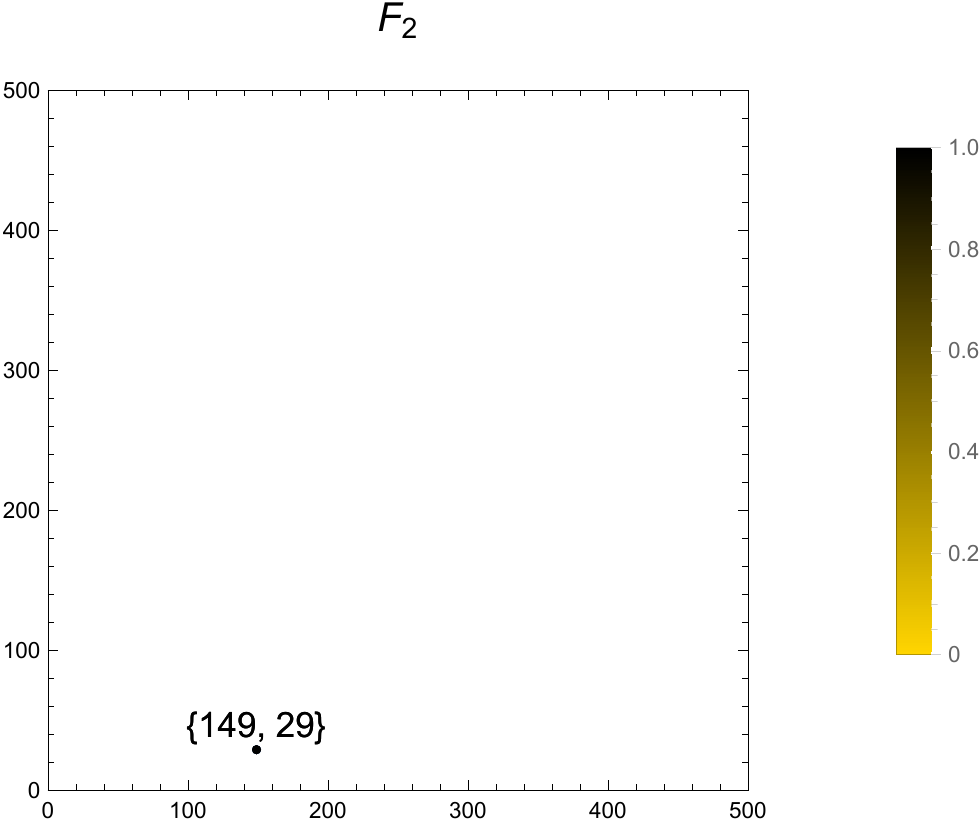}   \\\\
\includegraphics[height=6.5cm]{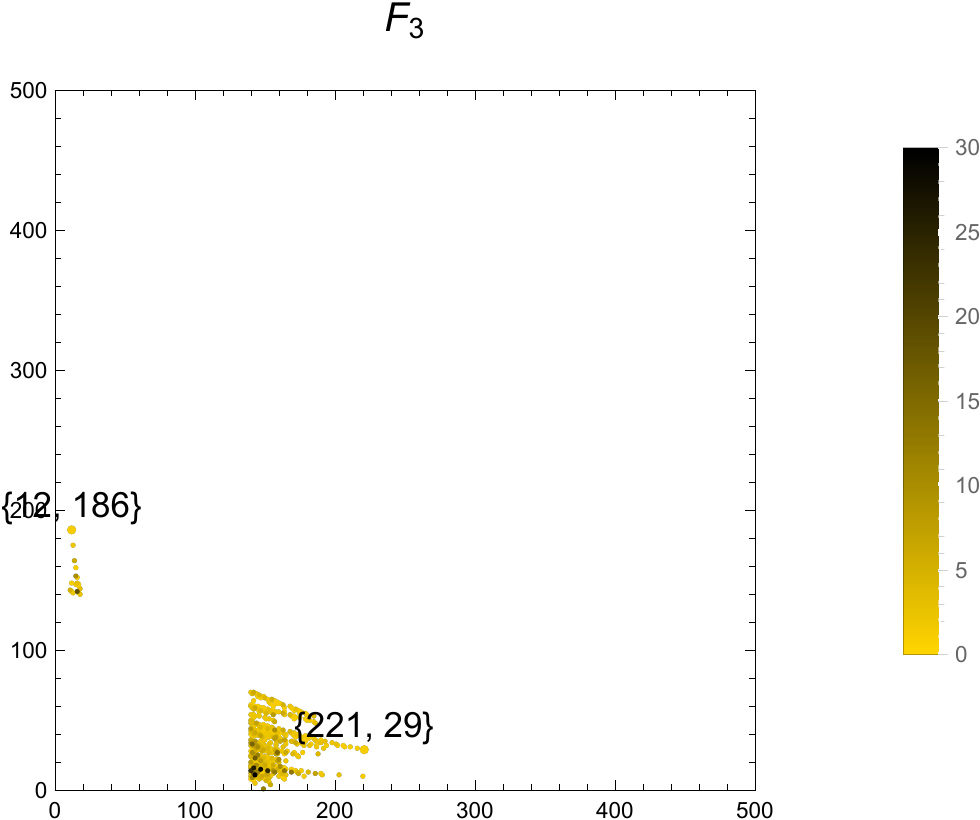}   & \includegraphics[height=6.5cm]{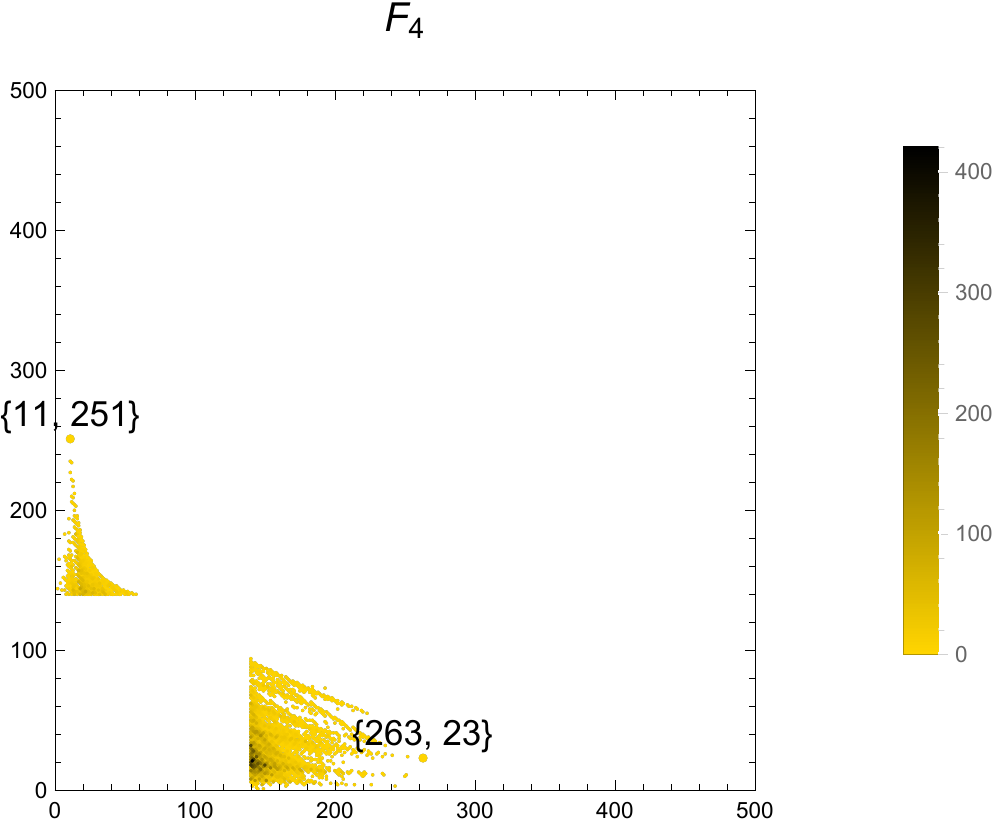}   
\end{tabular}
}

{\centering
  \begin{tabular}{cc}
\includegraphics[height=6.5cm]{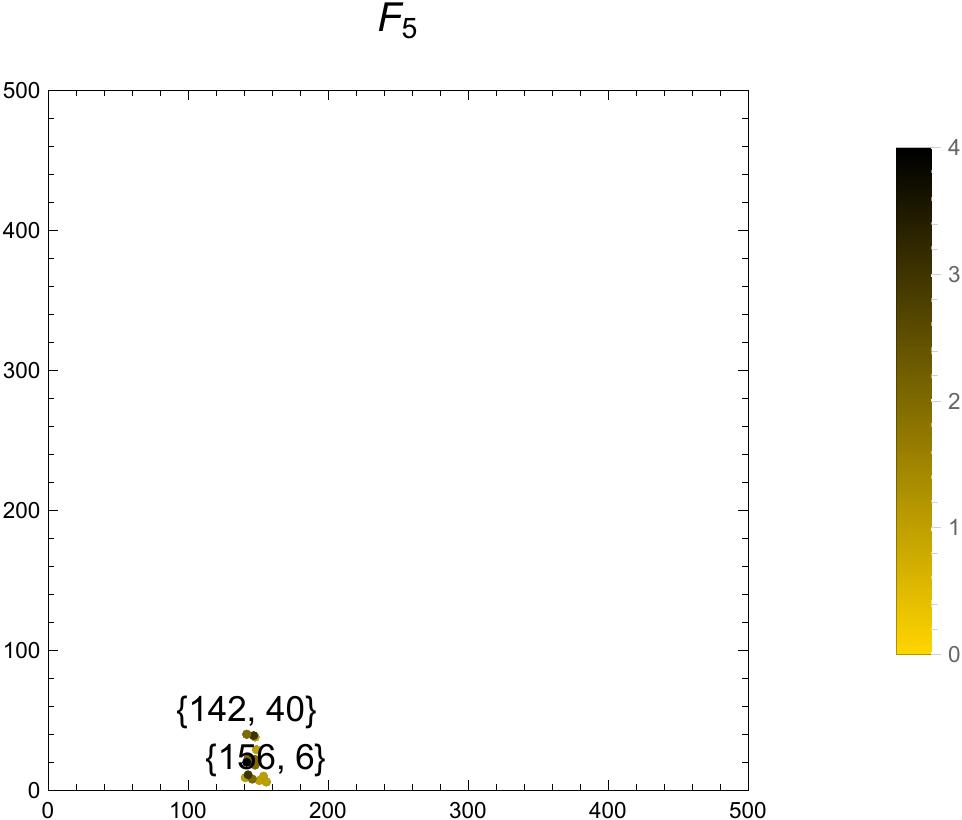}
&\includegraphics[height=6.5cm]{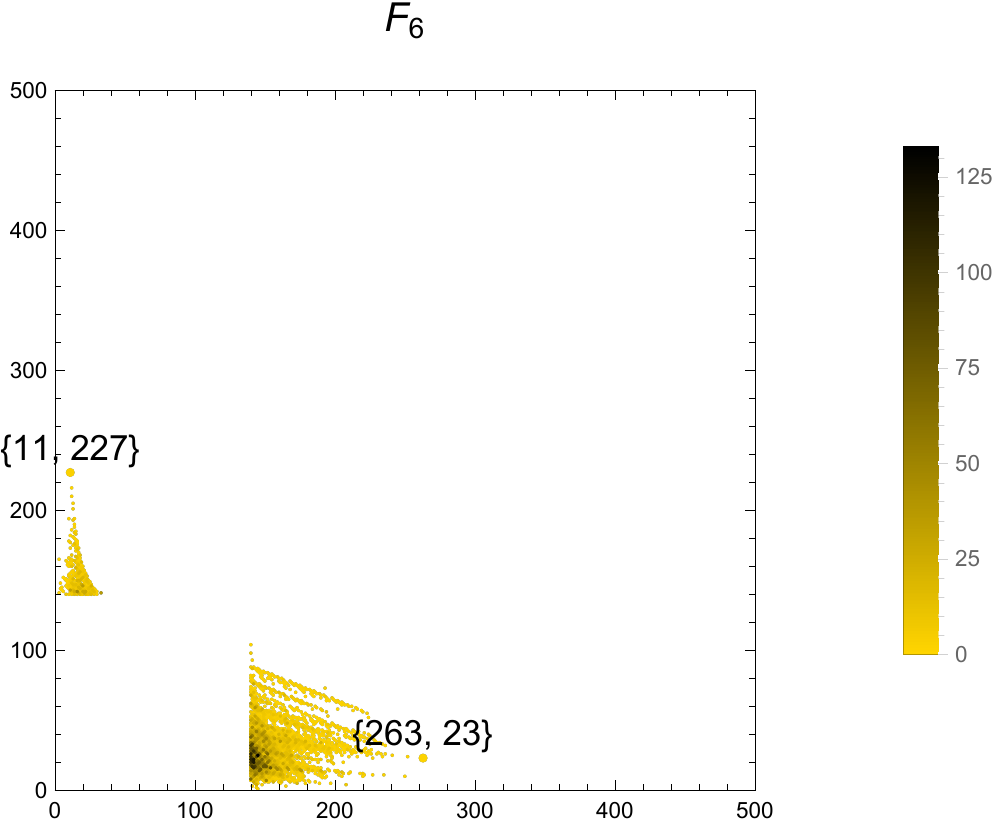}
\\\\
\includegraphics[height=6.5cm]{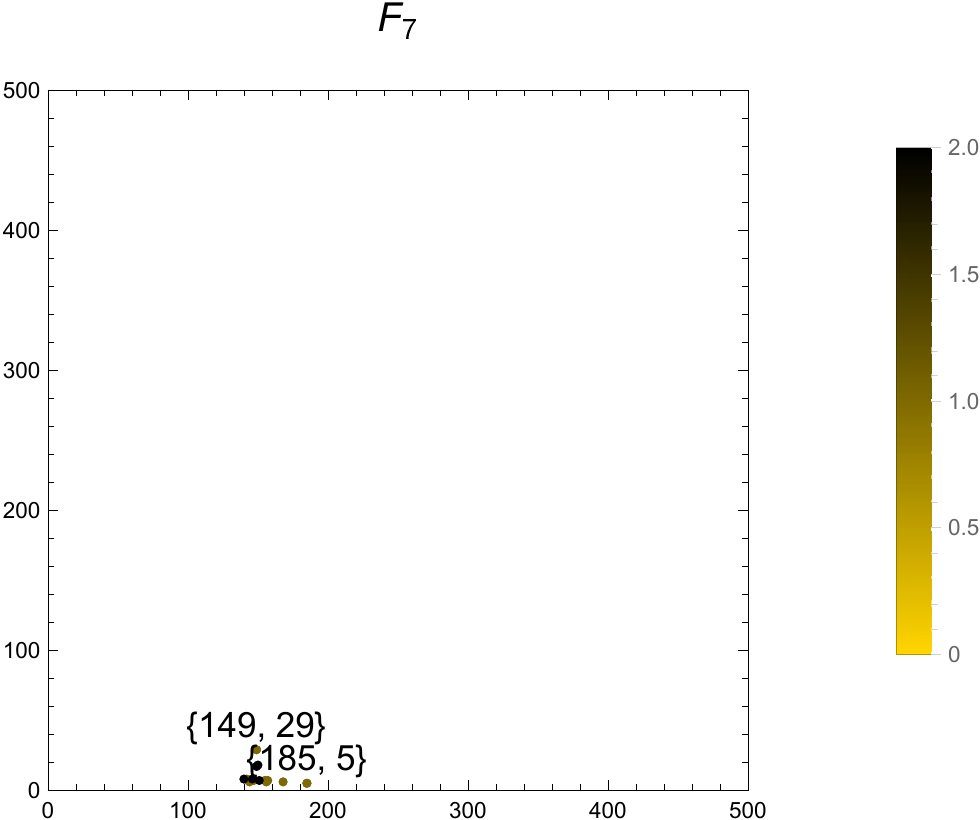}   & \includegraphics[height=6.5cm]{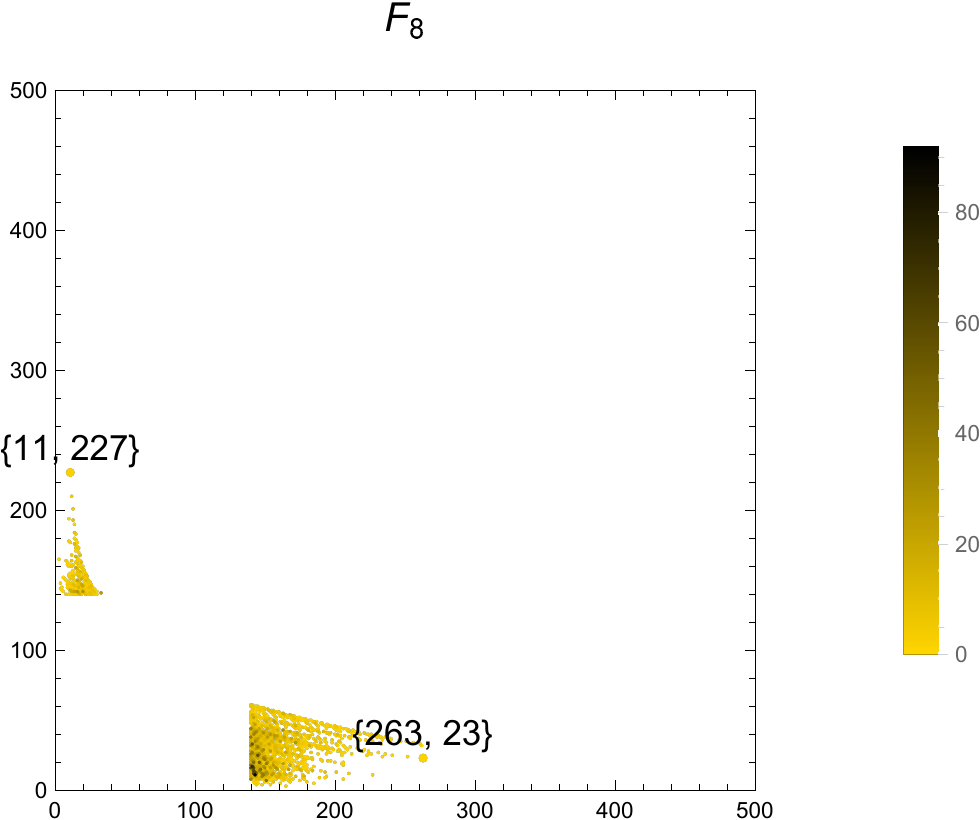}   \\\\
\includegraphics[height=6.5cm]{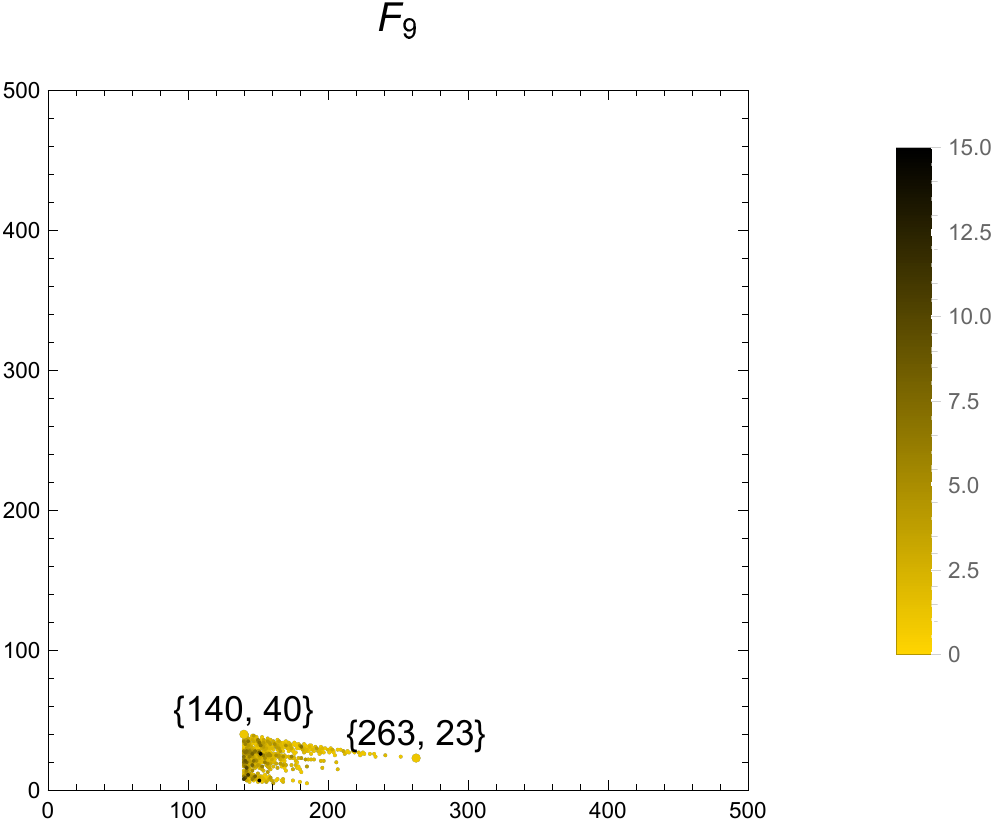}   & \includegraphics[height=6.5cm]{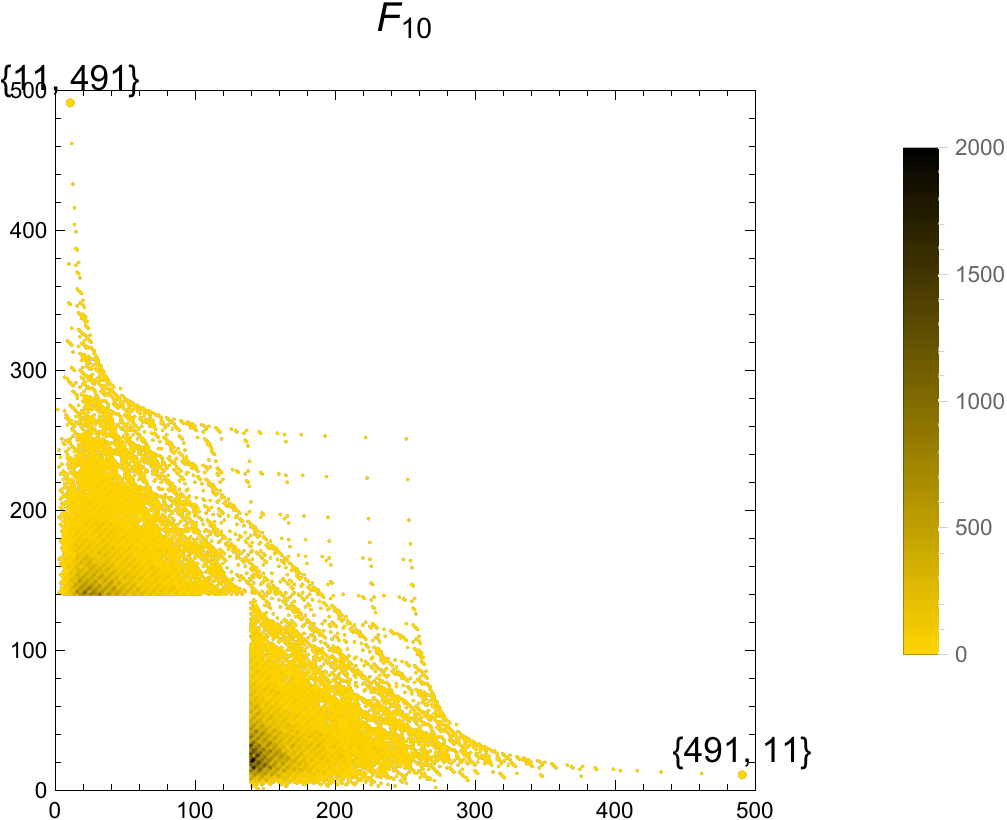}
\end{tabular}
}

{\centering
  \begin{tabular}{cc}
\includegraphics[height=6.5cm]{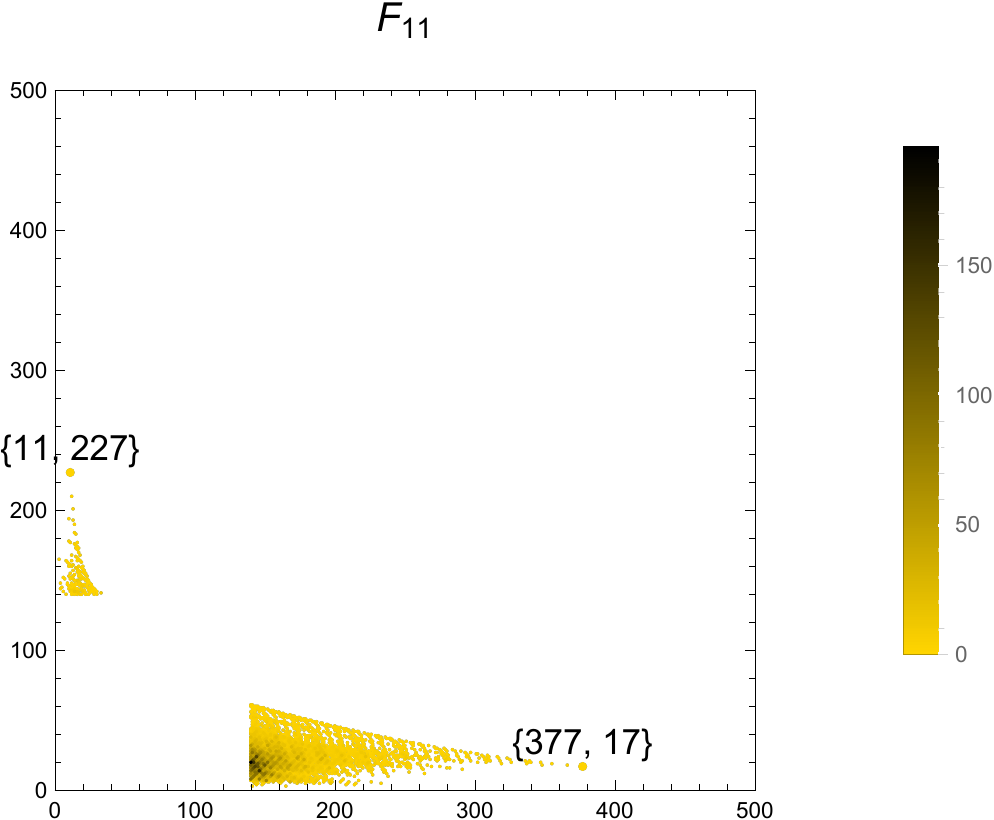}        &\includegraphics[height=6.5cm]{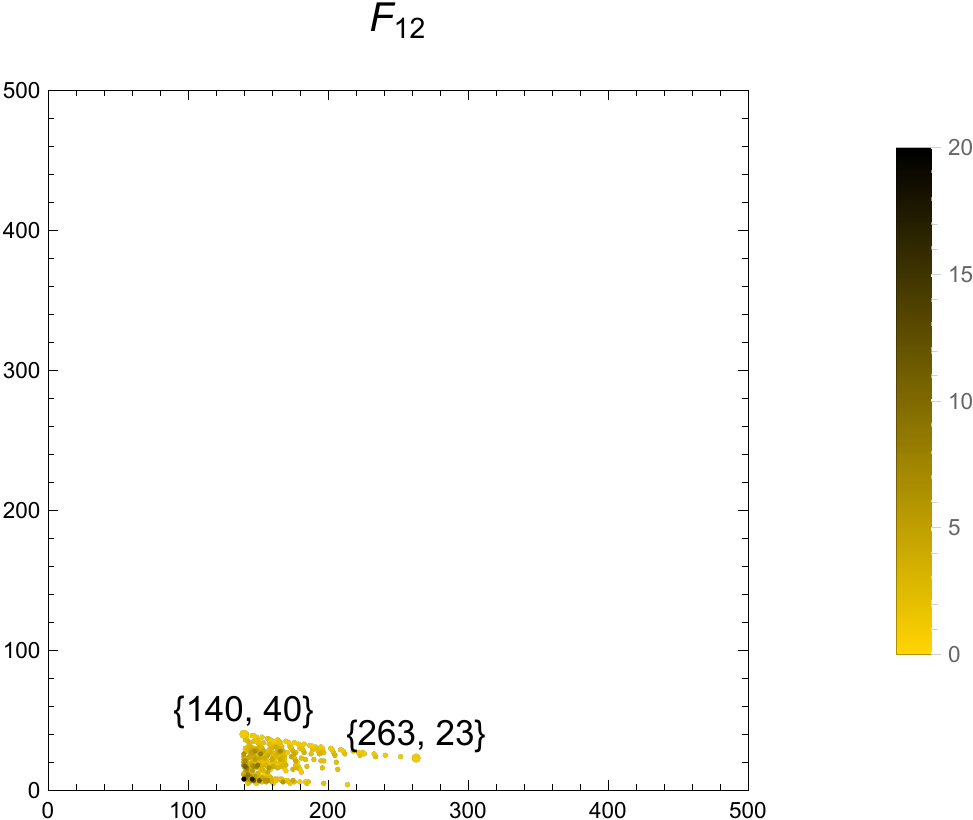}         \\\\
\includegraphics[height=6.5cm]{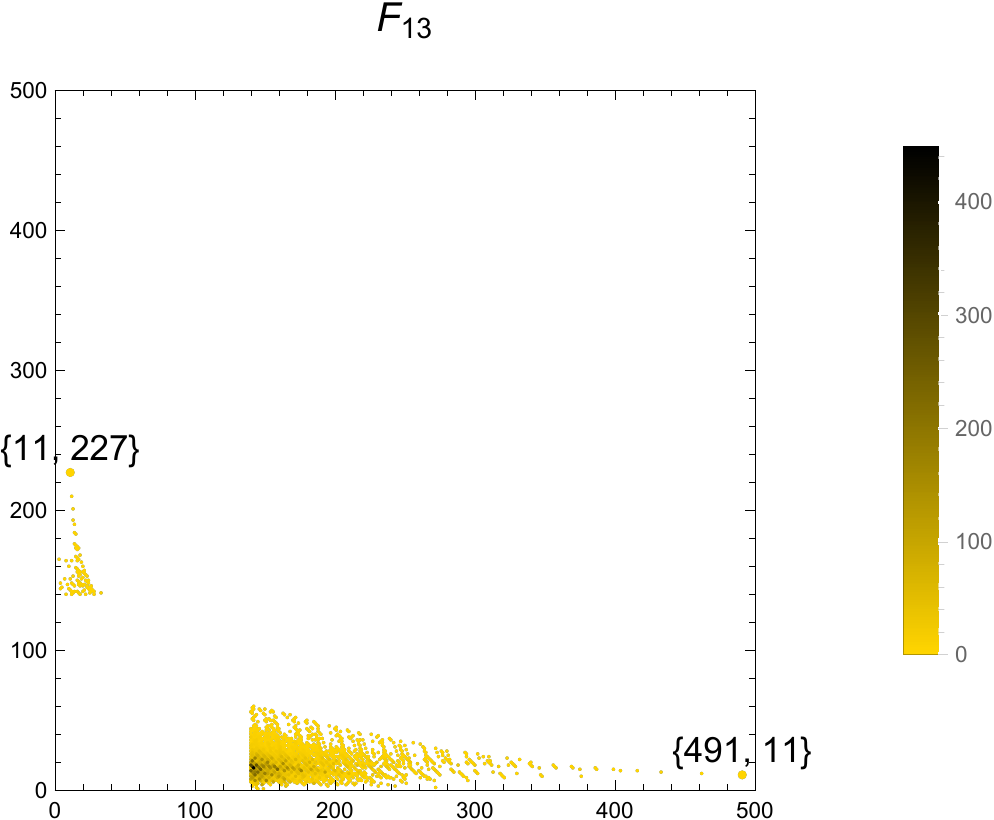}   & \includegraphics[height=6.5cm]{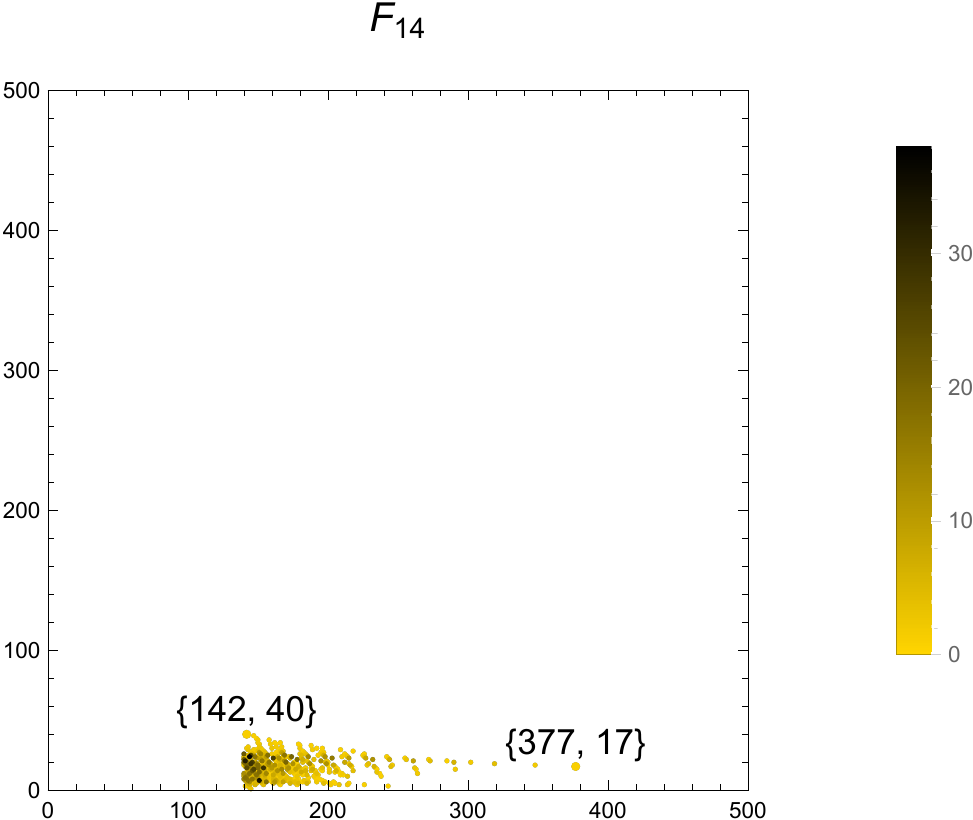}   \\\\
\includegraphics[height=6.5cm]{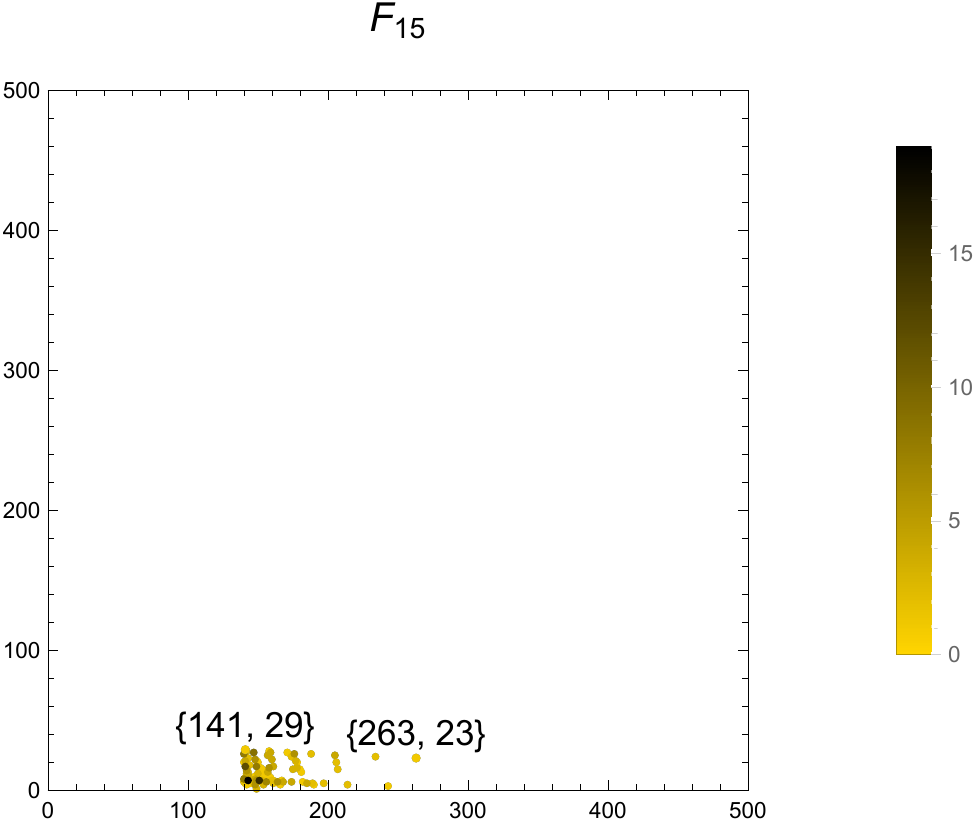}        &\includegraphics[height=6.5cm]{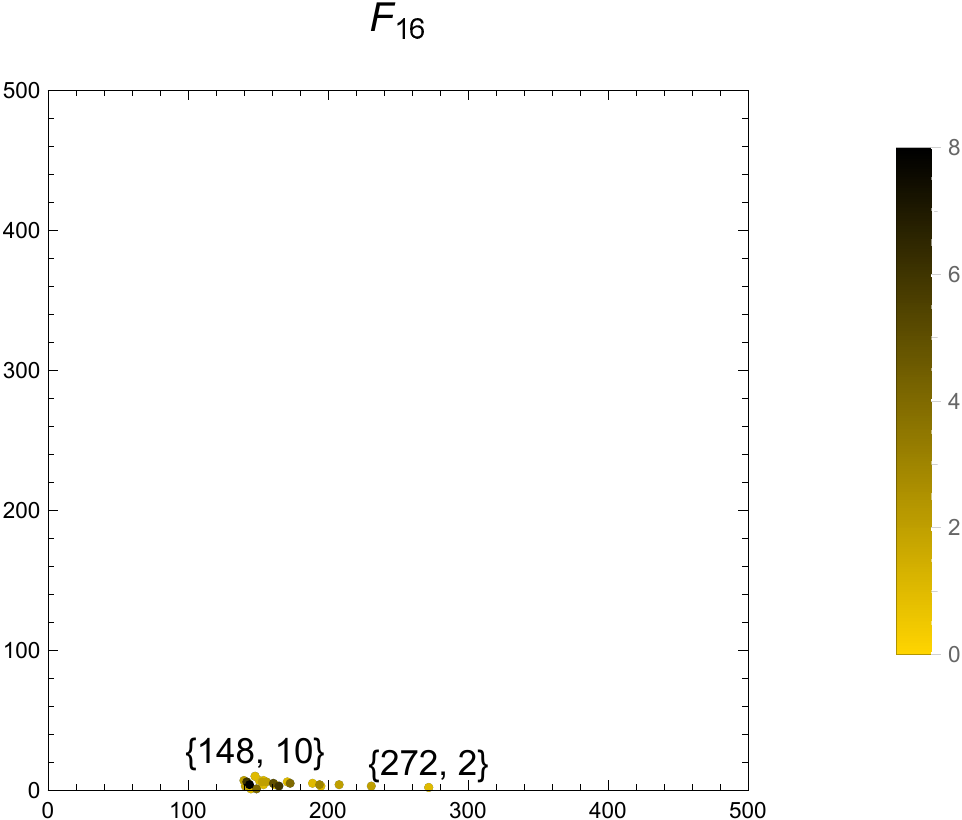}         
\end{tabular}
}

\begin{table}[]
\centering
\begin{tabular}{|c|c|c||l|c|}
\hline
$F$                       & $\vF_s$                                                                     & \begin{tabular}[c]{@{}c@{}}\{\#pts$_{\ds_2}(\mathcal{O}(-nK_B))\rvert$\\ $n=0,1,2,3,4,(5),6$\}\end{tabular} & \multicolumn{1}{c|}{fibered-polytope} & $B$       \\ \hline
$F_1$                     & $(1,0),(0,1),(-1,-1)$                                                                  & $\set{4, 3, 2, 1, 0, (0),0}$                                                                                & M:171 5 N:11 5 H:11,131               & $\F_6$    \\ \hline
$F_2$                     & $(1,0),(0,1),(0,-1),(-1,0)$                                                            & $\set{3,3,3,0,0,(0),0}$                                                                                     & M:117 8 N:11 6 H:9,93                 & $\F_4$    \\ \hline
\multirow{3}{*}{$F_3$}    & $(1,1)$                                                                                & $\set{2,3,4,0,0,(0),0}$                                                                                     & M:144 8 N:11 6 H:8,110                & $\F_4$    \\ \cline{2-5} 
                          & $(1,0),(0,1)$                                                                          & $\set{3,3,2,1,0,(0),0}$                                                                                     & M:170 9 N:14 7 H:11,131               & $\F_6$    \\ \cline{2-5} 
                          & $(-1,-1)$                                                                              & $\set{4,3,2,0,0,(0),0}$                                                                                     & M:90 8 N:11 6 H:10,76                 & $\F_4$    \\ \hline
\multirow{2}{*}{$F_4$}    & $(1,0)$                                                                                & $\set{5,3,1,0,0,(0),0}$                                                                                     & M:63 5 N:11 5 H:11,59                 & $\F_4$    \\ \cline{2-5} 
                          & $(-1,1),(-1,-1)$                                                                       & $\set{3,2,2,1,1,(0),0}$                                                                                     & M:311 5 N:15 5 H:11,227               & $\F_8$    \\ \hline
\multirow{3}{*}{$F_5$}    & $(1,0),(0,1)$                                                                          & $\set{3,3,2,0,0,(0),0}$                                                                                     & M:89 12 N:13 8 H:10,76                & $\F_4$    \\ \cline{2-5} 
                          & $(0,-1),(-1,0)$                                                                        & $\set{2,3,3,0,0,(0),0}$                                                                                     & M:116 12 N:13 8 H:9,93                & $\F_4$    \\ \cline{2-5} 
                          & $(-1,-1)$                                                                              & $\set{2,3,2,1,0,(0),0}$                                                                                     & M:169 13 N:17 9 H:11,131              & $\F_6$    \\ \hline
\multirow{4}{*}{$F_6$}    & $(1,0)$                                                                                & $\set{4,3,1,0,0,(0),0}$                                                                                     & M:62 9 N:13 7 H:11,59                 & $\F_4$    \\ \cline{2-5} 
                          & $(0,1)$                                                                                & $\set{2,3,2,1,0,(0),0}$                                                                                     & M:169 9 N:19 7 H:14,130               & $\F_6$    \\ \cline{2-5} 
                          & $(-1,1)$                                                                               & $\set{2,2,2,1,1,(0),0}$                                                                                     & M:310 9 N:19 7 H:11,227               & $\F_8$    \\ \cline{2-5} 
                          & $(-1,-1)$                                                                              & $\set{3,2,2,1,0,(0),0}$                                                                                     & M:158 9 N:16 7 H:11,131               & $\F_6$    \\ \hline
$F_7$                     & \begin{tabular}[c]{@{}c@{}}$(1,1),(1,0),(0,1),$\\ $(0,-1),(-1,0),(-1,-1)$\end{tabular} & $\set{2,3,2,0,0,(0),0}$                                                                                     & M:88 16 N:15 10 H:10,76$^{[3]}$       & $\F_4$    \\ \hline
\multirow{4}{*}{$F_8$}    & $(1,1)$                                                                                & $\set{2,2,3,0,0,(0),0}$                                                                                     & M:106 8 N:14 6 H:9,93                 & $\F_8$    \\ \cline{2-5} 
                          & $(1,0)$                                                                                & $\set{3,3,1,0,0,(0),0}$                                                                                     & M:61 9 N:16 7 H:12,58                 & $\F_4$    \\ \cline{2-5} 
                          & $(-1,1)$                                                                               & $\set{2,1,2,1,1,(0),0}$                                                                                     & M:296 9 N:22 7 H:11,227               & $\F_8$    \\ \cline{2-5} 
                          & $(-1,-1)$                                                                              & $\set{3,2,2,0,0,(0),0}$                                                                                     & M:79 8 N:14 6 H:10,76                 & $\F_4$    \\ \hline
\multirow{3}{*}{$F_9$}    & $(1,0)$                                                                                & $\set{3,3,1,0,0,(0),0}$                                                                                     & M:61 13 N:15 9 H:11,59                & $\F_4$    \\ \cline{2-5} 
                          & $(0,1),(0,-1)$                                                                         & $\set{2,3,2,0,0,(0),0}$                                                                                     & M:88 12 N:16 8 H:11,75$^{[2]}$        & $\F_4$    \\ \cline{2-5} 
                          & $(-1,1),(-1,-1)$                                                                       & $\set{2,2,2,1,0,(0),0}$                                                                                     & M:144 14 N:19 10 H:12,120             & $\F_6$    \\ \hline
\multirow{3}{*}{$F_{10}$} & $(1,0)$                                                                                & $\set{4,2,1,0,0,(0),0}$                                                                                     & M:63 5 N:11 5 H:11,59                 & $\F_4$    \\ \cline{2-5} 
                          & $(0,1)$                                                                                & $\set{3,2,1,1,0,(0),0}$                                                                                     & M:125 5 N:17 5 H:11,131               & $\F_6$    \\ \cline{2-5} 
                          & $(-3,-2)$                                                                              & $\set{2,1,1,1,1,(0),1}$                                                                                     & M:680 5 N:26 5 H:11,491               & $\F_{12}$ \\ \hline
\multirow{4}{*}{$F_{11}$} & $(1,0)$                                                                                & $\set{3,2,1,0,0,(0),0}$                                                                                     & M:51 9 N:16 7 H:11,59                 & $\F_4$    \\ \cline{2-5} 
                          & $(-1,2)$                                                                               & $\set{2,1,1,1,1,(0),0}$                                                                                     & M:257 9 N:24 7 H:11,227               & $\F_8$    \\ \cline{2-5} 
                          & $(0,-1)$                                                                               & $\set{2,3,1,0,0,(0),0}$                                                                                     & M:60 9 N:20 7 H:15,57$^{[2]}$         & $\F_4$    \\ \cline{2-5} 
                          & $(-1,-1)$                                                                              & $\set{2,2,1,1,0,(0),0}$                                                                                     & M:124 9 N:20 7 H:11,131               & $\F_6$    \\ \hline
\multirow{3}{*}{$F_{12}$} & $(1,0),(0,1)$                                                                          & $\set{2,3,1,0,0,(0),0}$                                                                                     & M:60 13 N:18 9 H:12,58                & $\F_4$    \\ \cline{2-5} 
                          & $(1,-1),(-1,1)$                                                                        & $\set{2,2,2,0,0,(0),0}$                                                                                     & M:78 12 N:16 8 H:10,76                & $\F_4$    \\ \cline{2-5} 
                          & $(-1,-1)$                                                                              & $\set{2,1,2,1,0,(0),0}$                                                                                     & M:145 13 N:21 9 H:11,131              & $\F_6$    \\ \hline
\multirow{2}{*}{$F_{13}$} & $(1,0)$                                                                                & $\set{3,1,1,0,0,(0),0}$                                                                                     & M:60 13 N:18 9 H:12,58                & $\F_4$    \\ \cline{2-5} 
                          & $(-1,2),(-1,-2)$                                                                       & $\set{2,1,1,0,1,(0),0}$                                                                                     & M:181 5 N:25 5 H:11,227               & $\F_8$    \\ \hline
\multirow{2}{*}{$F_{14}$} & $(2,-1),(-1,2)$                                                                        & $\set{2,1,1,1,0,(0),0}$                                                                                         & M:112 9 N:22 7 H:11,131               & $\F_6$    \\ \cline{2-5} 
                          & $(0,-1),(-1,0)$                                                                        & $\set{2,2,1,0,0,(0),0}$                                                                                         & M:50 9 N:19 7 H:12,58                 & $\F_4$    \\ \hline
$F_{15}$                  & $(-1,1),(1,1),(-1,-1),(1,-1)$                                                          & $\set{2,1,2,0,0,(0),0}$                                                                                         & M:68 8 N:17 6 H:10,76                 & $\F_4$    \\ \hline
$F_{16}$                  & $(2,-1),(-1,2),(-1,-1)$                                                                & $\set{2,1,0,1,0,(0),0}$                                                                                         & M:79 5 N:23 5 H:11,131                & $\F_6$    \\ \hline
\end{tabular}
\caption{\footnotesize Line bundles in the $v_s$ stacking $F$-fibered construction, with examples over Hirzebruch
  surfaces $\F_m$, where $-m$ saturates the negative curve bound in each case.}
\label{models}
\end{table}

\section{Automorphism symmetries and fibrations}
\label{automorphisms}
\subsection{Polytopes with non-trivial fibration orbits in the regions
  $h^{1,1}, h^{2,1}\geq 140$}
\label{automorphisms1}
The following table indicates the difference between the total number
of fibrations and the number of inequivalent fibration classes under
automorphisms in the relevant 16 cases.

\begin{center}
\begin{tabular}{|c|cccccccccccccccc|}
\hline
\multicolumn{1}{|l|}{\begin{tabular}[c]{@{}l@{}}polytope data (in the format\\  of the KS database)\end{tabular}} & \multicolumn{16}{l|}{\begin{tabular}[c]{@{}l@{}}\# fibrations for each of the 16 fibers\\  \# fibrations modulo the automorphism symmetry group\end{tabular}}  \\\hline
\multirow{2}{*}{M:12 5 N:348 5 H:251,5 [[492]]} & 0 & 0 & 0  & 0  & 0 & 0 & 0 & 0 & 0 & 3  & 0 & 0 & 1  & 0 & 0 & 0 \\
                    & 0 & 0 & 0  & 0  & 0 & 0 & 0 & 0 & 0 & 2  & 0 & 0 & 1  & 0 & 0 & 0 \\\hline
\multirow{2}{*}{ M:15 5 N:179 5 H:151,7 [[288]]}   & 0 & 0 & 0  & 2  & 0 & 0 & 0 & 0 & 0 & 1  & 0 & 0 & 1  & 0 & 0 & 0 \\
                    & 0 & 0 & 0  & 1  & 0 & 0 & 0 & 0 & 0 & 1  & 0 & 0 & 1  & 0 & 0 & 0 \\\hline
\multirow{2}{*}{M:14 5 N:196 5 H:161,5 [[312]]}   & 0 & 0 & 0  & 2  & 0 & 0 & 0 & 0 & 0 & 2  & 0 & 0 & 1  & 0 & 0 & 0 \\
                    & 0 & 0 & 0  & 1  & 0 & 0 & 0 & 0 & 0 & 2  & 0 & 0 & 1  & 0 & 0 & 0 \\\hline
\multirow{2}{*}{M:15 5 N:311 5 H:227,11 [[432]]}   & 0 & 0 & 0  & 0  & 0 & 0 & 0 & 0 & 0 & 2  & 0 & 0 & 3  & 0 & 0 & 0 \\
                    & 0 & 0 & 0  & 0  & 0 & 0 & 0 & 0 & 0 & 1  & 0 & 0 & 3  & 0 & 0 & 0 \\\hline
\multirow{2}{*}{M:17 5 N:177 5 H:151,7 [[288]]}   & 0 & 0 & 0  & 2  & 0 & 0 & 0 & 0 & 0 & 3  & 0 & 0 & 0  & 0 & 0 & 0 \\
                    & 0 & 0 & 0  & 2  & 0 & 0 & 0 & 0 & 0 & 2  & 0 & 0 & 0  & 0 & 0 & 0 \\\hline
\multirow{2}{*}{M:11 5 N:335 5 H:243,3 [[480]]}   & 0 & 0 & 0  & 0  & 0 & 0 & 0 & 0 & 0 & 3  & 0 & 0 & 3  & 0 & 0 & 0 \\
                    & 0 & 0 & 0  & 0  & 0 & 0 & 0 & 0 & 0 & 2  & 0 & 0 & 3  & 0 & 0 & 0 \\\hline
\multirow{2}{*}{ M:13 5 N:117 5 H:148,4 [[288]]}   & 0 & 0 & 0  & 1  & 0 & 0 & 0 & 0 & 0 & 2  & 0 & 0 & 3  & 0 & 0 & 0 \\
                    & 0 & 0 & 0  & 1  & 0 & 0 & 0 & 0 & 0 & 1  & 0 & 0 & 3  & 0 & 0 & 0 \\\hline
\multirow{2}{*}{M:13 5 N:267 5 H:208,4 [[408]]}   & 0 & 0 & 0  & 2  & 0 & 0 & 0 & 0 & 0 & 3  & 0 & 0 & 1  & 0 & 0 & 0 \\
                    & 0 & 0 & 0  & 1  & 0 & 0 & 0 & 0 & 0 & 2  & 0 & 0 & 1  & 0 & 0 & 0 \\\hline
\multirow{2}{*}{M:10 5 N:376 5 H:272,2 [[540]]}   & 0 & 0 & 0  & 0  & 0 & 0 & 0 & 0 & 0 & 4  & 0 & 0 & 3  & 0 & 0 & 1 \\
                    & 0 & 0 & 0  & 0  & 0 & 0 & 0 & 0 & 0 & 2  & 0 & 0 & 1  & 0 & 0 & 1 \\\hline
\multirow{2}{*}{M:12 5 N:131 5 H:165,3 [[324]]}   & 0 & 0 & 0  & 1  & 0 & 0 & 0 & 0 & 0 & 3  & 0 & 0 & 3  & 0 & 0 & 1 \\
                    & 0 & 0 & 0  & 1  & 0 & 0 & 0 & 0 & 0 & 1  & 0 & 0 & 1  & 0 & 0 & 1 \\\hline
\multirow{2}{*}{M:11 5 N:225 5 H:164,8 [[312]]}   & 0 & 0 & 0  & 2  & 0 & 0 & 0 & 0 & 0 & 5  & 0 & 0 & 3  & 0 & 0 & 0 \\
                    & 0 & 0 & 0  & 1  & 0 & 0 & 0 & 0 & 0 & 4  & 0 & 0 & 3  & 0 & 0 & 0 \\\hline
\multirow{2}{*}{M:10 5 N:196 5 H:143,7 [[272]]}   & 0 & 0 & 0  & 6  & 0 & 0 & 0 & 0 & 0 & 2  & 0 & 0 & 4  & 0 & 1 & 0 \\
                    & 0 & 0 & 0  & 3  & 0 & 0 & 0 & 0 & 0 & 1  & 0 & 0 & 3  & 0 & 1 & 0 \\\hline
\multirow{2}{*}{M:9 5 N:201 5 H:148,4 [[288]]}   & 0 & 0 & 0  & 2  & 0 & 0 & 0 & 0 & 0 & 10 & 0 & 0 & 6  & 0 & 0 & 0 \\
                    & 0 & 0 & 0  & 1  & 0 & 0 & 0 & 0 & 0 & 6  & 0 & 0 & 6  & 0 & 0 & 0 \\\hline
\multirow{2}{*}{M:8 5 N:225 5 H:165,3 [[324]]}   & 0 & 0 & 0  & 0  & 0 & 0 & 0 & 0 & 0 & 15 & 0 & 0 & 7  & 0 & 0 & 1 \\
                    & 0 & 0 & 0  & 0  & 0 & 0 & 0 & 0 & 0 & 4  & 0 & 0 & 3  & 0 & 0 & 1 \\\hline
\multirow{2}{*}{M:7 5 N:196 5 H:145,1 [[288]]}   & 0 & 0 & 0  & 6  & 0 & 6 & 0 & 0 & 0 & 12 & 0 & 0 & 9  & 3 & 0 & 1 \\
                    & 0 & 0 & 0  & 2  & 0 & 1 & 0 & 0 & 0 & 3  & 0 & 0 & 3  & 1 & 0 & 1 \\\hline
\multirow{2}{*}{M:7 5 N:201 5 H:149,1 [[296]]}   & 0 & 0 & 12 & 12 & 0 & 0 & 0 & 0 & 0 & 12 & 0 & 0 & 15 & 0 & 3 & 4 \\
                    & 0 & 0 & 1  & 1  & 0 & 0 & 0 & 0 & 0 & 1  & 0 & 0 & 3  & 0 & 1 & 1\\\hline\end{tabular}
\end{center}

\subsection{An example: the automorphism group of M:7 5 N:201 5 H:149,1 [[296]]}
\label{automorphisms2}
We give the details of the symmetry and fibration
structure for the polytope associated with the Calabi-Yau having Hodge
numbers (149, 1).  This is the polytope with the largest number of
 fibrations (including multiplicities in orbits of
automorphism symmetries).

The polytope $\dd$ in question has five vertices:
\begin{align}
 & A =(1, -1, -1, -1)\\
 & B =(-1, -1, -1, -1)\\
 & C =(-1, -1, -1, 7)\\
 & D =(-1, -1, 7, -1)\\
 & E =(-1, 7, -1, -1) \,.
\end{align}
  These vertices satisfy the linear condition
\begin{equation}
4A + B + C + D + E = 0 \,.
\label{eq:}
\end{equation}
The possible symmetries allowed by this equation include all
permutations on the vertices $B, C, D, E$.  The polytope is clearly
symmetric under all permutations on $C, D, E$, as these can be
realized by permutations on the axes 2, 3 and 4.  One can also check
that the polytope is symmetric under the linear transformation that
swaps $B$ and $C$ while leaving $D$ and $E$ fixed,
\begin{equation}
T =\left(\begin{array}{cccc}
1 & 0 & 0 & -4\\
0 & 1 & 0 & -1\\
0 & 0 & 1 & -1\\
0 & 0 & 0 & -1
\end{array} \right)\,.
\label{eq:}
\end{equation}
This matrix in $SL (2,\Z)$ satisfies  (acting on the right on row vectors)
\begin{equation}
B \cdot T = C,
C \cdot T = B,
A \cdot T = A,
D \cdot  T = D,
E \cdot T = E   \,,
\label{eq:}
\end{equation}
and is thus a symmetry of the polytope.  This shows that all 24
permutations on $B, C, D, E$ are symmetries.

Explicitly, let the column vectors $a, b, c, d, e$ be defined as
\begin{align}
 & a =(1, 0, 0, 0)^{\rm T}\\
 & b =(0, 1, 0, 0)^{\rm T}\\
 & c =(0, 0, 1, 0)^{\rm T}\\
 & d =(0, 0, 0, 1)^{\rm T}\\
 & e =(-4, -1, -1, -1)^{\rm T} \,,
\end{align}
which are the five vertices of the $\Delta$ polytope. The 24 linear
transformation matrices that leave the $\nabla$ polytope invariant are
\begin{eqnarray}\nonumber
\set{(a,b,c,d),(a,b,c,e),(a,b,e,d),(a,b,d,e),(a,b,e,c),(a,b,d,c),(a,c,b,d),(a,c,b,e),\\\nonumber(a,e,b,d),(a,d,b,e),(a,e,b,c),(a,d,b,c),(a,c,e,d),(a,c,d,e),(a,e,c,d),(a,d,c,e),\\\nonumber(a,e,d,c),(a,d,e,c),(a,c,e,b),(a,c,d,b),(a,e,c,b),(a,d,c,b),(a,e,d,b),(a,d,e,b)}.
\end{eqnarray}
The different fibrations go into orbits of this 24-element symmetry
group.  For example, there are 12 $F_3$ fibers; one of them is $\set{(0,-1,-1,0),(0,-1,1,-1),(0,0,2,-1),(0,1,-1,1)}$, and all the 12 fibers are
generated by 
\begin{eqnarray}\nonumber
\set{(a,b,c,d), (a,b,c,e),(a,c,b,d),(a,c,b,e),(a,d,b,c),(a,e,b,c),(a,b,d,c),(a,b,e,c),\\\nonumber(a,b,e,d),(a,b,d,e),(a,d,b,e),(a,e,b,d)}.
\end{eqnarray}

\end{document}